\documentclass[zpreprint,zbstdraft]{./LaTeX/zeus/zeus_paper}
\usepackage[english]{babel}
\chardef\usc=95
\chardef\til=126
\catcode`\@=11 
\DeclareRobustCommand\xdotspace{\futurelet\@let@token\@xdotspace}
\def\@xdotspace{%
  \ifx\@let@token.\else
  \ifx\@let@token\bgroup.\else
  \ifx\@let@token\egroup.\else
  \ifx\@let@token\/.\else
  \ifx\@let@token\ .\else
  \ifx\@let@token~.\else
  \ifx\@let@token!.\else
  \ifx\@let@token,.\else
  \ifx\@let@token:.\else
  \ifx\@let@token;.\else
  \ifx\@let@token?.\else
  \ifx\@let@token/.\else
  \ifx\@let@token'.\else
  \ifx\@let@token).\else
  \ifx\@let@token-.\else
  \ifx\@let@token\@xobeysp.\else
  \ifx\@let@token\space.\else
  \ifx\@let@token\@sptoken.\else
   .\space
   \fi\fi\fi\fi\fi\fi\fi\fi\fi\fi\fi\fi\fi\fi\fi\fi\fi\fi}
\catcode`\@=12 

\newcommand{\stru}[2]{%
   \relax\ifmmode\hbox{\vrule height#1 depth#2 width0pt}%
   \else\vrule height#1 depth#2 width0pt\fi}

\newcommand{\Ronum}[1]{\uppercase\expandafter{\romannumeral#1}}
\newcommand{\ronum}[1]{\expandafter{\romannumeral#1}}
\DeclareRobustCommand{\LaTeXZ}{%
  \LaTeX\kern-.05em4\kern-.1em
  {\raisebox{-0.2ex}{$\scriptstyle\text{ZEUS}$}}\xspace}

\DeclareMathAlphabet{\mathbf}{OT1}{cmr}{bx}{sl}
\newcommand{\eVdist}{\kern-0.06667em}

\newcommand{\Gev}{{\text{Ge}\eVdist\text{V\/}}}

\newcommand{\slashfrac}[2]{%
  \raisebox{0.5ex}{\ensuremath #1}\kern-0.12em/\kern-0.08em
  \raisebox{-.8ex}{\ensuremath #2}}

\newcommand{\sqr}[3]{%
    {\vcenter{\hrule height.#3ex\hbox{\vrule width.#2ex height#1ex
     \kern#1ex\vrule width.#3ex}\hrule height.#2ex}}}

\catcode`\@=11 
\newcommand{\parenbar}{\mathpalette\p@renb@r}
\def\p@renb@r#1#2{\vbox{%
  \ifx#1\scriptscriptstyle \dimen@.7em\dimen@ii.2em\else
  \ifx#1\scriptstyle \dimen@.8em\dimen@ii.25em\else
  \dimen@1em\dimen@ii.4em\fi\fi \offinterlineskip
  \ialign{\hfill##\hfill\cr
    \vbox{\hrule width\dimen@ii}\cr
    \noalign{\vskip-.3ex}%
    \hbox to\dimen@{$\mathchar300\hfil\mathchar301$}\cr
    \noalign{\vskip-.3ex}%
    $#1#2$\cr}}}
\catcode`\@=12 

\newcommand{\IP}{{\rm I$\kern-0.01667em$P}\xspace}

\newcommand{\F}{{\cal F}}

\mathchardef\qsm=63
\mathchardef\pls=43
\mathchardef\mns=512
\mathchardef\plm=518
\mathchardef\eql=61
\mathchardef\smallleft=300
\mathchardef\smallright=301
\mathchardef\les=316
\mathchardef\gre=318
\mathchardef\leq=532
\mathchardef\grq=533

\catcode`\@=11 
\newcounter{pict@width}
\newcounter{pict@height}
\newlength{\pict@scale}
\newcommand{\psfigadd}[4]{%
\setcounter{pict@width}{1*\ratio{#2+\pict@scale/2}{\pict@scale}}
\setcounter{pict@height}{1*\ratio{#3+\pict@scale/2}{\pict@scale}}
\setlength{\unitlength}{\pict@scale}
\hbox to #2{\hspace{-\fill}\begin{picture}(\thepict@width,\thepict@height)
\put(0,0){\psfig{figure=#1,width=#2,height=#3,clip=}}
\SetScale{0.283466457}
\SetWidth{1.763889}
{#4}
\end{picture}}
}
\newcounter{pict@widthfst}
\newcounter{pict@widthscd}
\newcounter{pict@widthtot}
\newcommand{\psfigaddtwo}[7]{%
\setcounter{pict@widthfst}{1*\ratio{#2+\pict@scale/2}{\pict@scale}}
\setcounter{pict@widthscd}{1*\ratio{#2+#4+\pict@scale/2}{\pict@scale}}
\setcounter{pict@widthtot}{1*\ratio{#2+#4+#6+\pict@scale/2}{\pict@scale}}
\setcounter{pict@height}{1*\ratio{#3+\pict@scale/2}{\pict@scale}}
\setlength{\unitlength}{\pict@scale}
\hbox{\hspace{-\fill}\begin{picture}(\thepict@widthtot,\thepict@height)
\put(0,0){\psfig{figure=#1,width=#2,height=#3,clip=}}
\put(\thepict@widthscd,0){\psfig{figure=#5,width=#6,height=#3,clip=}}
\SetScale{0.283466457}
\SetWidth{1.763889}
{#7}
\end{picture}}
}
\newcommand{\psfigror}[4]{%
\setcounter{pict@width}{1*\ratio{#2+\pict@scale/2}{\pict@scale}}
\setcounter{pict@height}{1*\ratio{#3+\pict@scale/2}{\pict@scale}}
\setlength{\unitlength}{\pict@scale}
\hbox{\begin{picture}(\thepict@width,\thepict@height)
\put(0,\thepict@height){\psfig{figure=#1,width=#3,height=#2,clip=,angle=270}}
\SetScale{0.283466457}
\SetWidth{1.763889}
{#4}
\end{picture}}
}
\newcommand{\psfigrol}[4]{%
\setcounter{pict@width}{1*\ratio{#2+\pict@scale/2}{\pict@scale}}
\setcounter{pict@height}{1*\ratio{#3+\pict@scale/2}{\pict@scale}}
\setlength{\unitlength}{\pict@scale}
\hbox{\begin{picture}(\thepict@width,\thepict@height)
\put(0,0){\psfig{figure=#1,width=#3,height=#2,clip=,angle=90}}
\SetScale{0.283466457}
\SetWidth{1.763889}
{#4}
\end{picture}}
}
\catcode`\@=12 
\newlength\listtextwidth

\catcode`\@=11 
\newlength{\@tabfninsert}
\newlength{\@tabfnwidth}
\newcommand{\tabfootnote}[2]{%
  \setlength{\@tabfninsert}{0.8em}
  \setlength{\@tabfnwidth}{\textwidth}
  \addtolength{\@tabfnwidth}{-\@tabfninsert}
  \addtolength{\@tabfnwidth}{-0.4em}
  \noindent\makebox[\@tabfninsert][r]{\footnotesize$^{#1}$\hfil}\hfill%
  \parbox[t]{\@tabfnwidth}{\footnotesize #2\hfill}}
\catcode`\@=12 

\newcommand{\newc}{\newcommand}

\newc{\ra}{\rightarrow}
\newc{\Ra}{\Rightarrow}

\newc{\pom}  {I\hspace{-0.2em}P}

\newc{\geva}{{\rm GeV}}

\newc{\rpv}{{\not \!\! R_p}}
\newc{\rpvm}{{\not  R_p}}

\newc{\gsim}{{\stackrel{>}{\sim}}}
\newc{\lsim}{{\stackrel{<}{\sim}}}
\newc{\sleq} {\raisebox{-.6ex}{${\textstyle\stackrel{<}{\sim}}$}}
\newc{\sgeq} {\raisebox{-.6ex}{${\textstyle\stackrel{>}{\sim}}$}}

\def\3{\ss}

\newc{\ETJ}{E^{{\rm jet}}_T}

\def\ET{E_{\rm T}}

\def\xgo{x_\gamma^{\rm obs}}

\def\xg{x_\gamma}
\def\xp{x_p}

\def\ptmin{\hat{p}_T^{\rm min}}

\def\pt{p_T}
\def\Et{E_T}

\def\kt{K_\perp~}

\def\ee{e^+ e^-}

\def\etgap{E_T^{\rm gap}}

\def\gp{\gamma p}
\def\q2{{\rm Q}^2}
\def\p2{{\rm P}^2}
\def\gev2{{\rm GeV}^{2}}
\def\f2c{F_2^{\rm charm}}
\def\f2{F_2}
\def\F2g{F_2^\gamma}

\def\lqcd{\Lambda_{\rm QCD}}

\def\d0{D^{0}}

\def\begr{\begin{flushright}}
\def\endr{\end{flushright}}

\def\begl{\begin{flushleft}}
\def\endl{\end{flushleft}}

\def\as{\alpha_s}

\def\herwig{{\sc Herwig}}
\def\pythia{{\sc Pythia}}

\begin{document}



\title{
High Energy Photoproduction
}
\author{J.M. Butterworth and M. Wing\\
Department of Physics and Astronomy\\
University College London\\
Gower St. London WC1E 6BT\\
England.}
%
\abstract{ The experimental and phenomenological status of high energy
photoproduction is reviewed. Topics covered include the structure of
the photon, production of jets, heavy flavours and prompt photons,
rapidity gaps, energy flow and underlying events. The results are
placed in the context of the current understanding of QCD, with
particular application to present and future hadron and lepton
colliders.  } \makezeustitle



\section{Introduction}
\label{sec:intro}

The photon is one of the earliest and most influential of the concepts
contributing to the so-called ``Standard Model'' which summarizes the
current understanding of particle physics. The debate as to whether
light has a wave- or particle-like nature was a key issue in the
physics of more than one century, and with both Newton and Einstein
numbered amongst the participants. The answer, within quantum
mechanics, is ``both!'', since quanta of light (as well as all other
quanta, including those usually thought of as particles) may exhibit
wave- or particle-like properties depending upon the experiments being
carried out. These quanta of light are, of course, photons.

In Quantum Electrodynamics (QED), photons arise from the breaking of
U(1) gauge symmetry and are responsible for electromagnetic
interactions. In the generalisation to the electroweak theory embedded
in the Standard Model, they still mediate the electromagnetic
interaction, but are now a linear combination of the neutral bosons
arising from the breaking of the U(1) hypercharge and SU(2) isospin
symmetries, and acquire massive companions in the $Z$ and $W$ bosons.

If wave/particle and electroweak double identities were not enough for
a single boson, yet another manifestation of the photon becomes
dominant in high energy photoproduction. In many aspects, the photon
appears very much like a hadron. That is, its interaction cross-sections 
behave (apart from a normalisation factor) very much like
hadronic cross-sections, and at the highest energies the photon even
appears to ``contain'' quarks and gluons, just as the proton or, more
specifically, a vector meson, does. Because of this, photoproduction
experiments have played an important role in the development of the
strongly interacting sector of the Standard Model, Quantum
Chromodynamics (QCD). This review thus contains much about QCD which
touches on general hadronic physics.

High energy photon beams are generally generated by high energy
electron or positron beams. High energy photoproduction was taken into
a new regime by the HERA electron-proton collider, which, as the first
lepton-hadron collider, dramatically extended the energy reach for
photon-hadron as well as electron-hadron interactions.

Since the photon always carries less energy than the parent lepton
beam, photoproduction experiments are rarely, if ever, at the energy
frontier in terms of searches for new physics. Nevertheless, the large
cross-sections and unique kinematics do mean that in several areas
they still set the best limits on physics beyond the standard
model. These cases are discussed in Section~\ref{sec:searches}.

Also because of the symbiotic relation with electron beams,
photon-photon collisions are an unavoidable feature of
electron-positron colliders, both past and future. Providing
information on the one hand on the general features of hadronic
collisions, and on the other on the interactions of the photon,
photoproduction data are useful, and sometimes critical, for
experimenters at the next generation of particle colliders,
specifically the Large Hadron Collider at CERN and the International
Linear electron-positron Collider currently under development.  These
connections, as well as those with cosmic-ray physics, are also
explored herein.

\section{Phenomenology of photoproduction}

In this chapter we introduce the language and concepts needed to
discuss high-energy photoproduction in term of the Standard Model (SM)
of particle physics. Within the SM, the interactions of the photon
should be describable using the language of fundamental fermions and
gauge bosons. However, as described below, the strong interaction
(QCD) plays an important part in photoproduction, and in QCD
perturbative calculations are only possible in the region of
asymptotic freedom, that is when a ``hard'' (short distance, high
energy) scale is present in the problem. This is not always, or even
usually, the case in high energy photoproduction, and thus other
phenomenological ideas must also be used.

\subsection{Total $\gp$ cross-section and ``elastic'' photoproduction}
\label{sec:totalxsec}

Photon-proton cross-sections at centre-of-mass energies above a few
GeV have for many years been related to hadron-hadron cross-sections
using the Vector Dominance model
(VDM)~\cite{anphy:11:1,pr:124:953}. The physics behind this is that
since the photon couples to charged-fermion pairs, a part of its
cross-section may be described by considering the interaction between
a hadron and a virtual fermion pair. This part of the cross-section
will on the face of it be suppressed by an additional coupling, since
if considered perturbatively it is a higher order correction. For
fermions which do not interact strongly this suppression is such that
they do not greatly affect the total cross-section. However, for a
$\gamma \ra q\bar{q}$ splitting, the interactions between the quarks
and a hadronic target are strong, the perturbative language does not
work, and in fact this configuration in general dominates the total
cross-section. A good picture of this is that the virtual $q\bar{q}$
pair radiates gluons and becomes a virtual, quasi-hadronic object with
the same quantum numbers as the photon. A vector meson is very close
to being such an object.

\begin{figure}[htp]
\begin{center}
~\epsfig{file=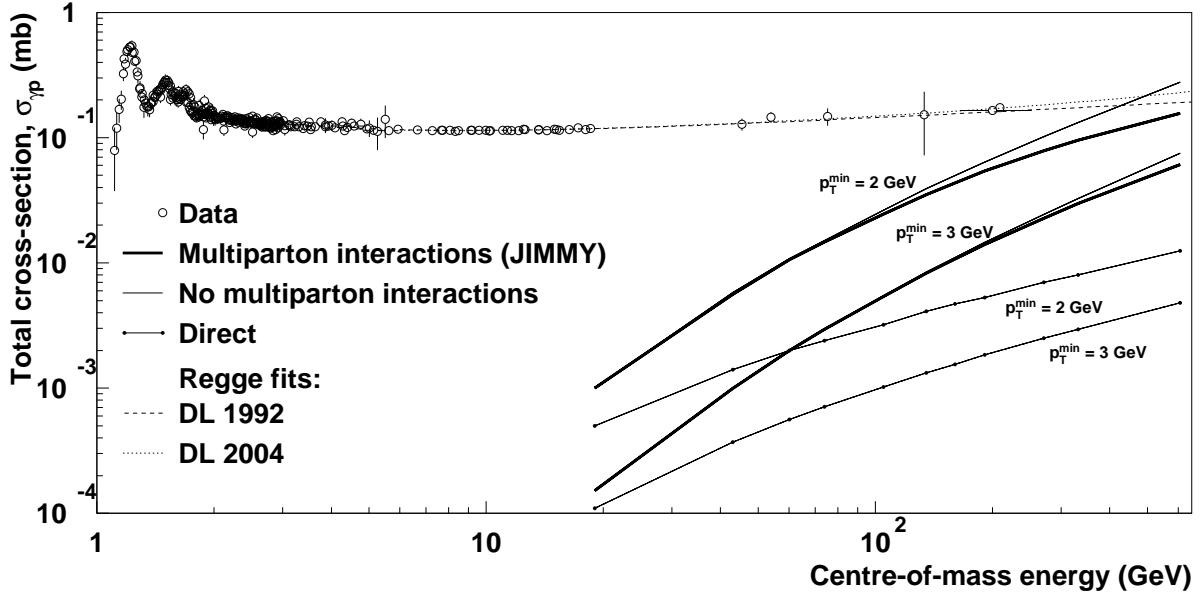,height=8cm} 
\caption[*]{\label{fig:gpxsec} The total $\gamma p$ 
cross-section~\cite{slac-pub-0653,*pr:d5:1640,*pl:b27:474,*prl:23:1187,*prl:23:1191,*pl:b33:189,*prl:39:737,*pr:d5:545,*pr:d7:1362,*np:b68:1,*pr:d8:1277,*prl:40:1222,*pam:66:565,*zfp:c69:27,*np:b627:3}
compared to the predictions of various models. The Regge fits shown
are those of Donnachie-Landshoff (1992~\cite{pl:b296:227} and
2004~\cite{Donnachie:2004pi}).}
\end{center}
\end{figure}

This picture is borne out in the data, shown in Fig.~\ref{fig:gpxsec}.
Hadronic cross-sections may be fitted to a high precision in Regge
phenomenology, in which the dominant contributions arise from Reggeon
exchange, which falls with increasing centre-of-mass energy, and
Pomeron exchange, which rises. Thus in this language, the high energy
photoproduction cross-section is dominated by Pomeron exchange. In
this model, the intercept of the Pomeron and Reggeon, which determine
the energy dependence of the cross-section, are the same as those
determined from hadron-hadron interactions. As an example, two such
fits from Donnachie and Landshoff are shown in the figure. The 1992
fit predates HERA data and is made over the range $6 < \sqrt{s}
< 20~\geva$. It has a single Pomeron term and a single Reggeon
term. The 2004 fit includes HERA data (photoproduction, as well as DIS
data for $\q2 < 45~\gev2$) and allows an additional second Pomeron
term. The difference in this figure is small. The other curves in the
figure are discussed in Section~\ref{sec:pqcd} below.

Another application of the VDM is seen in the exclusive
photoproduction of vector mesons, $\gamma p \ra V p$. In the language
of the VDM, this is a quasi-elastic process in which the photon
fluctuates into a virtual vector meson state which is then moved on
mass-shell by the collision with the proton. 

\begin{figure}[htp]
\begin{center}
~\epsfig{file=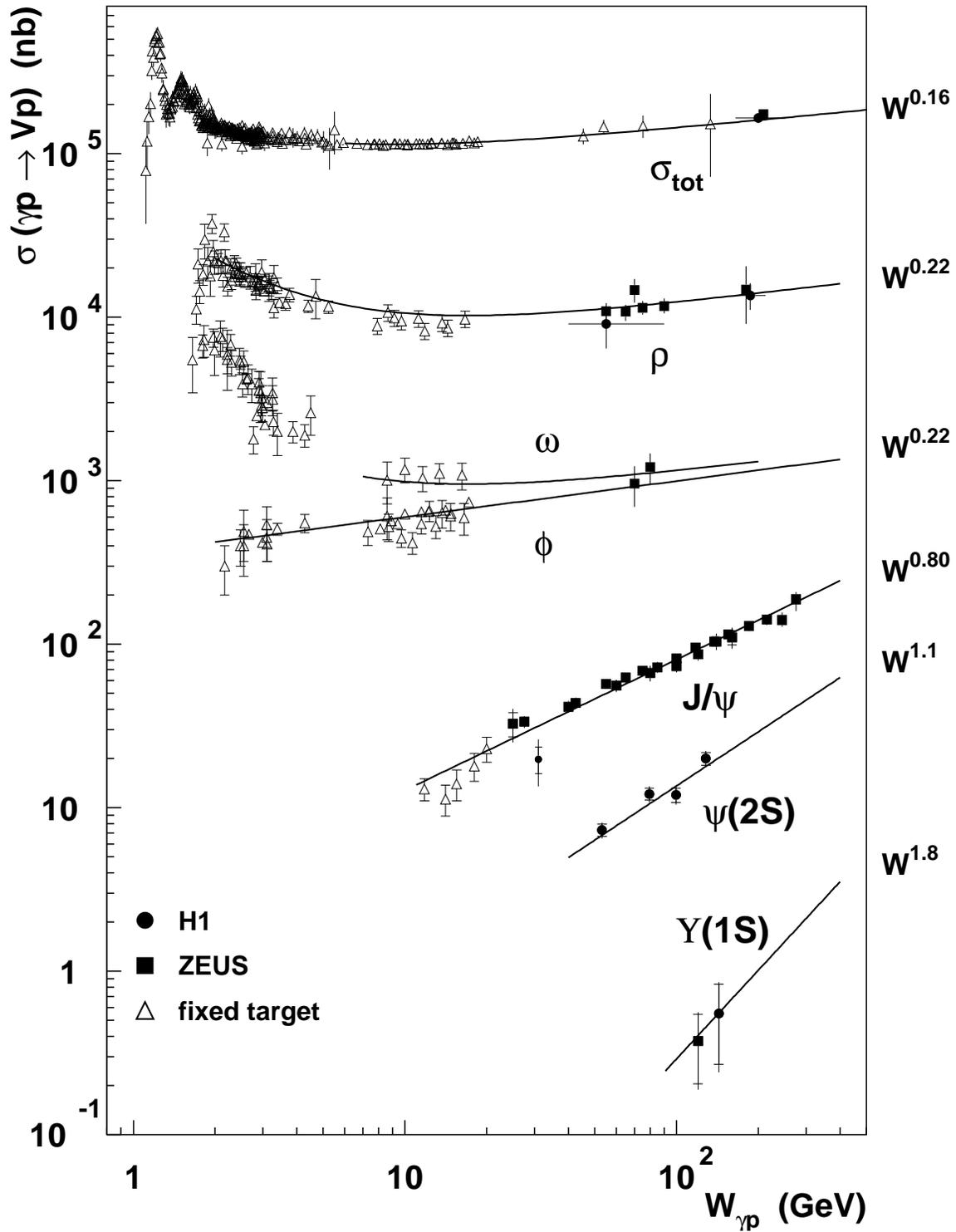,height=20cm} 
\caption[*]{\label{fig:vmw} The total $\gamma p$ cross-section
compared to the cross-sections~\cite{cern-hera-87-01,*np:b463:3,*zfp:c63:391,*zfp:c69:39,*zfp:c73:73,*epj:c2:247,*pr:d40:1,*pl:b377:259,*prl:48:73,*prl:52:795,*epj:c24:345,*pl:b541:251,*pl:b437:432,*pl:b483:23} for exclusive vector meson
production~\cite{levy:proc,*misc:meyer:private}.}
\end{center}
\end{figure}

The photoproduction cross-section for several vector meson states is
shown as a function of the photon-proton centre-of-mass energy, $W$,
in Fig.~\ref{fig:vmw}. The inclusive cross-section is also shown. All
the cross-sections rise with $W$. This is expected from Regge models
where the cross-section is driven by Pomeron exchange between the
vector meson and the proton. However, for high-mass vector mesons
($J/\psi, \Upsilon$) where a hard transverse scale is available due to
the vector meson mass, the cross-section rises more rapidly with
$W$. This is explained in perturbative
models~\cite{Ryskin:1992ui,*Brodsky:1994kf} where the cross-section is
partonic, and is dominated by gluon exchange. The strong rise is a
direct result of the rapid rise in the gluon density in the
proton~\cite{np:b407:515,*pl:b316:412}.

This process is amenable to analysis and measurement as a function of
several different energy scales (the vector meson mass, the photon
virtuality $\q2$, $W$) and is a sensitive probe of the transition
between soft and hard QCD. Information on the helicity structure of
the interaction may be gained by studying the angular distributions of
the decay products. The phenomenology of vector meson production in
both DIS and photoproduction has recently been reviewed
elsewhere~\cite{Ivanov:2005ax}.

Exclusive vector meson production is a particular case of a
diffractive process; that is a process where the proton loses a small
fraction of its energy and either remains intact, or dissociates into
a low-mass, forward-going, hadronic state. More inclusive diffractive 
cross-sections are discussed in Section~\ref{sec:eflow_rapidity_gaps}.

\subsection{Photoproduction and perturbative QCD}
\label{sec:pqcd}

Since there is good agreement between the VDM and the data for total
and low-mass-quasi-elastic cross-sections, it is clear that QCD
effects, {\it i.e.}  quarks and gluons, dominate the cross-section. 
Since the initial $\gamma \ra q\bar{q}$ splitting may in
general have very low virtuality, the interactions of the quarks will
not be entirely perturbative, even if a hard scale is present
somewhere in the process. This is a similar situation to that
pertaining in hadron-hadron collisions, and the method for dealing
with it is the same - the cross-section is factorised into a hard
partonic cross-section and a parton distribution function (PDF). The
soft (long distance) physics, as well as divergences associated with
collinear emission of quarks and gluons, is collected in an effective
photon structure, analogous to the parton distributions within the
proton\footnote{for a recent review of the theory of hard
photoproduction, see~\cite{Klasen:2002xb}}. A general schematic
formula for perturbative QCD calculations of photoproduction processes
may therefore be written;

\begin{equation}
 d\sigma_{\gamma p \rightarrow X } = 
 \displaystyle\sum_{a,b} \int_0^1
 dx_{\gamma} 
\displaystyle\int_0^1 dx_{p}
   f_{a/\gamma}(\xg,\mu_R,\mu_{F\gamma})\ 
 f_{b/p}(\xp,\mu_R,\mu_{Fp}) \;
   d\hat{\sigma}_{ab \rightarrow cd }(\xg,\xp,\mu_R),
\label{eq:pertxsec}
\end{equation}

where $\xg$ and $\xp$ are the longitudinal momentum fractions of the
parton $a$ in the photon and the parton $b$ in the proton,
respectively. The term $f_{a/\gamma}$ ($f_{b/p}$) represents the PDF
of partons with flavour $a$ ($b$) in the photon (proton). The
factorisation scale for the photon (proton) is denoted by
$\mu_{F\gamma}$ ($\mu_{Fp}$), and $\mu_R$ is the renormalisation
scale. The term $d\hat{\sigma}_{ab\rightarrow cd}$ is the hard
(partonic) cross-section. This perturbative cross-section diverges at
low values of the relative transverse momentum, $\pt$, of the scattered
partons ($c$ and $d$), and thus the whole expression is only valid for
$\pt$ above some minimum value $\ptmin$. This means that the cross-section 
in Eq.\ref{eq:pertxsec} is really only a part of the
photon-proton cross-section; that part in which a hard scale resolves
the parton structure of the proton and photon. Such processes are
known as \emph{resolved} processes. Soft scatters are not included.

Following on from the VDM, one might expect the parton densities
$f_{a/\gamma}(\xg,\mu_R,\mu_{F\gamma})$ to be essentially built out of
those of the vector mesons into which the photon may
fluctuate. Indeed, this is the starting point for most models of
photon structure. However, all such models must also take into account
the fact that for any evolution of the structure with $\mu_{F\gamma}$,
the $\gamma \rightarrow q\bar{q}$ splitting also plays a role.  This
term is of course absent in the vector meson case, and so gives rise
to an additional contribution to resolved processes. This is referred
to as the anomalous (or pointlike) component of the photon structure.

This component arises from the probability that the photon splits into
a $q\bar{q}$ pair at a scale too high for the approximation to a bound
vector meson state to work. Taking this further, the photon may in
fact not split into a $q\bar{q}$ pair at any scale below that of the
hard process being considered, and thus may enter directly into the
hard matrix element of the calculation.  To account for this
possibility, terms are included in Eq.\ref{eq:pertxsec} in which
parton $a$ is the photon, and
$f_{a/\gamma}(x_{\gamma},\mu_R,\mu_{F\gamma})$ is
$\delta(1-x_\gamma)$. Processes involving these terms are known a
direct photoproduction.

The separation between \emph{resolved} and \emph{direct} processes,
and the subdivision of resolved process into anomalous and VDM
components, has more to do with the limitations of our ability to
calculate QCD cross-sections than with fundamental physics. The
separations are not unique and depend upon the approximations being
employed in any given calculation. Nevertheless, as long as this is
borne in mind, the labels are useful tools for exploring the complex
world of photon physics.

In the above discussion, the transverse momentum of the scattered
partons is taken to define the hard scale. In general, this hard scale
can be provided by other aspects of the process, such as the mass of a
heavy quark. This is discussed further in Section~\ref{sec:hf}.

Two predictions using Eq.\ref{eq:pertxsec} are shown in
Fig.~\ref{fig:gpxsec}, for two different values of $\ptmin$ and using
the GRV LO~\cite{pr:d45:3986} PDF sets for the photon and the CTEQ5
LO~\cite{epj:c12:375} PDFs for the proton. The hard contribution to the total
cross-section depends very strongly on $\ptmin$. It is a small
fraction of the total for centre-of-mass energies below around
100~\geva, but rises quickly as the energy increases. The contribution
from direct processes is also shown separately.

\subsection{Eikonal models and the underlying event}
\label{sec:ue}

The effects of hadronic ``remnants'', {\it i.e.}  those parts of the
proton or resolved photon which do not participate in the hard
scatter, can be significant. Further interactions between the remnant
may lead to activity in the final state which can disrupt the
measurement of the short distance process of interest. The description
of such effects is in general beyond the scope of perturbative
QCD. However, perturbative ideas combined with other assumptions can
be used to build models of such effects. The basic ideas of the
eikonal model derive from the observation that for partonic scatters
above some minimum transverse momentum, $\ptmin$, the values of the
hadronic momentum fraction $x$ which are probed decrease as the
centre-of-mass energy, $\sqrt{s}$, increases. Since the proton
structure function rises rapidly at small
$x$
, high parton densities are probed.
Thus as discussed above and shown in Fig.~\ref{fig:gpxsec}, the
perturbatively-calculated cross-section grows rapidly with
$\sqrt{s}$. However, at such high densities, the probability of more
than one partonic scattering in a single hadron-hadron event may
become significant. Allowing such multiple scatters reduces the total
cross-section, and increases the activity in the final state of the
collisions.

An interesting historical aside is that some pQCD-based
calculations~\cite{Drees:1988wu,*Drees:1988ib,Gandhi:1990yh,Gandhi:1991ve}
of the total cross-section, which were made before HERA data, showed
that the photoproduction cross-section might, at high enough
centre-of-mass energies ($\approx 1$ TeV, equivalent to a 1~PeV photon
on a fixed target), be comparable to typical hadronic 
cross-sections~($\approx 1$~mb), and even to the cross-section for a photon
to initiate an electromagnetic shower in air ($\approx 100$~mb). This
was seen as a possible explanation for an observed
excess\cite{Samorski:1983zm,*Dingus:1988bx} of muons in
ultra-high-energy cosmic ray point sources.  However, eikonal
corrections were either absent or incorrectly implemented in these
calculations, and the highest of them also made use of very singular
parton densities in the photon. Most of them are inconsistent with the
HERA measurements at 200~GeV (40~TeV equivalent fixed target). With
more reasonable photon PDFs~\cite{Gandhi:1990yh} and a correct
application of eikonalisation~\cite{Forshaw:1991gd}, the models are
consistent with HERA data, but do not allow a fast enough rise to explain
the muon excess. In any case, the point source observations were not
confirmed by subsequent experiments~\cite{Hoffman:1999bm}. If the observed
showers were not from point sources, there is no reason why they could
not have been initiated by protons rather than hadrons, and so the
puzzle goes away. Nevertheless, the photoproduction cross-section
should be a significant input to simulations of high-energy cosmic-ray
air-shower simulations.

In any case, the widely used models assume some distribution of the
matter inside the hadron in impact parameter, $b$, space which is
independent of the momentum fraction, $x$. The multiparton interaction
rate is then calculated either
\begin{enumerate}
\item by using the cross-section for the hard subprocess, the
conventional parton densities, and the area overlap function, $A(b)$
\cite{Forshaw:1992yy,jimmy}; or 
\item by comparing the perturbative result with the Regge
fit to derive a mean number of scatters, and using Poisson statistics
to obtain the multiplicity distribution \cite{vanzijl}.
\end{enumerate}
In the former approach, impact-parameter correlations are built in via
the area overlap function. In the latter approach these are absent in
the simplest model, but may be introduced by using more complicated
matter distributions. More recent developments also fold in the parton
showering in a scale-ordered manner~\cite{sjskands}.

The approach which does not use the Regge fit as an input reduces the
hard scattering contribution to the total cross-section, as shown in
Fig.\ref{fig:gpxsec}. Comparisons of both types of model to data are
discussed in Sections~\ref{sec:res_dir} and \ref{sec:eflow_rapidity_gaps}.

\subsection{Jets and fragmentation}
\label{sec:jetpheno}

Due to colour confinement, quarks and gluons are never observed as
free, final-state particles.  When quarks or gluons are scattered with
high momentum transfers, the observable results are ``jets'' of
hadrons. Jets are defined and reconstructed by a variety of
algorithms, all of which in one way or another identify hadrons which
are close to each other in phase space, and whose bulk kinematic properties
reflect the kinematics of the short-distance parton scatter.

Thus, jets with a transverse momentum $\pt > \ptmin$ should be present
in a fraction of the photoproduction cross-section as discussed in the
previous section. In addition, in eikonal models some events will have
more than one jet pair. Several calculations exist of hard
cross-sections in photoproduction, including leading-logarithmic
parton-shower Monte Carlo
programs~\cite{jhep:01:010,*Corcella:2002jc,Sjostrand:2003wg,Engel:1994vs,cpc:86:147,Jung:2001hx}
and next-to-leading order (NLO) partonic calculations of
jets~\cite{epj:c17:413,np:b507:315,*np:b467:399,*np:b507:295,zfp:c76:67,pr:d56:4007,Gordon:1992tw},
hadrons~\cite{Fontannaz:2002nu,Binnewies:1995pt} and prompt
photons~\cite{Krawczyk:2001tz,Fontannaz:2001ek,Gordon:1994sm}. The Monte
Carlos give in general a more realistic final state, but are very
unreliable in normalisation due to the fact that the matrix elements
are currently leading order (LO). The NLO calculations are more accurate in
normalisation but in general need to have corrections applied for
hadronisation effects, and are less reliable in some event properties
due to the fact that they only allow at most one parton radiation in
addition to the primary jet pair.

Many measurements of photoproduction jet cross-sections have been
made. These have led to many extensions in the understanding of QCD,
the photon and the proton. These measurements and their consequences
are discussed in detail in the subsequent sections.

\subsection{Virtual photons}
\label{sec:virtual1}

As mentioned previously, the highest energy photoproduction studies
are all carried out with quasi-real, rather than real, photon beams. In
this case the photoproduction regime is defined by some more-or-less
arbitrary requirement on the maximum virtuality of the
photon. Somewhere on the other side of this cut is the deep inelastic
scattering regime. There is a lot to be learned by studying the
transition between these two regimes.

In the hard photoproduction regime, the photon is probed at some scale
$\mu$ by a hard process, for example jet production.  As the photon
virtuality approaches $\mu$, it cannot develop structure on a scale
large enough to be resolved by the hard process, and the photon
structure therefore reduces to a perturbative splitting into a
collinear $q\bar{q}$ pair (see for example~\cite{Klasen:2002xb}). The
development of photon structure between these extremes, and what it
might tell us, is discussed in Section~\ref{sec:virtual2}.

In the proton rest frame, the $q\bar{q}$ system is developed from the
photon a long time before the interaction with the proton and may be
considered as a dipole (see~\cite{forshaw:1997:pomeron} and references
therein). As photon virtuality increases, the average size of the
interacting dipole decreases~\cite{pr:d60:074012}. Therefore the
probability of reabsorption, or rescattering, of any forward going
particle (generally a proton or a neutron) is expected to decrease as
$\q2$ increases. This is discussed in
Section~\ref{sec:eflow_rapidity_gaps}.

\section{The dual nature of the photon}
\label{sec:dual}


\subsection{Resolved and direct} 
\label{sec:res_dir}

Photoproduction occurs in a lepton-proton collider such as HERA when a
quasi-real photon, emitted from the incoming positron, collides with a
parton from the incoming proton. At HERA, the majority of
photoproduction processes are peripheral collisions at low transverse
momentum where the products of the collision continue in the general
direction of travel of the incoming particles.  However, as discussed
in the previous chapter, in a subset of events short distance
processes occur which involve a \emph{hard} scatter and the production
of jets. In \emph{direct} photoproduction, the simplest such process,
the photon acts as a pointlike object and couples to the quarks in the
proton. As discussed in Section~\ref{sec:pqcd}, the photon can also
act as a source of partons, one or more of which interacts with a
parton in the proton; this is called the \emph{resolved} process.

\begin{figure}[htb]
\begin{center}
~\epsfig{file=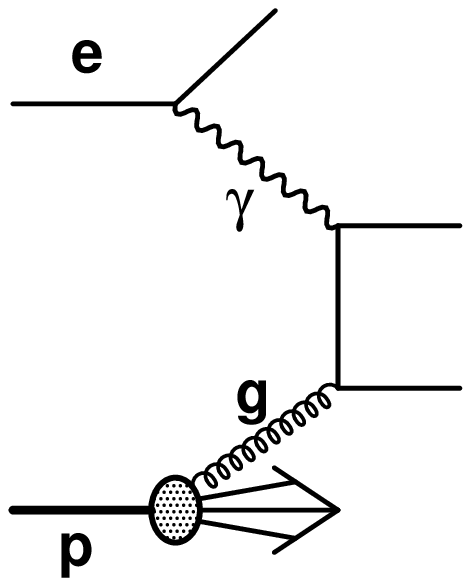}
\hspace{1cm}~\epsfig{file=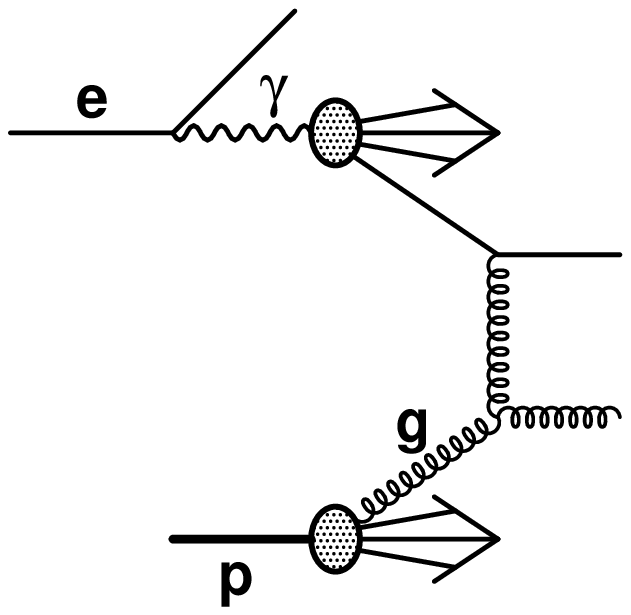}
\put(-305,6){\makebox(0,0)[tl]{\large (a)}}
\put(-95,6){\makebox(0,0)[tl]{\large (b)}}
\end{center}
\caption{Examples of direct and resolved dijet photoproduction
         diagrams in LO QCD. (a) shows a boson-gluon fusion (BGF)
         process. The other direct process is QCD Compton
         scattering. (b) shows one of many possible resolved photon
         diagrams, which may involve quarks and gluons from the photon
         and the proton.}
\label{fig:feyn}
\end{figure}

Examples of hard scattering processes are shown in
Fig.~\ref{fig:feyn}.  High transverse momentum photoproduction process
are calculated to next-to-leading order in QCD. The photon structure
describes the probability of finding a parton in the photon with a
momentum fraction, $x_\gamma$, at a scale $E_T$. Parametrisations of
the photon structure incorporate both the VDM and anomalous
components. At high enough scales, the VDM component is negligible and
the photon structure function can be fully calculated in
QCD~\cite{np:b120:189}.

To test this perturbative picture, the direct component was searched
for in photon-hadron collisions by comparing the data with those from
hadron-hadron collisions. The WA69 collaboration at CERN performed
measurements of the charged particle distribution in photon-proton,
pion-proton and kaon-proton collisions~\cite{zfp:c43:63}. Fixed-target
collisions for the different types of incident beam used the same
experimental apparatus to measure the final state. The energy of the
photon beam was varied between 110 and 170 GeV corresponding closely
to the incident pion and kaon beam energy of 140 GeV. To compare the
photon-beam data with that of hadron-beam data, the relative fraction
of pion- and kaon-beam data was taken to be that expected for the
strange and non-strange quark fraction in the photon. This corresponds
to 40\% kaon-beam and 60\% pion-beam data.

\begin{figure}[htp]
\begin{center}
~\epsfig{file=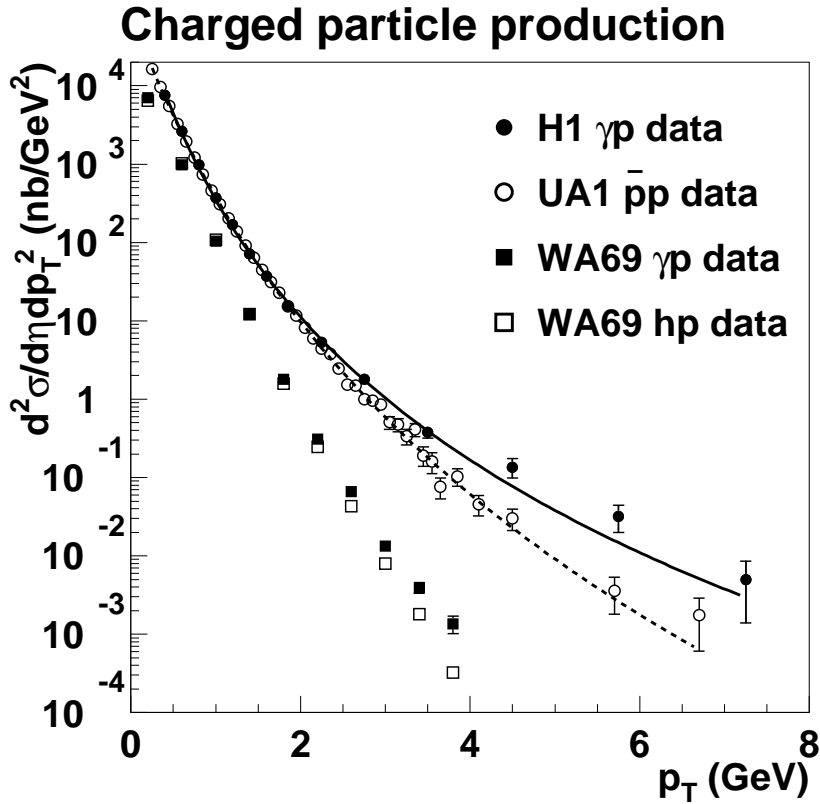,height=11cm}
\end{center}
\caption{Differential cross-section for the production of charged
         particles at a centre-of-mass energy $\sqrt{s_{\gamma p}}
         \approx 200$~GeV and $\sqrt{s_{\bar{p} p}} \approx 200$~GeV
         for the H1 and UA1 data, respectively.  Curves of the
         functional form $A \left( 1 + p_T/p_{T, 0} \right)^{-n}$ are
         fitted to the data.  Also shown are data from WA69 at
         $\sqrt{s_{\gamma p}} \approx 16$~GeV and $\sqrt{s_{{\rm
         hadron} p}} \approx 16$~GeV. }
\label{fig:charged_particles}
\end{figure}

The differential cross-sections for the production of charged
particles as measured by the WA69 collaboration are shown in
Fig.~\ref{fig:charged_particles}. The hadron-beam data is shown
divided by a VDM factor of 215 and only the relative normalisations
(not the absolute) are kept for the two sets of data in this
figure. Both sets of data fall rapidly over about 7 orders of
magnitude in the cross-section in the range of $p_T$ between 0 and 4
GeV. At low momentum, the data sets show similar behaviour. However,
the photon-proton data fall less rapidly than the hadron-proton data.
This indicates that the photon-beam data contain a component in
addition to the VDM contribution, the relative importance of which
increases with increasing $p_T$. A similar measurement of the charged
particle distribution in photoproduction was made by the H1
collaboration~\cite{pl:b328:176}. The measurement was performed at
much larger centre-of-mass energies than those at WA69:
$\sqrt{s_{\gamma p}} \approx 200$~GeV rather than $\sqrt{s_{\gamma p}}
\approx 16$~GeV. The H1 measurement is compared to data from
$\bar{p} p$ collisions measured by the UA1 experiment at a similar
centre-of-mass energy of $\sqrt{s_{\bar{p} p}} \approx 200$~GeV. The
two sets of data are also shown in Fig.~\ref{fig:charged_particles}
with the UA1 data normalised to the H1 data at $p_T = 1.5$~GeV. Both
$\gamma p$ and $\bar{p} p$ data fall over many orders of magnitude in
the cross-section and extend up to higher values of $p_T$ than the
WA69 data due to the larger centre-of-mass energy. The increased
centre-of-mass energy also produces harder $p_T$ spectra. As with 
the WA69 data, the $\gamma p$ data fall less rapidly with increasing
$p_T$ than the hadron-hadron collision data. This is again indicative
of the QCD expectation that direct and anomalous processes are present
in photoproduction, but absent in hadron-hadron collisions.

Evidence for hard photon scattering, in quantitative agreement with
QCD calculations~\cite{Aurenche:1983eq}, had been observed in
fixed-target
experiments~\cite{zfp:c43:63,pl:b168:163,*pl:b174:458,*pl:b182:409,*pl:b100:433,*zfp:c36:527}.
Observing hard scattering in photoproduction was one of the first
investigations performed at
HERA~\cite{pl:b297:205,*pl:b297:404}. Measurements of the total
transverse energy in the event and the production of charged particles,
similar to that in Fig~\ref{fig:charged_particles}, were made. The
data were compared with theoretical predictions of soft interactions
only, as well as predictions which included hard interactions, as
implemented in the Monte Carlo models {\sc Pythia} and {\sc
Herwig}. The predictions for soft interactions were concentrated at
low transverse energy of both the event and the charged
particles. With increasing transverse energy the prediction fell
rapidly below the data by many orders of magnitude.  With inclusion of
the hard scattering component, both {\sc Pythia} and {\sc Herwig}
described the event and particle distributions well at high transverse
energy.

The observation of hard-scattering led to the reconstruction of two
back-to-back jets within events as expected from two-body
scattering. The reconstruction of jets allows the kinematics and
properties of the events to be investigated in more detail. A dijet
sample was used to separate the contributions of direct and resolved
photon processes~\cite{pl:b322:287}.

In two-to-two parton scattering, the momenta of the incoming partons
can be calculated from the momenta of the outgoing partons. Let $x_p$
and $x_\gamma$ be the fraction of momentum carried by the partons from
the proton and photon, respectively. Conserving energy and momentum
gives

\[ x_p = \frac{\sum_{\rm partons}(E+p_z)_{\rm partons}}{2E_p}, 
\ \ \ \ x_\gamma = \frac{\sum_{\rm partons}(E-p_z)_{\rm partons}}{2E_\gamma} \]

where $E_\gamma$ is the initial photon momentum and the sum is over
the two final state partons. For direct processes, $x_\gamma = 1$,
whereas for resolved processes, $x_\gamma < 1$.

\begin{figure}[htp]
\begin{center}
~\epsfig{file=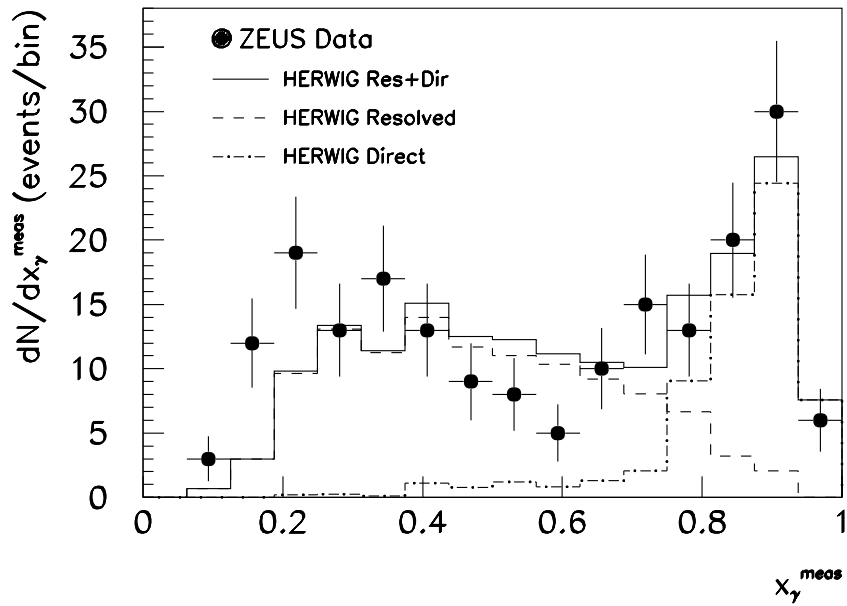,height=8cm}
\end{center}
\caption{Measurement of $x_\gamma^{\rm meas}$ for events with two or
         more jets. The Monte Carlo prediction is shown fitted to the
         data. Taken from publication~\protect\cite{pl:b322:287}.}
\label{fig:first_xgamma}
\end{figure}

Obviously, partons cannot be measured, but hadronic jets provide a
good approximation of their four-momenta. Experimentally, therefore,
the quantity $x_\gamma$ was reconstructed as

\[ x_\gamma^{\rm meas} = \frac{\sum_{\rm jets}(E-p_z)_{\rm jets}}{\sum_i (E-p_z)_i} \]

where $E_\gamma \approx y E_e \approx y_{\rm JB} E_e$ and the sum
in the denominator runs over all energy deposits in the
calorimeter. The first measurement of this quantity is shown in
Fig.~\ref{fig:first_xgamma} compared with expectations from the {\sc
Herwig} Monte Carlo programme. The data exhibit a two-peak structure
at high and low values of $x_\gamma^{\rm meas}$. The Monte Carlo
prediction gives a reasonable representation of the data when direct
and resolved photon processes are added together. The direct and
resolved components in the Monte Carlo have very different shapes. The
resolved component describes the low $x_\gamma^{\rm meas}$ region
reasonably well but cannot describe the data at high $x_\gamma^{\rm
meas}$. This data at high $x_\gamma^{\rm meas}$ can only be described
with the inclusion of the component from direct processes. Hence these
data constitute the first observation of direct processes in
photoproduction.

Resolved photon processes, like hadron-hadron collisions, have a more
complicated structure than direct or deep inelastic scattering
processes. As resolved photon processes are the effective collision of
two composite objects, they are subject to extra unknowns from the
structure of the photon (discussed in
Sections~\ref{sec:photon_structure}~and~\ref{sec:nature_of_photon}),
from the presence of the remnant of the photon and proton (left after
the interacting partons have scattered), and from remnant-remnant
interactions (or more generically the underlying event discussed in
Section~\ref{sec:ue}).

Although the point-like coupling of the photon to $q \bar{q}$ pairs is
included in the parametrisations of the photon structure function,
Monte Carlo simulations model the resolved photon as a hadron, with
collinear incoming partons. This results in a photon remnant with low
transverse momentum with respect to the
beam-axis. Next-to-leading-order contributions of fluctuations of the
photon into a $q \bar{q}$ pair with high virtuality may lead to a
``photon remnant'' which has sizeable transverse momentum with respect
to the incident photon
direction~\cite{proc:aspen:1994:110,*proc:twophoton:1994:163,*pl:b320:186}.

The photon remnant and its properties were studied by the ZEUS
collaboration~\cite{pl:b354:163}.  The photon remnant was isolated in
two-jet events by requiring a cluster of low energy in the direction
of the incident photon (electron) beam. The cluster exhibited a
collimated energy flow with a limited transverse energy with respect
to the cluster axis, characteristic of a jet structure. Indeed the
cluster was compared with the other hard jets (assumed to come from
the two-to-two scatter) in the event. Although the origin of these two
types of jets may be different within statistics and the kinematic
range studied, they exhibit similar properties for the energy flow and
the transverse and longitudinal energy with respect to the jet
axis. The $\eta, \pt$ and energy distributions for this jet are shown
in Fig.~\ref{fig:photon_remnant}. The measured transverse momentum
relative to the photon axis is $p_T = 2.1 \pm 0.2$~GeV, and is larger
than that seen in the default {\sc Herwig} distribution, which
generates the transverse momentum only via parton showers. Both {\sc
Pythia} (which includes intrinsic transverse momentum in the photon
by default) and {\sc Herwig} with 1~GeV of such momentum added, give a
better description of the data.

\begin{figure}[htp]
\begin{center}
~\epsfig{file=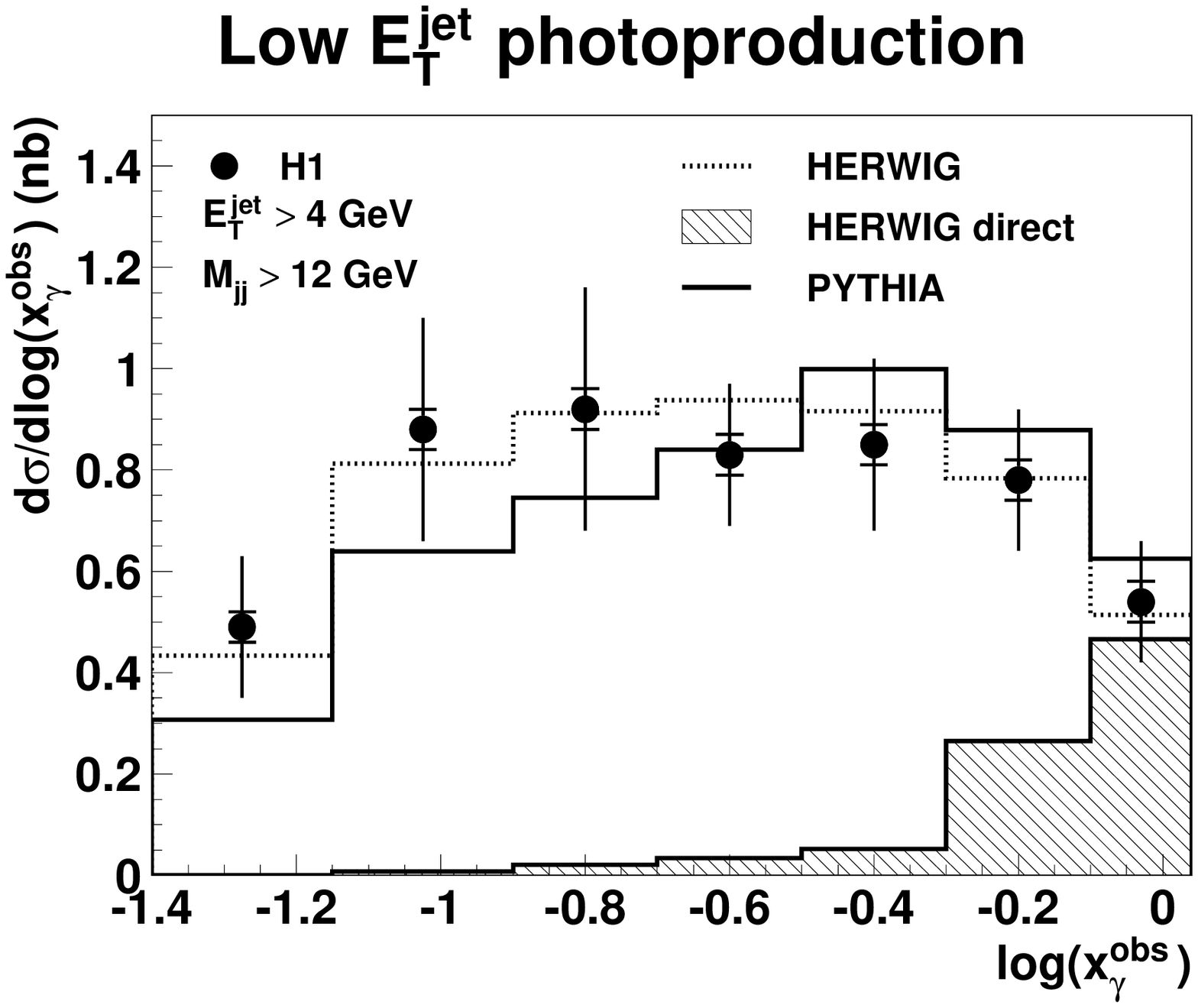,height=6.5cm}
~\epsfig{file=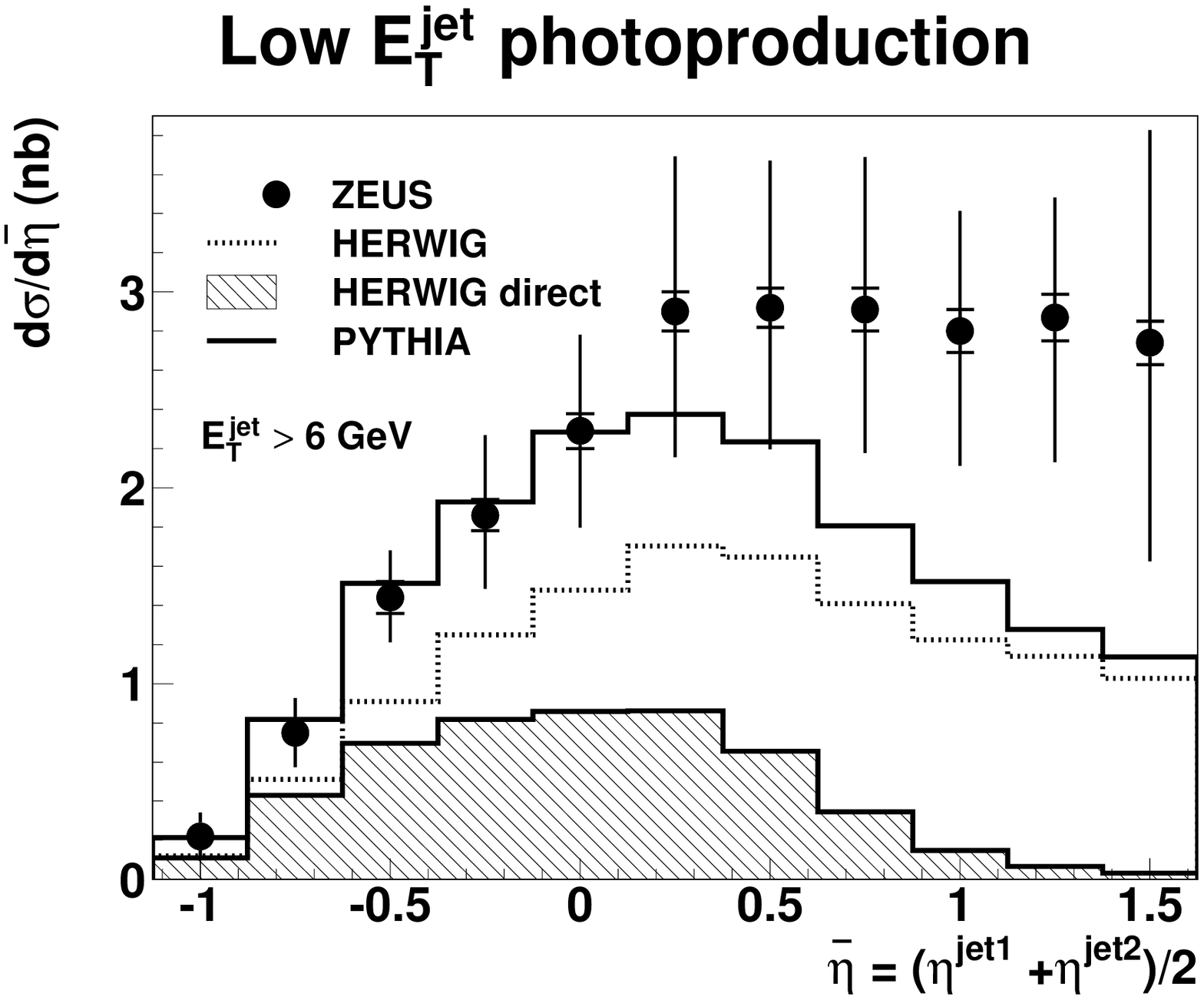,height=6.5cm}
\end{center}
\caption{H1 measurement of $d\sigma/d\xgo$ at low $\ET$, and a ZEUS
         measurement of $d\sigma/d\bar{\eta}$ compared to
         various Monte Carlo models.}
\label{fig:low_et}
\end{figure}

Early cross-section measurements~\cite{pl:b483:36,epj:c1:109} showed
discrepancies with predictions from Monte Carlo models, which were
reduced with the inclusion of the possibility of secondary
scatters. This underlying event (UE) or multiparton interaction (MPI)
results from the collision of two composite objects and is a feature of
hadron-hadron collisions (see Section~\ref{sec:ue}). Examples of
measured data compared with Monte Carlo predictions are shown in
Fig.~\ref{fig:low_et}. The first measurement of a cross-section as a
function of $\xgo$ is shown to be described by predictions which
incorporate models for secondary scatters. The variable
$\xgo$~\cite{pl:b348:665} may be calculated to any order in QCD and
for any final state, using energy deposits in a calorimeter, final
state hadrons in a Monte Carlo simulation or partons in a QCD
calculation. The quantity is defined as:

\[ \xgo = \frac{E_T^{\rm jet1}e^{-\eta^{\rm jet1}} + E_T^{\rm jet2}e^{-\eta^{\rm jet2}}}
               {2 y E_e}. \]

The additional cross-section from the Monte Carlo MPI prediction is
concentrated at low $\xgo$ and at high $\bar{\eta}$. However, in the
MC models shown, it is not sufficient to describe the data in
Fig.~\ref{fig:low_et}b.

\begin{center}
~\epsfig{file=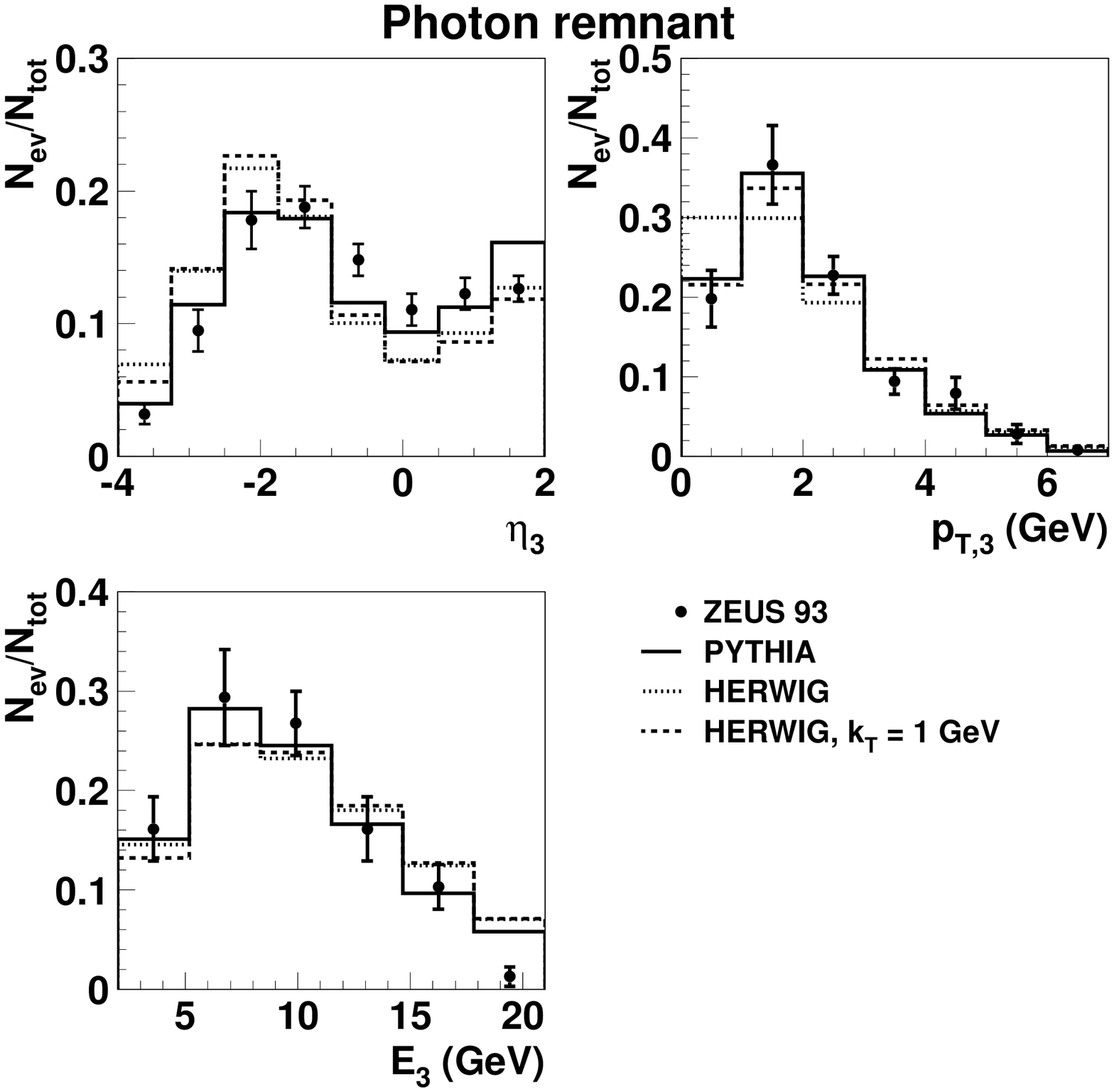,height=13cm}
    \put(-215,335){\makebox(0,0)[tl]{\large (a)}}
    \put(-35,335){\makebox(0,0)[tl]{\large (b)}}
    \put(-215,160){\makebox(0,0)[tl]{\large (c)}}
\end{center}
\begin{figure}[!htp]
\caption[*]{Properties of the photon remnant~\cite{pl:b354:163}. (a) 
shows the the rapidity for the third jet (ordered in transverse energy) 
in events reconstructed using the $\kt$ cluster algorithm. For (b) 
transverse momentum and (c) energy, the jet is required to have pseudorapidity $\eta < -1$,
{\it i.e.} to be close to the photon direction. The default {\sc Herwig}
distribution has no intrinsic $\kt$ in the photon or proton, whereas
{\sc Pythia} has 1~GeV (Gaussian width) by default, as does the second
{\sc Herwig} model (dashed lines).}
\label{fig:photon_remnant}
\end{figure}

At the low transverse energies measured here, the corresponding $\xg$
value is low and hence the data are sensitive to the gluon
distribution in the photon. However, to access information on the
photon structure, the nature of these secondary scatters has to be
understood. Although their models generally improve the description of
the data, there is as yet no definitive evidence for their occurrence
in photoproduction. Investigations of these phenomena are discussed
further in Section~\ref{sec:eflow_rapidity_gaps}. A solution to
extracting information on the photon structure without being affected
by the underlying event is to make measurements at higher transverse
energy where the effect is minimised.

\subsection{Photon structure} 
\label{sec:photon_structure}

\begin{figure}[htp]
\begin{center}
~\epsfig{file=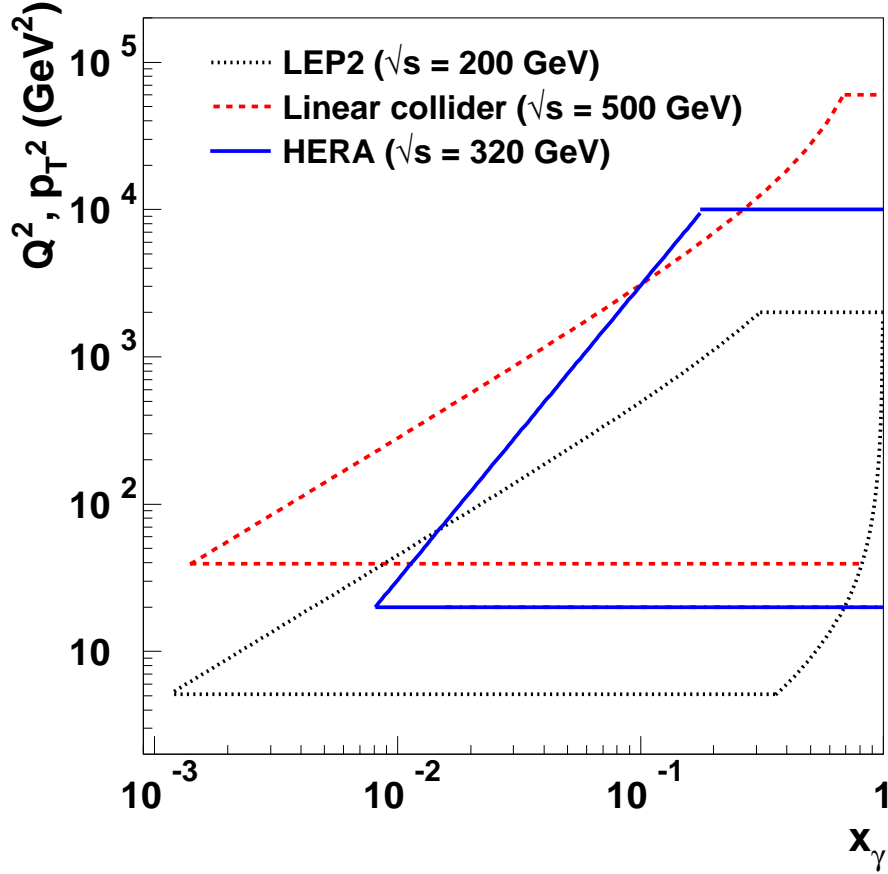,height=12cm}
\end{center}
\caption{The kinematic plane in $\q2$ and x, showing the region of
         sensitivity of previous, current and possible future $\ee$
         and $\gp$ experiments.}
\label{fig:q2_vs_xgamma}
\end{figure}

The structure of the photon has traditionally been extracted from deep
inelastic $e\gamma$ scattering at $e^+e^-$ colliders (see
\cite{physrep:332:165} for a review). In this reaction, a virtual
photon, emitted from the incoming electron or positron, probes the
structure of the target photon. The range in scale of the interaction,
$\q2$, depends on the centre-of-mass energy, $\sqrt{s}$ of the
collider, starting at a few GeV$^2$ and rising up to about
2000~GeV$^2$ for later LEP data taking. The corresponding range in
momentum fraction of the parton, $x_\gamma$, is
$0.001<x_\gamma<1$. The kinematic plane for LEP ($\sqrt{s} =$~200~GeV)
in $x$ and $\q2$ is shown in Fig.~\ref{fig:q2_vs_xgamma}.

As discussed previously, photoproduction data from HERA are sensitive
to the structure of the photon. Although the processes at the two
colliders are different, the kinematic regions significantly
overlap. The HERA data also extend to higher scales and, for low
scales, to higher $x$ as shown in Fig.~\ref{fig:q2_vs_xgamma}.  Note
that, in the case of HERA, where the virtuality of the exchanged
photon is small, the scale of the interaction is given by the $p_T$ of
the produced jets. Also shown in Fig.~\ref{fig:q2_vs_xgamma} is the
kinematic region from a future $e^+e^-$ linear collider. A similar
detector coverage to those at LEP is assumed, hence the same minimum
$x_\gamma$, but higher $\q2$. The kinematic reach to high scales of
the linear collider is significantly above that of LEP with HERA
midway between. This demonstrates for the next 10~years the
significance of HERA data in understanding the structure of the photon
at high scales and its potential impact for studies at the linear
collider.

\begin{figure}[htp]
\begin{center}
~\epsfig{file=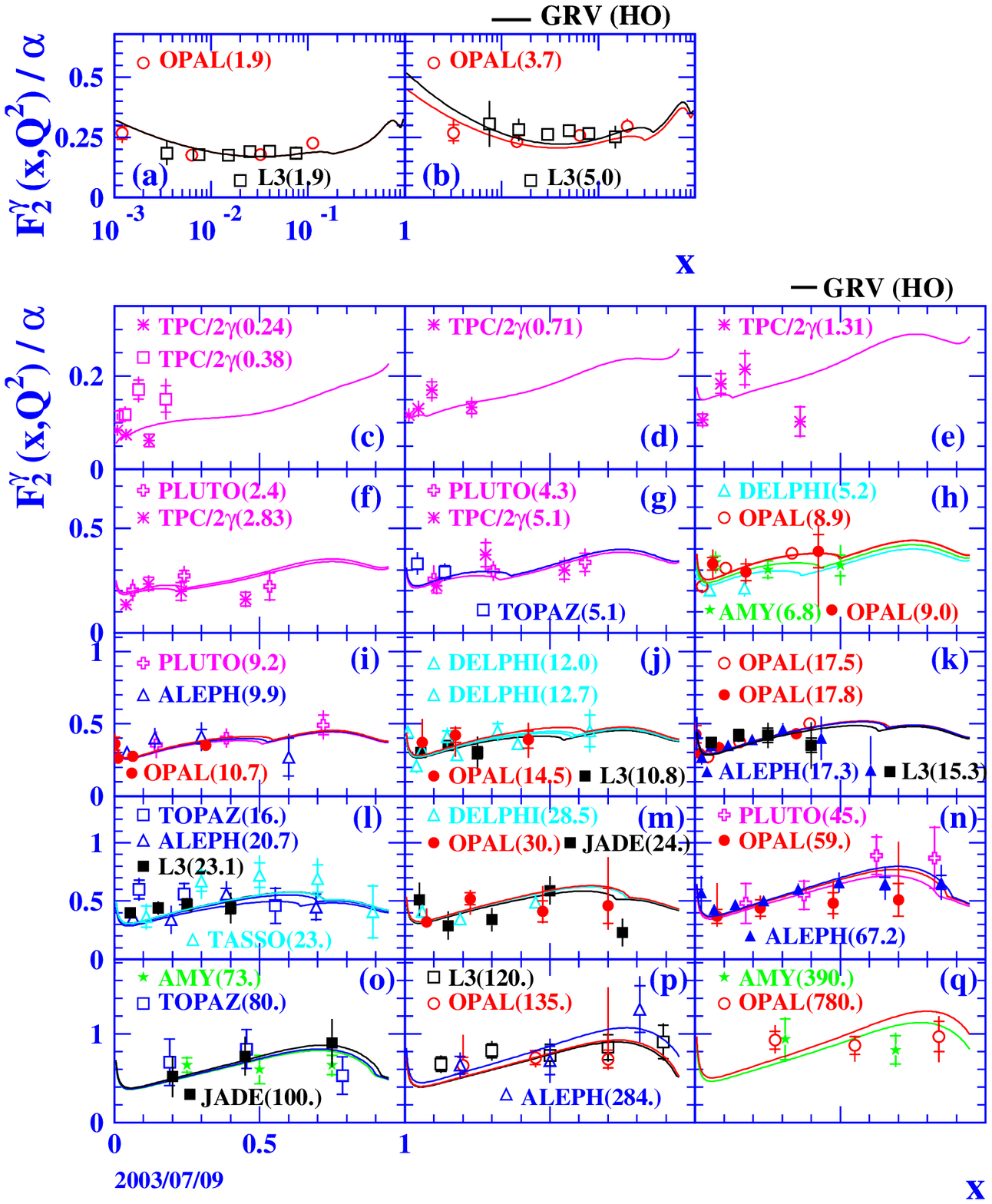,height=8.5cm}
~\epsfig{file=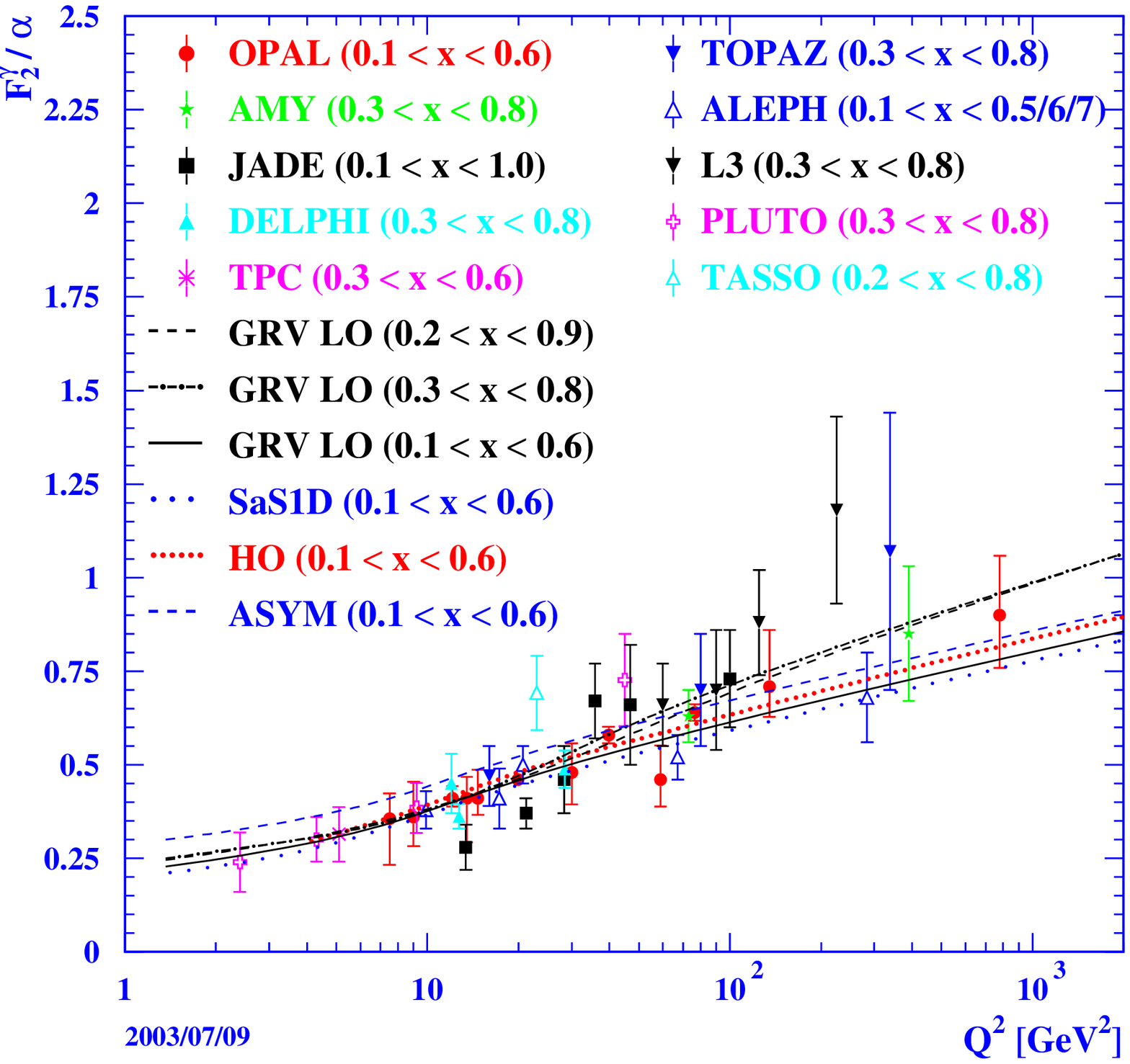,height=8.5cm}
\put(-355,6){\makebox(0,0)[tl]{\large (a)}}
\put(-130,6){\makebox(0,0)[tl]{\large (b)}}
\end{center}
\caption{Measurements of the photon structure function, $\F2g$, as a
function of (a) $x$ for fixed values of $\q2$ and (b) $\q2$ for fixed
values of $x$. Data from many experiments are shown compared with
theoretical parametrisations derived from fits to a subset of the
data. Taken from~\protect\cite{physrep:332:165}.}
\label{fig:f2gamma}
\end{figure}

The current status of measurements of $\F2g$ from $e^+e^-$ collisions
is shown in Fig.~\ref{fig:f2gamma}~\cite{physrep:332:165}. The data in
Fig.~\ref{fig:f2gamma}a are generally flat as a function of $x$ and
the data from different experiments and colliders are broadly
consistent with each other. At sufficiently low $x$, the gluon
density is expected to dominate resulting in a rapid rise of the
structure function with decreasing $x$. Unfortunately due to the
precision of the measurements and coverage of the detector to
investigate lower values of $x$ this has not been demonstrated by
these measurements, although they are consistent with the hypothesis
as shown by the data at very lowest $\q2$ and $x$. The data in
Fig.~\ref{fig:f2gamma}b are shown to rise logarithmically as a
function of $\q2$ as expected from QCD~\cite{np:b120:189}. This
so-called asymptotic solution is shown to describe the rise in
$\q2$. Such comparisons have lead to a rather competitive measurement
of the strong coupling constant $\as$~\cite{Albino:2002ck}.
However, due to the accuracy of the data, the structure of the
photon is not well constrained and has a significant uncertainty when
extrapolated to the energies at a future linear collider.

\begin{figure}[htp]
\begin{center}
~\epsfig{file=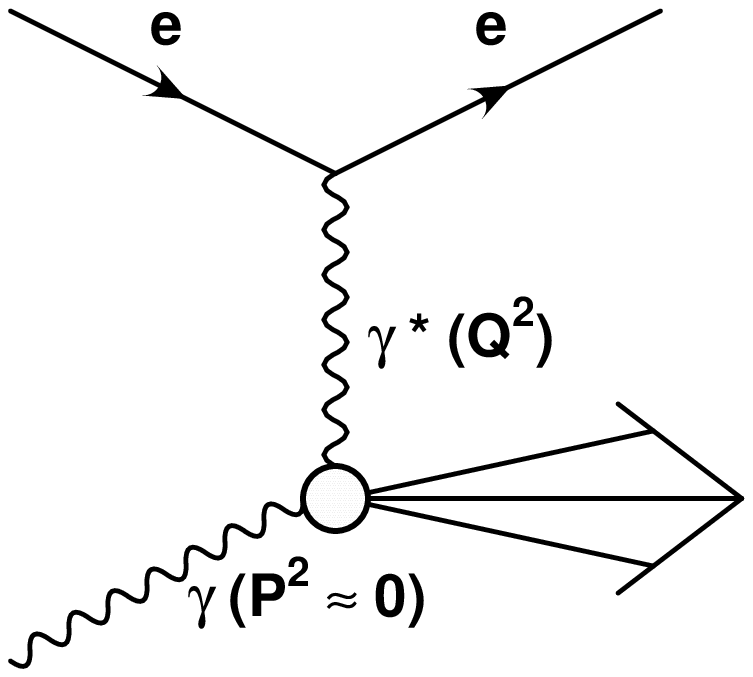,height=5.75cm}
\hspace{1cm}~\epsfig{file=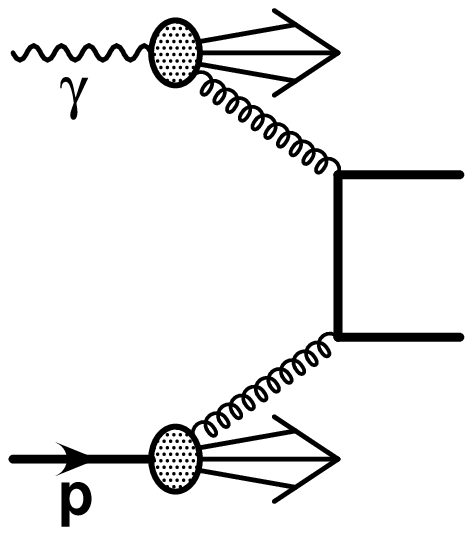,height=5.75cm}
\put(-285,6){\makebox(0,0)[tl]{\large (a)}}
\put(-95,6){\makebox(0,0)[tl]{\large (b)}}
\end{center}
\caption{Feynman diagrams of (a) deep inelastic $e\gamma$ scattering
in $e^+e^-$ collisions and (b) gluon-gluon scattering in
photoproduction.}
\label{fig:feyn_ee_gg}
\end{figure}

Measurements of photoproduction processes also provide valuable
information on the structure of the photon because the gluon density
of the photon is probed directly (see Fig.~\ref{fig:feyn_ee_gg})
rather than just the quark density as in the case of deep inelastic
$e\gamma$ scattering. In fits to the photon structure function,
$\F2g$, the gluon density is constrained via its contribution to the
scaling violations.

In photoproduction $\F2g$ is not measurable, since it is defined in
terms of deep inelastic lepton-photon scattering. Sensitivity to the
photon parton density functions (PDFs) is present in several
measurable cross-sections, however. Thus, precision measurements in
high-energy photoproduction can provide a test of QCD factorisation
for the photon structure and constraints on the photon PDFs.

\subsubsection{Jet photoproduction and photon structure}
\label{sec:pdfjets}

Jet cross-sections are sensitive to the structure of the photon and
can be used to distinguish between parametrisations, although they are
not as easily interpretable as $\F2g$ since they are a convolution of the
photon and proton parton densities and the short-distance matrix
elements (see Eq.\ref{eq:pertxsec}). Alternatively the jet cross-sections 
could be fit along with $\F2g$ in a fully consistent
theoretical approach as has recently been performed for the proton, as
discussed in Section~\ref{sec:proton_structure}.

\begin{figure}[htp]
\begin{center}
~\epsfig{file=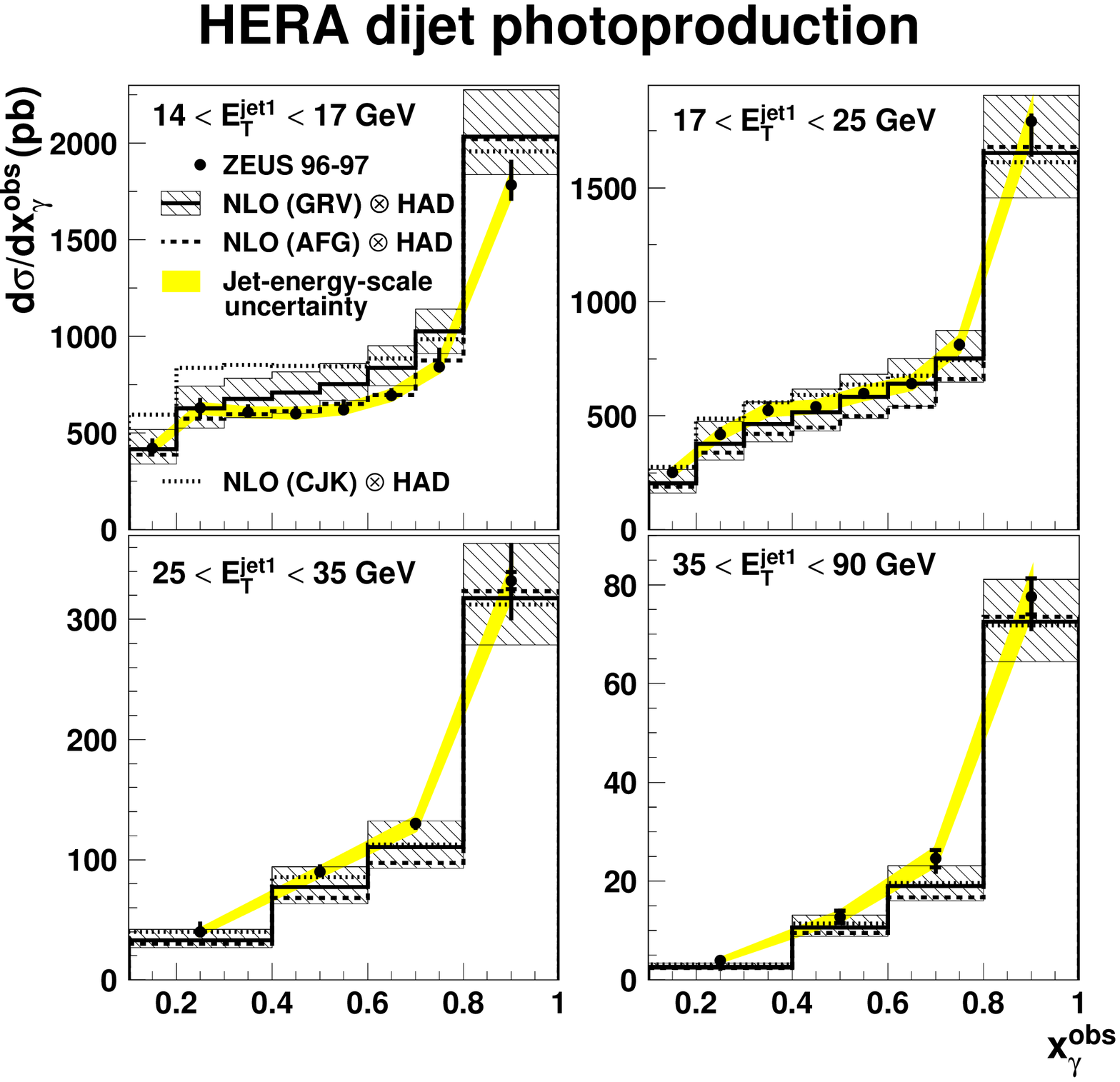,height=11cm}\\
\vspace{-1cm}
~\epsfig{file=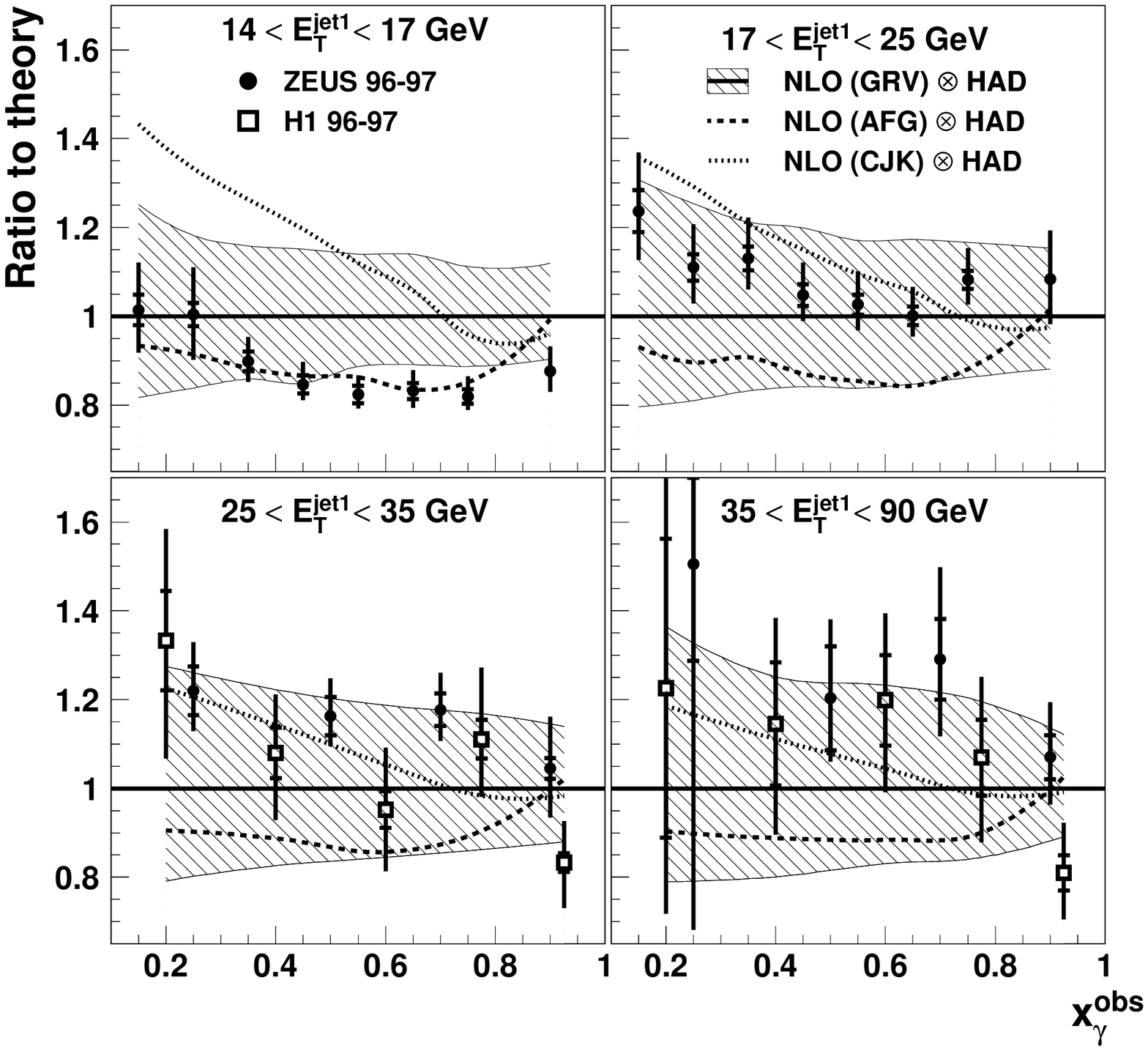,height=11cm}
\end{center}
\caption[*]{H1 and ZEUS measurements of $d\sigma/d\xgo$ at high
        $\ETJ$. The data~\cite{epj:c23:615,Adloff:2002au} are compared
        to the NLO QCD program of Frixione and
        Ridolfi~\protect\cite{np:b507:315,*np:b467:399,*np:b507:295}.}
\label{fig:xg_high_et}
\end{figure}

A further complication in understanding jet cross-sections is the
effect of the underlying event. This affects jet cross-sections most
strongly at low transverse energies as discussed in
Section~\ref{sec:res_dir}. As the size and indeed nature of this
effect is relatively unknown, constraining it and then being able to
distinguish between different parametrisations of the photon structure
is difficult. Therefore measurements are made in which sensitivity to
the underlying event is drastically reduced; this is achieved by
requiring jets of large transverse energy. This however has its
disadvantages: the maximum possible range in scale, $\ETJ$, and $x$ is not
explored; and the elimination of data at low $\ETJ$ and low $x$
reduces sensitivity to the gluon density in the photon which is
expected to dominate at low $x$. A solution to this problem would be
to make measurements highly to sensitive to the underlying event but
weakly dependent on the photon structure. Using these constraints obtained 
for the underlying event, measurements at low $E_T^{\rm jet}$ which are 
sensitive to the photon structure could then be reinvestigated. However, 
this has not yet been performed so the best information currently attainable 
from HERA data is the comparison of NLO calculations with jet cross-sections at
high transverse energies, above about 14~GeV as shown in
Fig.~\ref{fig:xg_high_et}~\cite{epj:c23:615,Adloff:2002au}.

The quantity $\xgo$ is shown in Fig.~\ref{fig:xg_high_et}, measured
for different regions of transverse energy of the leading jet. The
data are compared with an NLO calculation using the three currently
available parametrisations of the photon PDFs. It can be seen that
within the large theoretical uncertainties at lower $E_T^{\rm jet1}$
and both theoretical and experimental uncertainties at higher
$E_T^{\rm jet1}$ data and theory agree. However they exhibit a
different shape as a function of $E_T^{\rm jet1}$; the central theory
prediction lies above the data at low $E_T^{\rm jet1}$ and below the
data at high $E_T^{\rm jet1}$. This results in preferring one photon
PDF in one region of phase space and another PDF in a different region
of phase space. The parametrisations AFG~\cite{zfp:c64:621} and
GRV~\cite{pr:d45:3986} were fitted to similar data sets and used
similar assumptions in their fit and hence give similar results. The
result from CJK~\cite{Cornet:2004nb,*Cornet:2002iy} is the newest\footnote{At 
the time of writing, two new parametrisations of the photon PDFs have become 
available~\cite{hep-ph-0503259,*hep-ph-0504003} but have not yet been 
interfaced to the relevant QCD calculations.}
photon PDF using all data on $\F2g$ and has a more careful treatment
of heavy quarks. This parametrisation has a gluon density that rises
faster to low $x$; this is reflected in the larger cross-section at
low $\xgo$ for low $E_T^{\rm jet}$. As all three photon PDFs give a
good description of the $\F2g$ data, this demonstrates how poorly the
data constrain the gluon density in the photon and how well the HERA
data could constrain it. The average scale of the interaction measured
here is also significantly larger than at LEP. The average $E_T^2$ for
the highest transverse-energy region is about
2000\,GeV$^2$. Measurements in bins of transverse energy extend up to
about 5000\,GeV$^2$. This and future HERA data could prove crucial in
constraining the PDF at high scales and improve knowledge of the
structure of the photon for physics at a linear collider.

The different trend with transverse energy in the data compared to
that in the theory can be understood in terms of the effect of
additional QCD radiation in the event. It should be noted that Monte
Carlo predictions, which include multiple parton branching in the
leading-logarithmic approximation, reproduce the transverse-energy
dependence~\cite{epj:c23:615}. This can be explored by studying the
requirement on the minimum transverse energy of the second jet. This
is always 11\,GeV, which means that in these distributions the
difference between the cuts on the two jets varies from 3\,GeV to
24\,GeV. The difference in the transverse energies of the two jets is
due to QCD radiation. The effect of the cut on the second jet was
studied by fixing the cut on the first jet to be 25\,GeV and varying
that on the second jet from 11\,GeV to 25\,GeV. The (total)
cross-section was then measured as a function of this
cut~\cite{epj:c23:615}. The predictions of the shape of this
distribution from $\mathcal{O}$($\alpha \alpha_s^2$) QCD are therefore
the lowest non-trivial order predictions. The QCD predictions fall
less rapidly at low $E_{T}^{\rm jet2, cut}$ and more rapidly at high
$E_{T}^{\rm jet2, cut}$ than the data. This has been investigated
further by considering the cross-section for regions of high and low
$\xgo$. For $\xgo~>~0.8$, the data and NLO QCD converge for low
$E_{T}^{\rm jet2, cut}$, both being reasonably insensitive to the cut
and similar in shape. For the region $\xgo<0.8$, the predictions lie
below the data at low $E_{T}^{\rm jet2, cut}$, but within the
theoretical uncertainties. By adjusting $E_{T}^{\rm jet2, cut}$
separately in each $E_{T}^{\rm jet1}$ range, it would be possible to
achieve agreement between the NLO prediction and the data. However,
this seems to be a somewhat arbitrary procedure.  This issue
complicates the comparison with the photon PDFs and improved
higher-order, or resummed, calculations would make the constraints
more stringent.

\subsubsection{Other final states as a probe of photon structure}
\label{sec:pdfothers}

Processes at HERA other than jet production can be measured which are
sensitive to the structure of the photon. Of particular interest is
the photoproduction of prompt photons which has been measured 
by both H1 and ZEUS collaborations~\cite{pl:b472:175,pl:b511:19,*epj:c38:437}. 
The process is
dominated by the production of hard photons radiated by a quark from
the proton. This is a well-understood process as the quark density in
the proton is measured to high precision and the hard scatter is
governed by QED; its measurement thereby provides a useful test of the
assumed factorisation of the cross-section and quark density in the
proton. The resolved processes, as with jet production, are a similar
convolution of the proton and photon PDFs with the short-distance
cross-section, but with fewer possible diagrams because of the
requirement of a photon in the final state which is radiated by a
quark. This advantage and also that of a cleaner experimental
signature are outweighed by the small cross-section of this process.
Current measurements are not of a high enough precision to distinguish
between different parametrisations of the photon PDFs or constrain the
quark density in the proton. The increased statistical sample expected
at HERA II will prove invaluable.

Charged particle production also has a strong sensitivity to photon
structure. Measuring charged particles~\cite{pl:b328:176,zfp:c67:227} 
is experimentally simpler than
making measurements of jet cross-sections, but the price paid is an
increased sensitivity to non-perturbative physics, in particular the
fragmentation of quarks and gluons into hadrons. The sensitivity to
the underlying event also remains a problem at low track momentum. An
attempt has been made to estimate the gluon density in the photon at
LO in QCD from such measurements~\cite{epj:c10:363}. A more
sophisticated phenomenological technique and higher statistics at
higher $\pt$ could lead to interesting results in the future.

Photoproduction of lepton pairs via the Drell-Yan process has been
discussed for some time as a process with the potential to provide
information on photon
structure~\cite{Jones:1979wa,*Kang:1979xe,*Irving:1980qa,*Vourdas:1980bp}. However,
the cross-section is rather small, and to date no measurement has been
made. Within the full HERA I + II data sample, some few tens of events
are expected per experiment in which both leptons are detected, and it
seems likely that the background from Bethe-Heitler processes can be
controlled, for example by requiring the presence of a photon
remnant~\cite{Bussey:1996vq}.  Thus after many years of waiting there
is a real possibility that HERA could make the first measurement of
this process. Such a measurement would at very least be a clear test
factorisation. At best it would provide information on photon
structure with systematics very different from jets; in particular,
the absence of the hadronisation corrections associated with
low $\ETJ$ jets may mean that this is the best method for studying
low $\xg$ in photoproduction.

\subsubsection{Photon structure and future experiments}

Accurate constraints on the structure of the photon are vital for
discovering new physics at a future linear collider. To be able to
discover new physics at future colliders, the effects of conventional
processes need to be understood. The discovery potential of both the
Large Hadron Collider (LHC) and International Linear Collider (ILC) is
enormous, but being able to see rarely-produced exotic events amongst,
in particular, the copious events governed by QCD requires precise
knowledge of the strong force. Current colliding-beam experiments can
provide accurate measures of the ingredients of a QCD process: the
structure of the incoming particles; the coupling strength of the
strong force; the dynamics of the outgoing particles; and, more
challengingly, ``soft'' physics not directly associated with the hard
partonic scatter.  In the absence of exact predictions from QCD, Monte
Carlo generators are used to model the production of such processes
and predict their rate at future colliders. The models often have many
free parameters which can be constrained by tuning to measurements
over a large phase space and, preferably, with different colliding
objects.

\begin{figure}[htp]
\begin{center}
~\epsfig{file=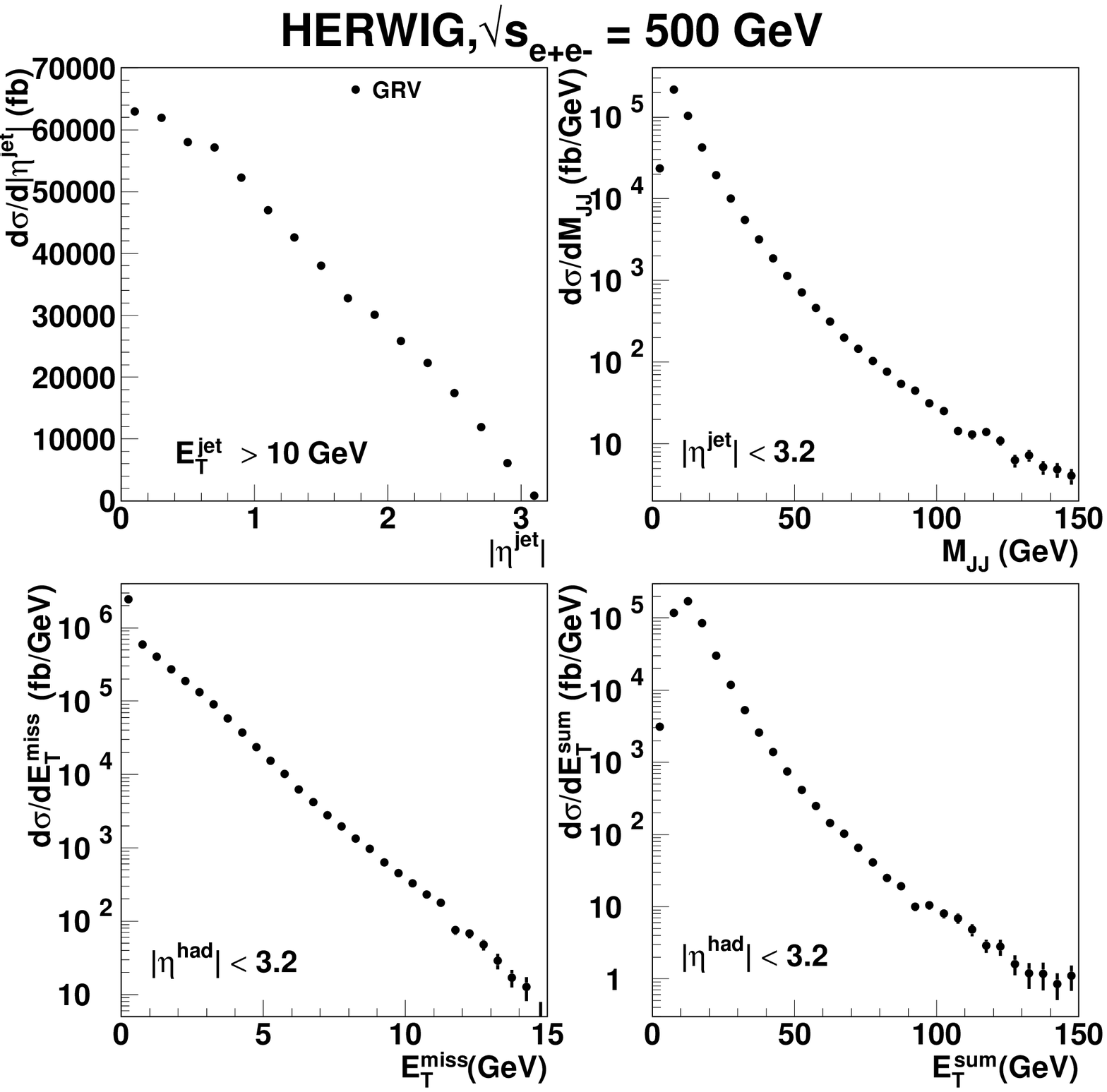,height=10.5cm}
~\epsfig{file=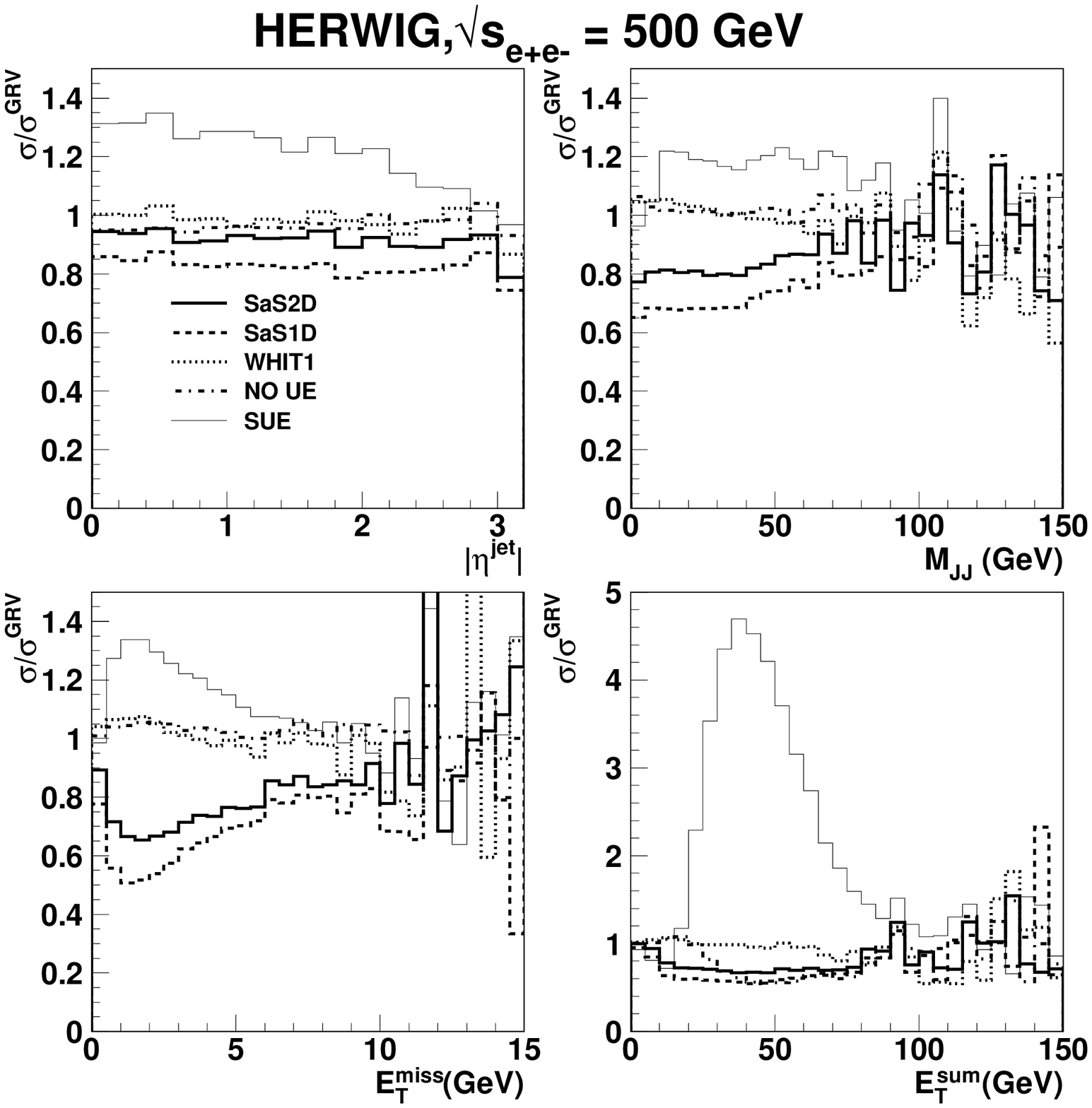,height=10.5cm}
\end{center}
\caption{Predictions of cross-sections in $\gamma \gamma$ collisions at a future 
$e^+e^-$ collider using the {\sc Herwig} Monte Carlo program (upper) and ratios 
to the default prediction using different parametrisations of the photon PDF or 
underlying event (lower).}
\label{fig:xsec_ilc_pdf}
\end{figure}

Predictions for the production of jets of high transverse energy in
$\gamma \gamma$ collisions at an $e^+e^-$ linear collider are shown in
Fig.~\ref{fig:xsec_ilc_pdf} using the {\sc Herwig} Monte Carlo program. 
The background to new physics at high energy is large; about ($10^4-10^5$) 
QCD events will be produced for a two-jet invariant mass above 100~GeV for a 
TESLA run of 500 fb$^{-1}$.  Also shown are the ratios of the cross-sections 
for predictions using different PDFs or models of the underlying event to the 
default prediction. Differences between the four predictions from different 
PDFs are observed, although the SaS1D and SaS2D parametrisations are similar. 
However, particularly in the distribution of total transverse energy, 
$E_T^{\rm sum}$, and missing transverse energy, $E_T^{\rm miss}$, differences 
of over a factor of two exist between different PDFs. Since much of the data 
from LEP has been finalised, future constraints on the photon PDF, and hence 
reductions in the uncertainty of these backgrounds, can only come from HERA. 
Use of a different underlying event (see Section~\ref{sec:ue} for a discussion 
of the phenomenology): the {\sc Jimmy} model for secondary hard scatters 
(default prediction); a soft underlying event; and no underlying event lead to 
much more significant differences. The cross-section in $E_T^{\rm sum}$ varies 
by up to almost an order of magnitude between models. Current understanding of 
such models from present data is discussed further in 
Section~\ref{sec:eflow_rapidity_gaps}.

\subsection{Where the photon structure comes from} 
\label{sec:nature_of_photon}

Measurements of the cross-section of the angle between the jet-jet
axis and the beam axis in the dijet centre-of-mass system, $\theta^*$,
have given insight into the nature of the sub-process and hence the
``content'' of the photon (and proton). The dependence of the dijet
cross-section on the cosine of the angle, $\cos\theta^*$, can be
understood in terms of the propagator in the hard scatter. For direct
photon processes where the dominant diagram is BGF, the angular
dependence of the cross-section is approximately $\propto (1 - |\cos
\theta^*|)^{-1}$ due to the spin-1/2 quark propagator. For resolved
photon processes, which are dominated by processes with a spin-1 gluon
propagator such as $qg \to qg$ and $gg \to gg$, the angular dependence
is $\propto (1 - |\cos \theta^*|)^{-2}$.

\begin{figure}
\begin{center}
~\epsfig{file=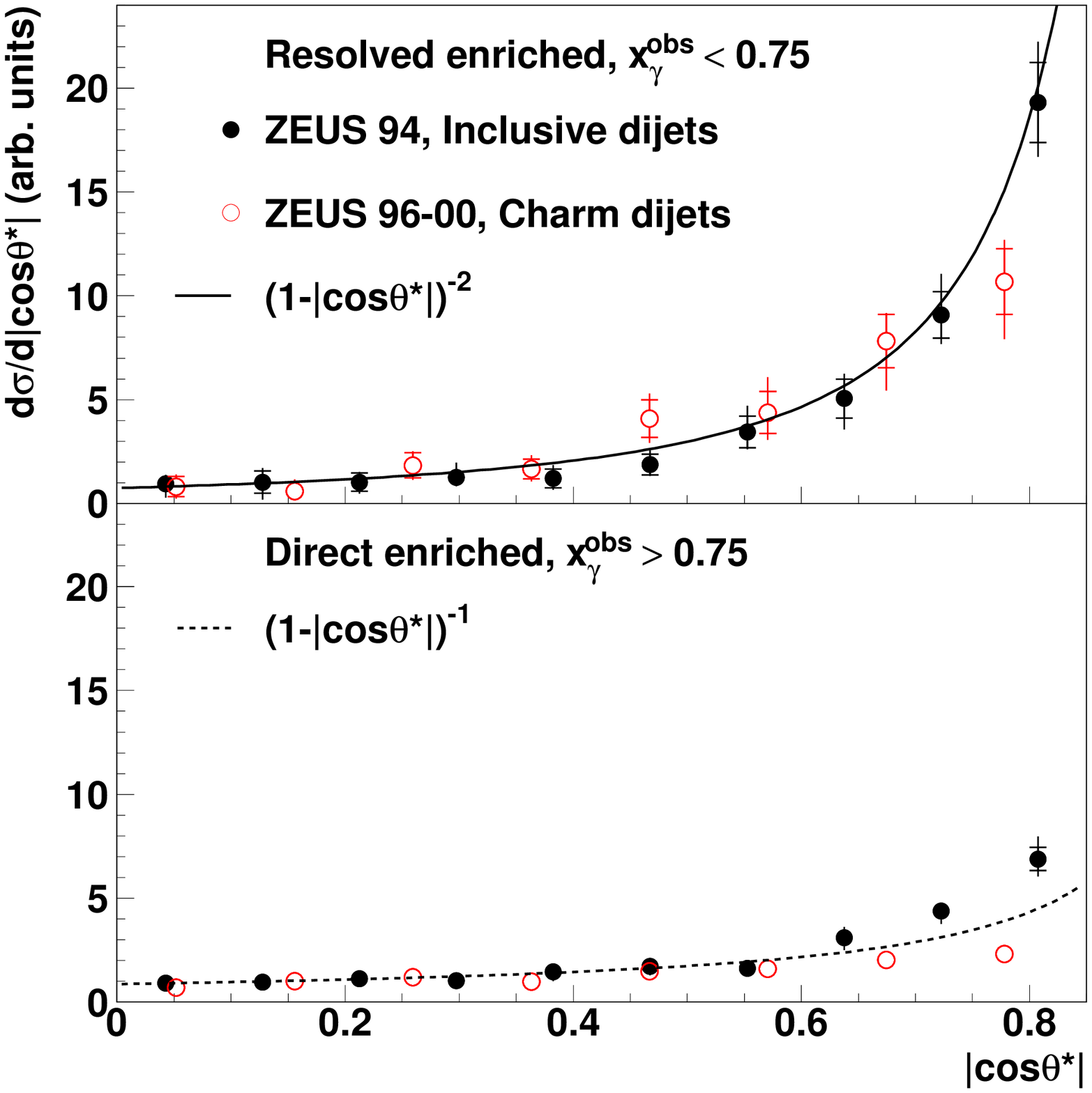,height=7.5cm}
~\epsfig{file=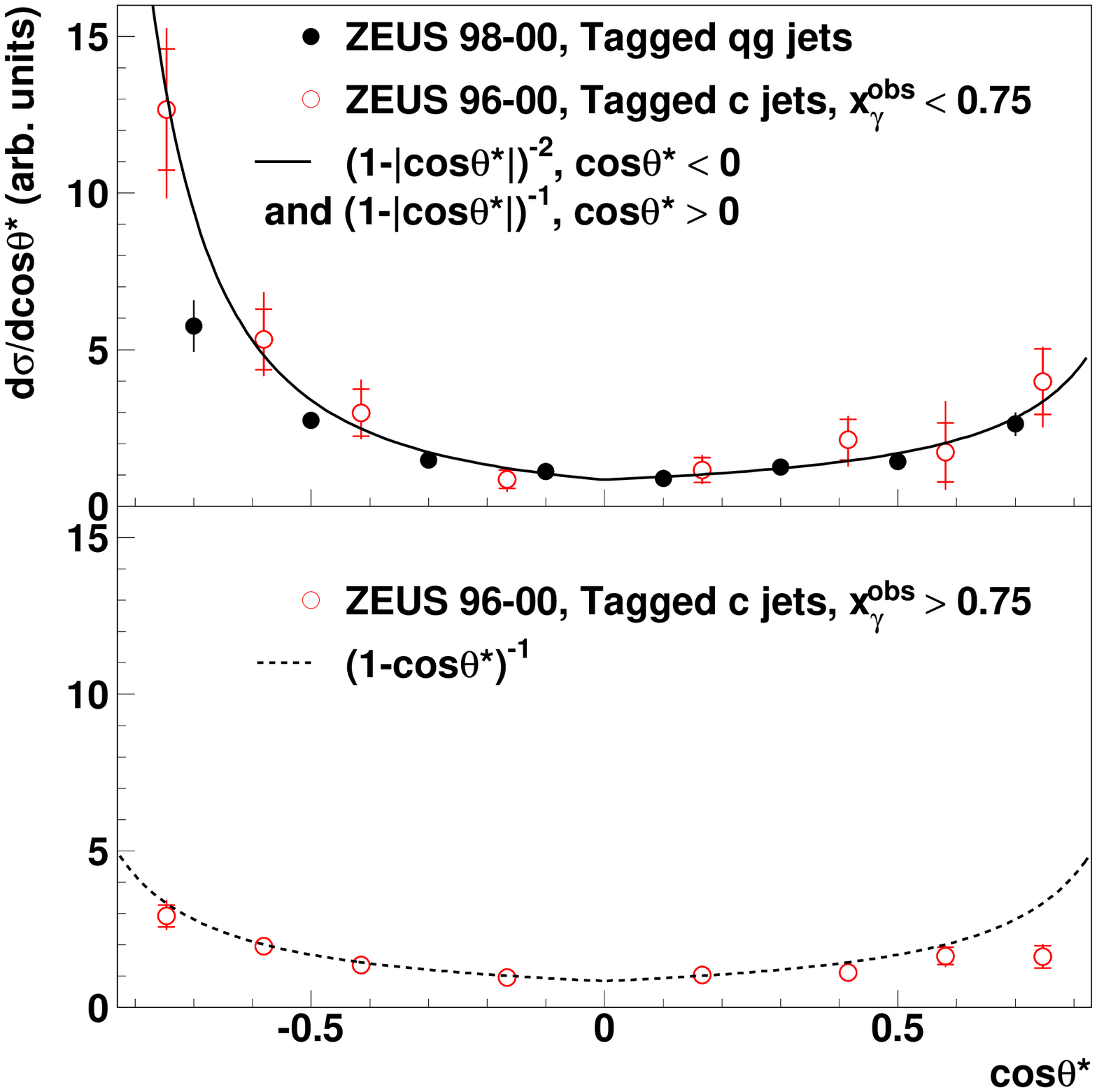,height=7.5cm}
\put(-332,6){\makebox(0,0)[tl]{\large (a)}}
\put(-107,6){\makebox(0,0)[tl]{\large (b)}}
\end{center}
\caption{Angular distributions inclusively (solid points) and for charm 
(open points) for high and low $x_\gamma^{\rm obs}$ compared to simple 
analytic functions.}
\label{fig:costheta}
\end{figure}

The angular dependence of the cross-section was first studied
inclusively in dijet photoproduction~\cite{pl:b384:401}, where due to
the indistinguishability of the two jets the modulus, $|\cos
\theta^*|$, was measured. The measurement was performed at relatively
low jet transverse energy, $E_T^{\rm jet} > 6$~GeV, although an
additional requirement on the mass of the dijet system, $M_{\rm jj} >
23$~GeV, ensured that the effective energy was higher. The data were
split into two regions, enhanced in direct photon ($x_\gamma^{\rm obs}
> 0.75$) and resolved photon ($x_\gamma^{\rm obs} < 0.75$) events. The
cross-sections were found to be well described by LO and NLO QCD
calculations~\cite{pl:b384:401} and are shown in
Fig.~\ref{fig:costheta}a compared with the simple analytic functions
expected from the dominant propagator. The cross-section for the
sample enriched in resolved photon events rises more rapidly than that
for the sample enriched in direct photon events. The function $(1 -
|\cos \theta^*|)^{-2}$ describes the low $x_\gamma^{\rm obs}$ data
well.  For high $x_\gamma^{\rm obs}$, the data are reasonably well
described by the function $(1 - |\cos \theta^*|)^{-1}$. The somewhat
steeper rise of the data could be due to contamination of resolved
photon events. The data clearly demonstrate the dominance of gluon and
quark propagators in resolved and direct photon events,
respectively. These trends were also seen at higher transverse
energies and masses~\cite{epj:c23:615}.

A measurement of dijet production in the presence of
charm~\cite{epj:c6:67} demonstrated a significant cross-section at low
$x_\gamma^{\rm obs}$. Within the framework of Monte Carlo predictions,
this was explained as arising from a large resolved-photon component
dominated by charm in the photon's structure function. This
interpretation was investigated further by measuring dijet angular
distributions in events containing a charm quark reconstructed by
tagging a $D^*$ meson~\cite{pl:b565:87}.  The dijet angular
distributions were measured in a similar energy range ($E_T^{\rm jet}
> 5$~GeV and $M_{\rm jj} > 18$~GeV) to that for inclusive dijet
production and the data were again split into samples enriched in
direct and resolved photon events by applying the same cut on
$x_\gamma^{\rm obs}$. The data are shown in Fig.~\ref{fig:costheta}a
compared with the inclusive data and analytic functions. The charm
data exhibits similar behaviour to that of the inclusive data and is
well described by the two functions. The differences at high $|\cos
\theta^*|$ between the data and the function can be explained by a
small bias due to angular cuts on the $D^*$ meson which are well
reproduced by QCD calculations~\cite{pl:b565:87}. In fact for high
$x_\gamma^{\rm obs}$, the charm data agrees better with the function
than does the inclusive data. The charm data in this region
constitutes a purer sample of BGF processes, since QCD Compton events
are heavily suppressed by the fact that the charm content of the
proton is much smaller than the gluon content. Calculations performed at NLO
in QCD describe the data in this region well.  The data at low
$x_\gamma^{\rm obs}$ indicate that the dominant process has a
gluon propagator. The data at low $x_\gamma^{\rm obs}$ are not well
described by an NLO calculation which does not have a charm component
in the photon PDF. This suggests that the photon has significant
component in which the photon splits into a $c \bar{c}$ pair.

By tagging one jet as arising from a charm quark, the sign of the
angle can be measured.  The tagged $D^*$ meson is associated with one
of the jets and the scattering angle of this jet defined with respect
to the proton direction. The data are shown in Fig.~\ref{fig:costheta}b
in the two regions of $x_\gamma^{\rm obs}$. The angular distribution,
enriched in direct photon processes ($x_\gamma^{\rm obs} > 0.75$)
exhibits a symmetric distribution with a shallow rise to high values
of $\cos\theta^*$. This is indicative of the exchange of a quark in
the hard sub-process with the charm produced via the boson-gluon
fusion process. At low $x_\gamma^{\rm obs}$, where the sample is
enriched in resolved photon processes, the data are asymmetric,
exhibiting a rapid rise to negative $\cos\theta^*$. This demonstrates
that the charm comes from the photon and exchanges a gluon in the hard
process. The shallow rise to positive $\cos\theta^*$ is consistent
with the exchange of the charm quark itself.

Also shown compared to the charm jets at low $x_\gamma^{\rm obs}$ are
dijets tagged as a quark and a gluon~\cite{Chekanov:2004kz}. No flavour
requirement is made for this quark which is defined with respect to
the proton direction. This measurement has been performed at higher
transverse energy and mass to reduce the effect of the underlying
event. The jets are tagged by considering the different jet
substructure for quark and gluon jets.  As gluons radiate more than
quarks, gluon-initiated jets are expected to be broader than those
initiated by quarks. Therefore a sample of events enriched in one
quark and one gluon jet is achieved by requiring one broad and one
narrow jet (see Section~\ref{sec:substructure}).  The purity is 52\%
with the largest contamination from $q \bar{q}$ events. It should be
noted that for the tagged $qg$ events, there is no cut on
$x_\gamma^{\rm obs}$; such a cut would improve the purity of the
sample but also reduce statistics. The data show similar behaviour to
the charm data and are reasonably well described by the function,
although the large background results in a shallower rise at high
negative $\cos\theta^*$. The consistency of the two sets of data
confirms that the charm is coming from the photon.

\subsection{Virtual photon structure}
\label{sec:virtual2}

The evolution of the jet cross-section with the photon virtuality, $\q2$, can 
provide information~\cite{epj:c37:141,epj:c35:487,epj:c13:397,pl:b479:37} on the 
structure of the photon. As discussed in Section~\ref{sec:virtual1}, the region 
above $\q2$ of about 1\,GeV$^2$ is (somewhat arbitrarily) defined as the domain 
of deep inelastic scattering. If no other hard scales are present, the photon 
acts as a pointlike probe.  However, since the jets also provide a hard scale, 
there is a complex situation between the limit 
$(\ETJ)^2 \gg \Lambda_{\rm QCD}^2 > \q2$ (definitely photoproduction) and 
$\q2 \gg (\ETJ)^2, \q2 \gg \Lambda_{\rm QCD}^2$ (definitely deep inelastic 
scattering). In particular, for the configuration 
$(\ETJ)^2 \gg \q2 \gg \Lambda_{\rm QCD}^2$, the virtuality of the photon 
regulates the collinear singularity in the photon structure and in principle 
the process is perturbatively calculable. In practice, large logarithms in the 
ratio of the scales given by $\ETJ$ and ${\rm Q}$ may be present and a 
resummation may be required to restore convergence of the perturbative series. 
It seems reasonable to suppose that such a resummation could take the form of 
a completely perturbative virtual photon structure. Amongst other things, this 
could offer a rather accurate determination of $\as$, one of the original goals 
of real photon structure measurements. Unfortunately, despite some advances in 
this direction~\cite{cpc:133:105}, no complete NLO calculation is currently 
available and the data are compared either to LO ans\"{a}tze~\cite{pl:b376:193} 
or to conventional NLO calculations for deep inelastic scattering or 
photoproduction.

\begin{figure}[htp]
\begin{center}
~\epsfig{file=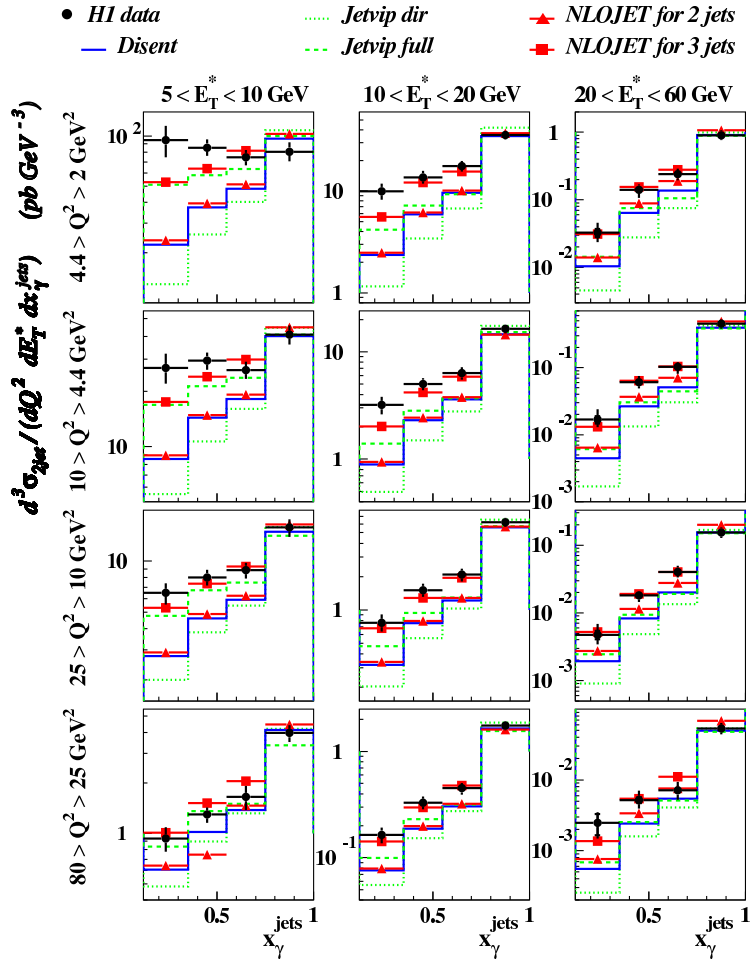,height=13.cm} 
\end{center}
\caption[*]{Triple differential cross section~\cite{epj:c37:141} compared to NLO 
         calculations containing only direct photon processes, DISENT (full line) 
	 and JETVIP (dotted line), the sum of direct and resolved photon processes, 
	 JETVIP (dashed line) and for NLOJET for two- (triangles) and three-jet 
	 (squares) production, NLOJET. Taken from~\cite{epj:c40:469}.}
\label{fig:virtual_photon}
\end{figure}

The H1 collaboration have measured triple differential cross-sections
as a function $\q2$, $\ETJ$ and $x_\gamma^{\rm obs}$ shown in
Fig.~\ref{fig:virtual_photon}.  A similar measurement by the ZEUS
collaboration~\cite{epj:c35:487} was also presented as the ratio of
the dijet cross-sections for low and high $x_\gamma^{\rm obs}$ as a
function of $\q2$. This ratio was also measured in the photoproduction
region {\it i.e.} down to $\q2 \sim 0$. The fraction of the low
$x_\gamma^{\rm obs}$ cross-section decreases with increasing $\q2$ for
all values of $\ETJ$. However, the fall becomes less rapid as $\ETJ$
increases. These trends are generally reproduced by
leading-logarithmic parton-shower Monte Carlo models (not shown) which
introduce a virtual photon structure both transversely and
longitudinally~\cite{pl:b488:289} polarised, suppressed with
increasing $\q2$.  NLO photoproduction calculations are consistent
with the low $\q2$ data~\cite{epj:c35:487}. However, the cross-section
is not well-reproduced by deep inelastic scattering NLO calculations
for $\q2 > 1-2~\gev2$. These calculations, which assume no explicit
photon structure, have large uncertainties at low $\q2$ and generally
underestimate the data. Such effects may well be a sign of the large
logarithms mentioned above. A NLO calculation,
JETVIP~\cite{cpc:133:105}, incorporating processes with a virtual
photon is also shown in Fig.~\ref{fig:virtual_photon}. This improves
the description of the data significantly. However, the calculation is
sensitive to a technical parameter in the calculation and so is not
reliable. The best description~\cite{epj:c40:469} of the data is
achieved by a NLO calculation for three-jet production, which has a
significant cross-section at low $x_\gamma^{\rm obs}$. This
demonstrates the sensitivity of this cross-section to higher-order
effects and the need for a full calculation including a virtual
photon.

\begin{figure}[htp]
\begin{center}
~\epsfig{file=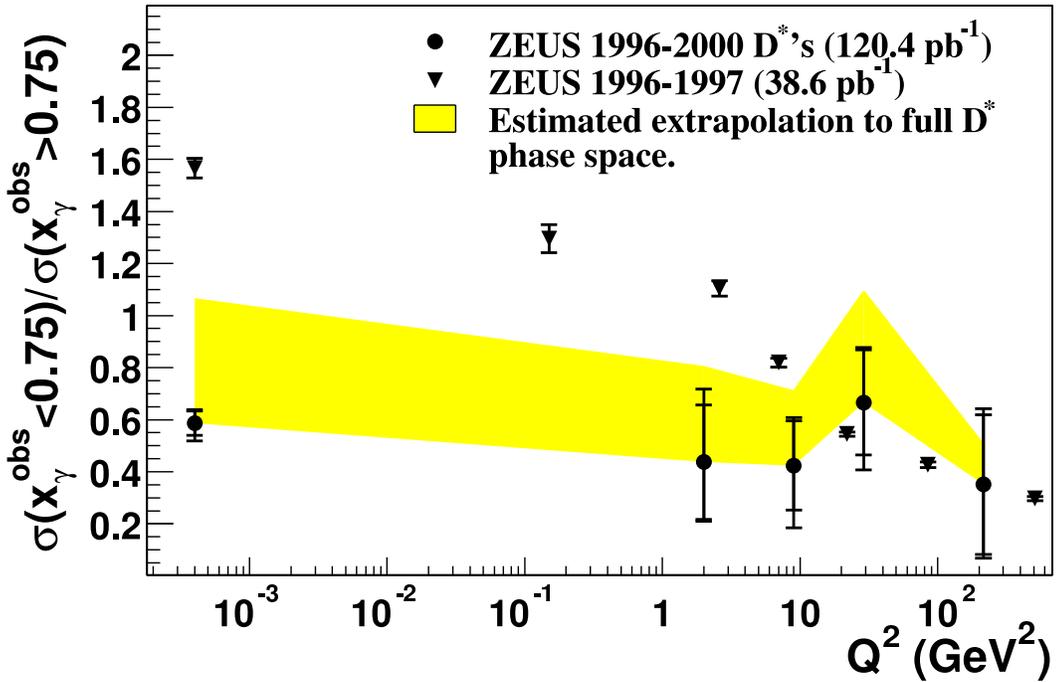,height=9cm}
\end{center}
\caption{Dependence of the ratio of low- to high-$x_\gamma^{\rm obs}$
cross-sections on the virtuality of the incoming photon for inclusive
(triangles) and charm (dots) dijet production (taken
from~\protect\cite{thesis:lightwood:2004}).}
\label{fig:virtual_photon_c}
\end{figure}

The ratio of the cross-sections for low and high $x_\gamma^{\rm obs}$ has also 
been studied~\cite{thesis:west:2001,thesis:lightwood:2004} for a more exclusive 
process where the event contains at least one charm quark, by tagging a $D^*$ 
meson. The result for charm is then compared to the previous inclusive 
({\it i.e.} all quark flavours) result in Fig.~\ref{fig:virtual_photon_c}. The 
result is, however, complicated by the $D^*$ meson which adds an extra kinematic 
constraint on one of the jets not present in the inclusive sample. This can only 
be corrected for using a Monte Carlo model; after this correction, the 
conclusions still remain the same. A difference is evident in the presence of 
charm; the ratio is flat as a function of $\q2$ and in the photoproduction region 
lies well below the inclusive data. This can be understood in terms of the 
effect of the extra scale, the charm mass, which in addition to ${\rm Q}$ and 
$E_T$ leads to a suppression of the photon structure.

\section{Photoproduction and QCD}

Given the double identity of the photon as a fundamental gauge boson
and a hadron-like object, studies of photoproduction reactions provide
a number of unique opportunities for exploring QCD.  Combining data
from photon-proton collisions with deep inelastic scattering (DIS),
photon-photon and proton-(anti)proton collisions can be a particularly
fruitful exercise. In this chapter we explore the major areas in QCD
where photoproduction data have had such an impact.

\subsection{Proton structure} 
\label{sec:proton_structure}

The parton densities in the proton constitute one of the best windows
onto the physics of the strong interaction. They encapsulate the
behaviour of quarks and gluons over a huge range of distance scales
and densities, and their scale dependence provides one of the most
precise measurements of the strong coupling, $\as$. In addition,
parton densities are a crucial input to all cross-section calculations
at proton colliders. Finally, measurements of a range of different
cross-sections at hadron colliders, all of which make use of the same
parton densities, provide one of the most stringent tests of QCD, and
in particular of factorisation.

As with photon structure (see Section~\ref{sec:photon_structure}), the
principle measurements used to extract information of the proton's
parton density function (PDF) are measurements of the structure
functions in DIS, in particular the structure function $\f2$. These
data are fitted by several
groups~\cite{Kretzer:2003it,Martin:2004dh,epj:c42:1,Adloff:2003uh,Alekhin:2002fv,epj:c5:461}
using the conventional DGLAP evolution equations. Put simply, this
approach makes no prediction as to the $x$ dependence of the PDFs
(apart from information indirectly derived from the conservation of
momentum) but for $\q2$ sufficiently above $\lqcd^2$ the $\q2$
dependence ({\it i.e}. the scaling violation) is accurately predicted. Fits
are currently carried out to NLO accuracy. However, the full
calculation of the DGLAP evolution to NNLO has recently been
performed~\cite{Vogt:2004mw,*Moch:2004pa,*Moch:2004xu,*Vermaseren:2005qc}. This
shows that the perturbative series converges well and that in most
cases the NLO approach is rather a good approximation.

Measurements of $\f2$ are only directly sensitive to the quark
densities. The gluon density is obtained from the fits since gluon
radiation from quarks, as well as gluon splitting to quarks, drives
the scaling violations. Nevertheless, measurements which are more
directly sensitive to the gluon density are important, partly to test
the QCD picture but also because they potentially have a
sensitivity to higher $x$ than do extractions based on scaling
violations. Since high-$x$ gluons are responsible for a big fraction
of the highest energy collisions in the world (currently at the
Tevatron and soon at the LHC), the high-$x$ gluon density is a very
important object to pin down.

Prompt photons and high transverse energy ($\ETJ$) jets are two
processes which have been used in fits to help constrain the
gluon. However, theoretical uncertainties (in the case of prompt
photons) and experimental uncertainties (in the case of jets) have
limited the impact of these data. In addition, high-$\ETJ$ jet 
cross-sections at the Tevatron are sensitive to new physics, and much of
this sensitivity is lost if the data is simply used to fit the gluon
density~\cite{Stump:2003yu}.

High $\ETJ$ jet photoproduction is also directly sensitive to the
gluon density in the proton. For direct photon events, this is via the
boson-gluon fusion diagram, Fig.~\ref{fig:feyn}a. Resolved events are
also sensitive (Fig.~\ref{fig:feyn}b) but the situation here is
complicated by the photon PDFs, which are less well known than those
of the proton. At high transverse energies (which also means high $x$)
the uncertainties due to hadronisation and underlying events are also
rather small.

\begin{figure}[!htp]
\begin{center}
~\epsfig{file=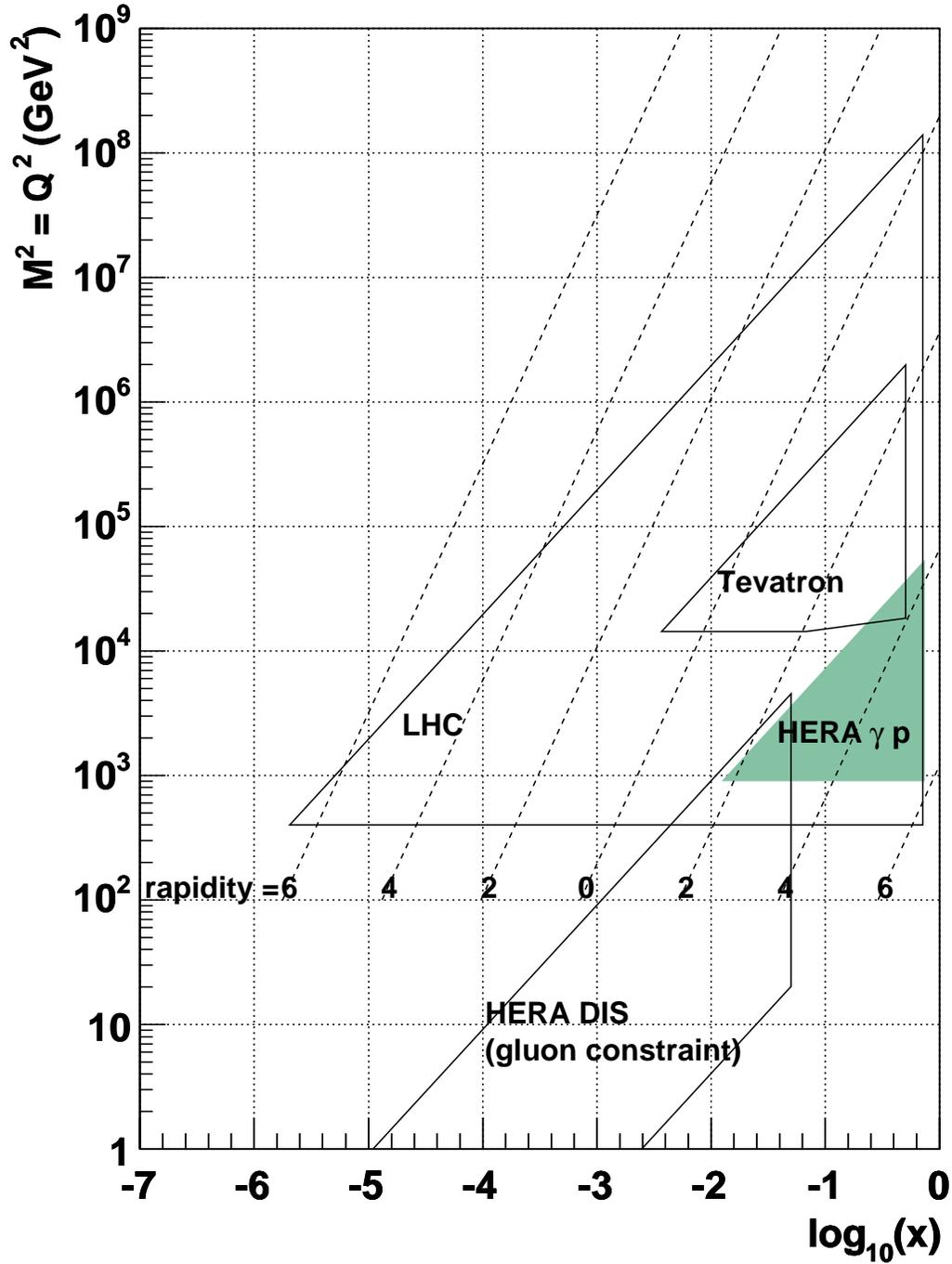,height=18cm}
\caption[*]{\label{fig:kinematicplane} The parton kinematic plane. The
approximate regions in which the gluon in the proton is constrained by
various experiments are shown, as well as the whole region of
sensitivity at the LHC. The variable $M^2 = \q2$ is the dijet mass for
LHC, Tevatron and photoproduction, and the negative
four-momentum-transfer-squared for DIS. The dashed lines show the
rapidity of the dijet system at the LHC. The assumptions are that
precise jet measurements are possible down to $\Et = 60\,\Gev, \eta =
2.0$ at Tevatron. The HERA jet region is taken from
\cite{epj:c23:615}.}
\end{center}
\end{figure}

Figure~\ref{fig:kinematicplane} shows the parton kinematics for
several experiments. The region over which jet measurements in
photoproduction at HERA constrain the gluon density is shown, as well
as the regions where Tevatron and HERA DIS measurements provide
constraints. The HERA DIS data, as well as fixed target data, extend
to higher $x$ and $\q2$ than shown, but give little constraint on the
gluon in that region. The region over which the LHC is sensitive to
the gluon density is also shown. However over much of this region,
actual measurement of the gluon at the LHC will be extremely
difficult since the main sensitivity is at low $\ETJ$, where
other uncertainties due to hadronisation and the underlying event are
important.

\begin{figure}[!thb]
\begin{center}
~\epsfig{file=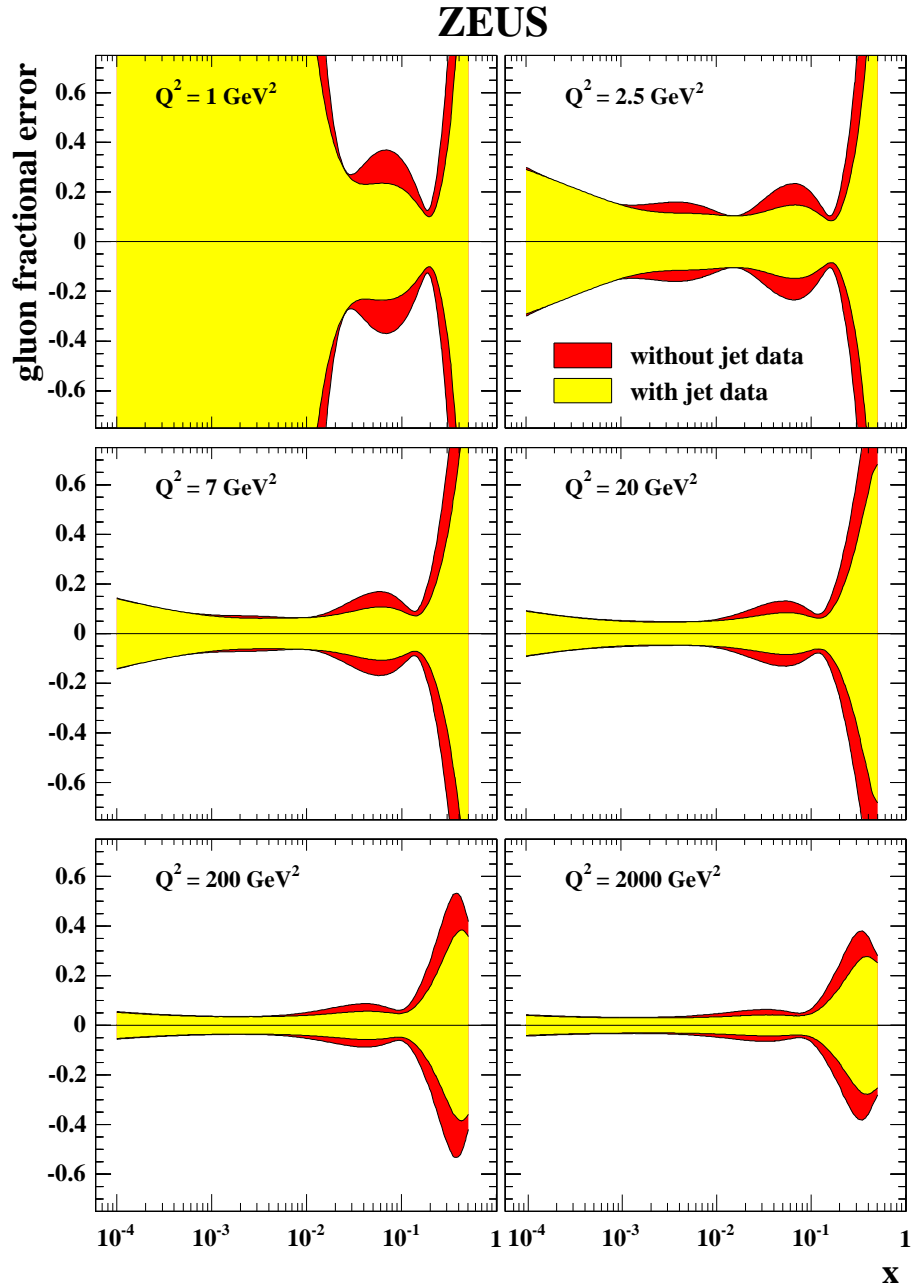,height=17cm}
\caption[*]{The fractional uncertainty in the gluon PDF before
(outer+inner shaded areas) and after (inner shaded area) the inclusion of DIS and
photoproduction jet data in the QCD fit~\cite{epj:c42:1}.
\label{fig:fiterrors}}
\end{center}
\end{figure}

The ZEUS collaboration has published~\cite{epj:c42:1} the first QCD
fits which incorporate jet photoproduction data~\cite{epj:c23:615}
(as well as DIS jet data~\cite{Chekanov:2002be}). The inclusion
significantly reduces the uncertainties on the gluon density at
intermediate and high $x$, as shown in Fig.~\ref{fig:fiterrors}.  This
is a graphic demonstration of the power of high energy photoproduction
data, but is not the full story. The data set used in these results
was taken in 1996-1997, and is statistically limited at high $\ETJ$,
the kinematic region corresponding to high $x$. This is only around
a third of the HERA I data set; HERA II should increase this again by a
factor of around 5. 

As well as simply using more data, the constraint on the gluon density
could also be improved by optimising the cross-section for sensitivity
to the high $x$ gluon by (for example) extending to the forward
region.  In such studies~\cite{cta,*Butterworth:2004is} it has also
been observed that resolved photon cross-sections exhibit a large
sensitivity to the gluon in the proton. For the existing
cross-sections, including the low-$\xgo$ cross-sections in the fit
reduces the statistical errors, but increases the systematic errors
due to the uncertainties associated with the photon structure, so that
there is no net benefit~\cite{epj:c42:1}. Currently no NLO fit for the
photon structure with full uncertainties exists, and so even this
statement is an estimate. Obviously, a combined NLO QCD fit to the
proton and the photon PDFs would bring great benefits. Finally, it is
also possible to determine $\as$ from such fits, as is discussed in
Section~\ref{sec:as}.

It is to be hoped that these possibilities will be vigorously pursued
by the fitting groups and experimental collaborations during HERA II.
The whole topic of parton distributions and their uncertainties is a
very active area at present. Reviews and recent developments may be
found elsewhere~\cite{heralhc,Devenish:2004pb}.

\subsection{Measurement of the strong coupling constant} 
\label{sec:as}

\begin{figure}[htp]
\begin{center}
~\epsfig{file=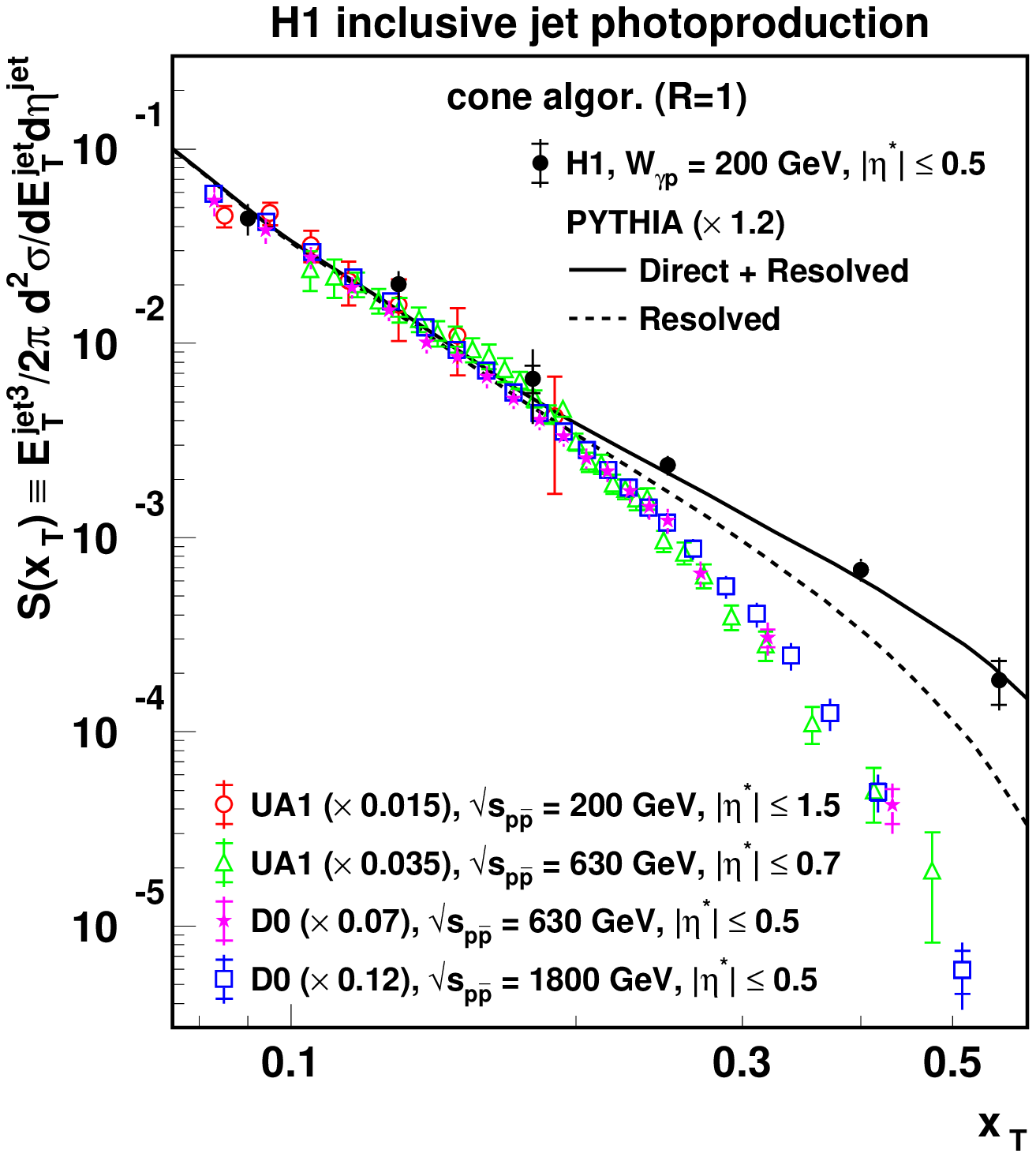,height=7cm}
\put(-25,132){\makebox(0,0)[tl]{\large (a)}}
\put(174,172){\makebox(0,0)[tl]{\large (b)}}
~\epsfig{file=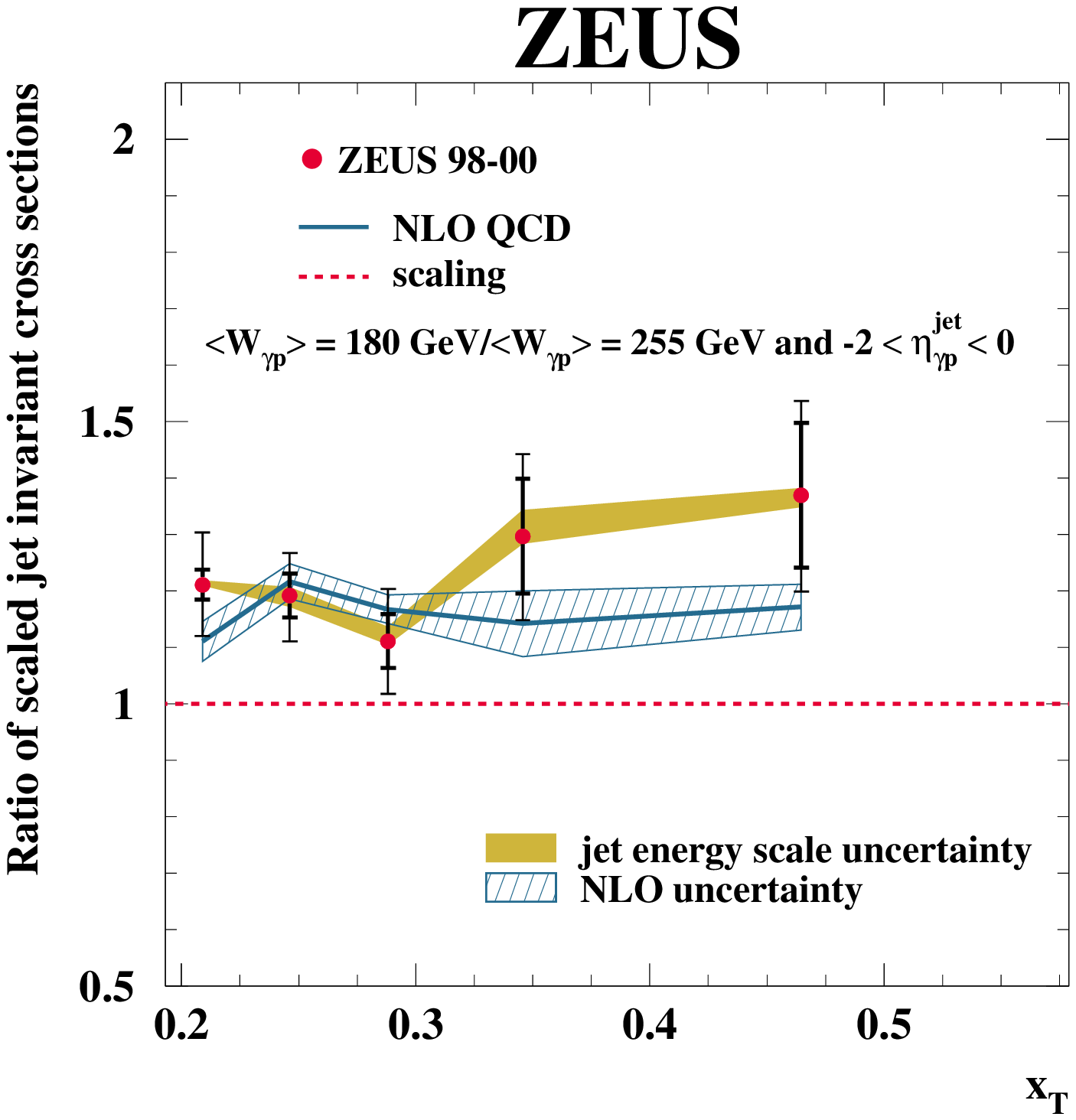,height=7cm}
\end{center}
\caption{(a) Scaled jet cross-sections at one value of $W_{\gamma p}$ compared 
with results from $p\bar{p}$ collisions and {\sc Pythia}. (b) Ratio of scaled jet 
cross-section at two different $W_{\gamma p}$ values compared with NLO QCD 
predictions.}
\label{fig:scaling_violations}
\end{figure}

According to QCD, jet cross-sections should exhibit a non-scaling
behaviour due to both the evolution of the structure functions of the
colliding beams and to the running of $\as$. Scaling violations
have been seen both in measurements of the inclusive deep inelastic
scattering cross-section, $F_2$ and for jet cross-sections in
$p\bar{p}$ collisions~\cite{prl:70:1376,prl:86:2523}. In the absence
of scaling violations, the ratio of jet invariant cross-section,
$(E_T^{\rm jet})^4 E^{\rm jet} d^3\sigma/dp_x^{\rm jet}dp_y^{\rm
jet}dp_z^{\rm jet}$, at one $\gamma p$ centre-of-mass energy,
$W_{\gamma p}$ to that at any other energy would be unity for all $x_T
\equiv 2 E_T^{\rm jet}/W_{\gamma p}$. Measurements of inclusive
jet photoproduction~\cite{Adloff:2003nr} are shown in
Fig.~\ref{fig:scaling_violations}a compared directly with data from
$p\bar{p}$
collisions~\cite{prl:86:2523,prl:82:2451,*pl:b172:461,*np:b309:405}. The
shape of the distributions for $\gamma p$ and $p\bar{p}$ data are
compatible for $x_T < 0.2$.  For larger values of $x_T$, the harder
distribution in $\gamma p$ collisions arises from the increasing
proportion of direct processes and the increasing proportion of the
quark density in the resolved photon relative to that in the
proton. These results are similar to the inclusive particle production
discussed in Section~\ref{sec:res_dir} which led to the observation
of the direct component. To test the scaling hypothesis, the ratio of
the scaled jet invariant cross-sections as a function of $x_T$ are
shown in Fig.~\ref{fig:scaling_violations}b for two chosen values of
$W_{\gamma p}$~\cite{pl:b560:7}. A clear deviation from unity, in
agreement with NLO QCD predictions, is seen. This is the first
observation of scaling violations in $\gamma p$ cross-sections.

\begin{figure}[htp]
\begin{center}
~\epsfig{file=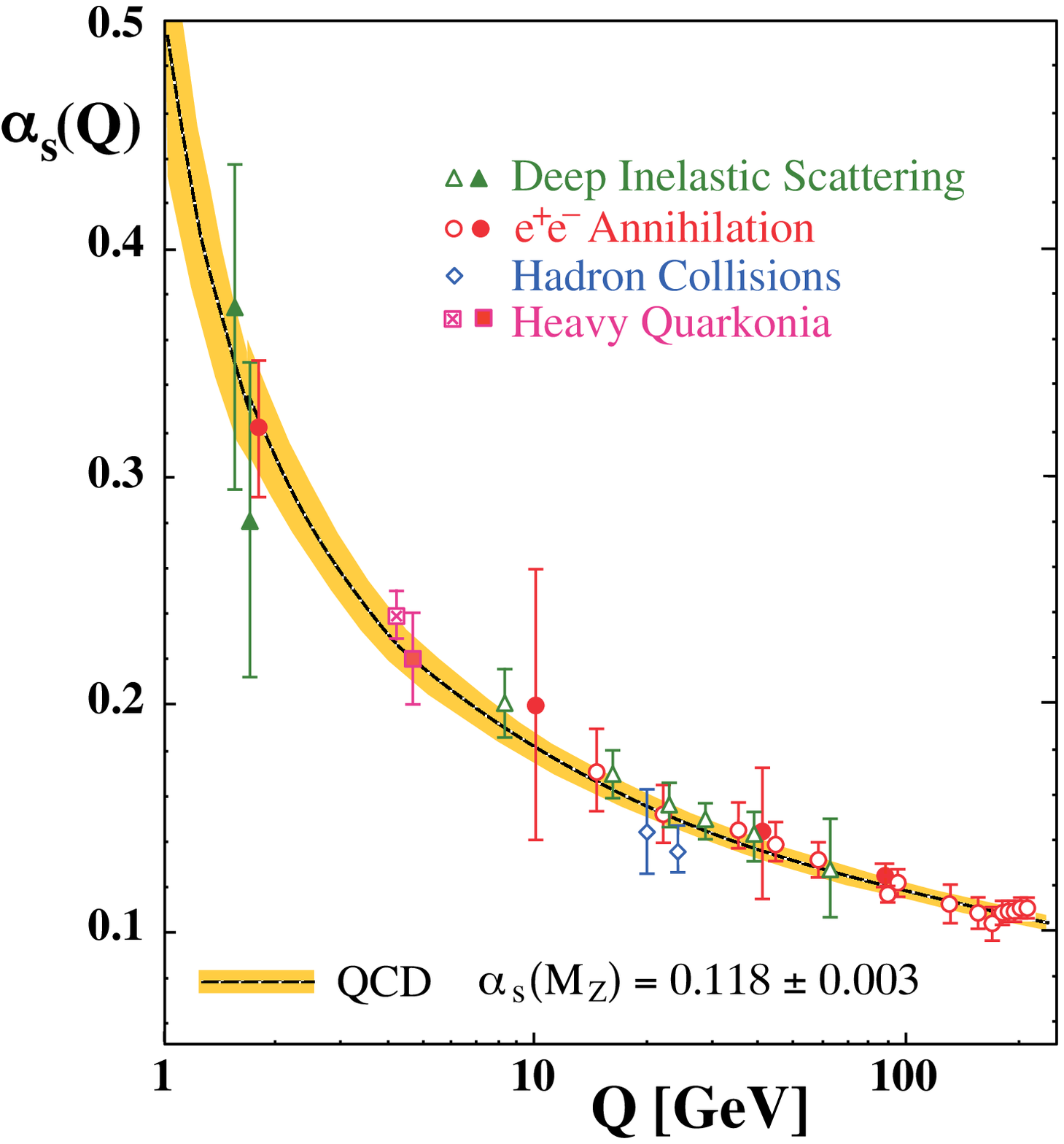,height=7cm}
\put(-27,192){\makebox(0,0)[tl]{\large (a)}}
~\epsfig{file=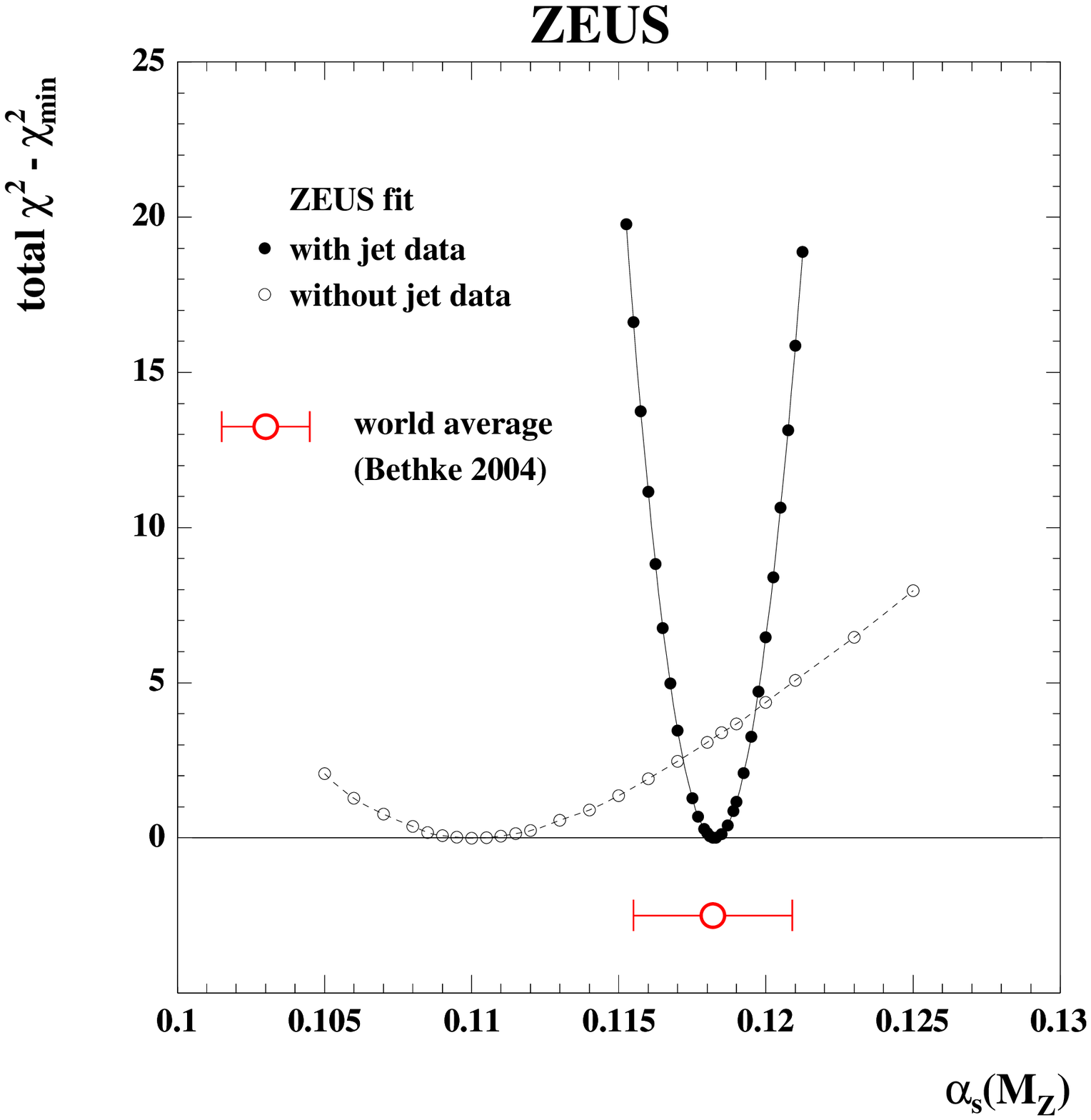,height=7.25cm}
\put(-27,192){\makebox(0,0)[tl]{\large (b)}}
\caption[*]{(a) The running of $\as$ from all 
experiments~\cite{proc:zinnowitz:2004:bethke}. The HERA data in both 
photoproduction and deep inelastic scattering are shown as the open triangles. (b) 
The $\chi^2$ profile for $\as$ before and after the inclusion of DIS and photoproduction 
jet cross-sections into the QCD fit~\cite{epj:c42:1}.}
\label{fig:fitalphas}
\end{center}
\end{figure}

The inclusive jet cross-section over a wide range of $E_T^{\rm jet}$ also allows a 
determination of $\as$ and its energy scale dependence. The high precision 
of the data, in particular the well-understood energy scale of the calorimeter, 
yielded a value of $\as$ with a total experimental uncertainty of less than 2\%. 
The limitation of the analysis arises from the theory which has an uncertainty of 4\%. 
This measurement of $\as$ and the many in deep inelastic scattering have also 
clearly demonstrated the energy scale dependence or ``running'' of $\as$. The 
current collection~\cite{proc:zinnowitz:2004:bethke} of all data is shown in 
Fig.~\ref{fig:fitalphas}a. The consistency of the data with the running hypothesis 
is clearly seen and the impact of the HERA data is significant. 

The theoretical calculation used to fit the jet data also uses a parametrisation of 
the proton structure from global fits to deep inelastic scattering and other data. 
However, combining jet and inclusive $F_2$ data as described in the previous section, 
the full correlation of the gluon density and $\as$ can be accounted for and these 
both accurately extracted from HERA data alone~\cite{epj:c42:1}. The impact of the 
jet data can be seen in Fig.~\ref{fig:fitalphas}b, where the $\chi^2$ profile is shown 
for the fit to inclusive deep inelastic scattering data with and without the jet 
measurements. The extracted value is $\as(M_Z) = 0.1183 \pm 0.0028 {\rm (exp.)} 
\pm 0.0008 {\rm (model)}$.

\subsection{Open heavy-quark production}
\label{sec:hf}

Photoproduction of heavy quarks is again governed by QCD within the
Standard Model. 
Figure~\ref{fig:feyn_ee_gg}b shows the production of heavy quarks in a
resolved $\gamma p$ collision. In fact, a similar diagram would apply
for resolved $\gamma \gamma$ collisions, or hadron-hadron
process. Thus knowledge accumulated at HERA, LEP and the Tevatron will
directly benefit future programmes such as the LHC and a future linear
collider where heavy quarks will be produced by the same
mechanism. Since heavy quarks feature in many signatures for new
physics, understanding the QCD production mechanisms becomes even more
important.

Theoretically, heavy quarks provide ideal tools for probing QCD due to
their relatively large mass, $m_Q \gg \Lambda_{\rm QCD}$, which entails
a fast convergence of the perturbative expansion of the cross-section. 
The production of heavy quarks is also {\em directly}
sensitive to the gluon density in the colliding hadron (see
Fig.~\ref{fig:feyn_ee_gg}b). The gluon density is usually determined
in the DGLAP-evolution fits to measurements of structure functions in
inclusive deep-inelastic scattering, and more recently by including
jets (Section~\ref{sec:proton_structure}).  In the case of jets, the
non-zero photon virtuality (in DIS) or the $\xgo$ cut (in
photoproduction) reduces the uncertainties associated with photon
structure. Requiring charm in the final state is another method of
achieving the same effect, with a different set of associated benefits
and problems.

Several measurements of charm photoproduction have been made at HERA,
some of which were discussed in Section~\ref{sec:dual} in the context
of the information they provide on the structure of the
photon. Measurements of charm production in deep inelastic scattering
provide information on the charm contribution to the proton structure
function. This subject, although related, is outside the remit of this
review and the interested reader is referred to the most recent
relevant publications~\cite{pr:d69:012004,*epj:c40:349}.  There are
also several results on charm production from the HERA experiments
which are preliminary and not yet published. Only the published
results on charm and beauty production will be discussed here.

Much of the recent interest in heavy-quark production arose from
discrepancies between data from the Tevatron
experiments~\cite{prl:71:500,*prl:71:2396,*pr:d53:1051,*pr:d55:2546,*pl:b487:264},
CDF and D0, and NLO QCD predictions. The data, measured in many decay
channels, was a factor of $2-3$ above NLO QCD for all measured
regions. This was in contrast to earlier UA1
results~\cite{pl:b256:121} at the $Sp\bar{p}S$ collider and led to a
mini-crisis in QCD. Many explanations were put forward as to reasons
for this discrepancy: the input parameters to the QCD calculations
such as the structure function of the proton and the beauty-quark
mass; inaccurate values for the extrapolation of the data outside the
measured region; and extensions beyond the Standard Model such as the
presence of a light supersymmetric
particle~\cite{prl:86:4231}. Results from HERA have contributed to the
understanding of heavy-quark production in QCD as discussed below.
Measurements from other colliders are also briefly discussed.

A measurement of inclusive $D^*$ mesons currently provides the largest
data sample for charm production at HERA. Experimentally, the tagging
of $D^*$ mesons via the decay chain $D^* \to D^0 \pi_s \to K \pi
\pi_s$ is very clean and although the branching ratio is small
($\approx 2\%$) high-precision physics with about 60~000 events is
possible~\cite{epj:c38:29}. Published results on $D^*$ photoproduction
cross-sections use a smaller sample but already are of sufficient
precision for an informative comparison with QCD calculations. Data
can be compared to different types of calculations: the ``massive''
and ``massless'' scheme. In the fixed-order, or ``massive'', scheme,
$u$, $d$ and $s$ are the only active quarks in the proton and
photon; charm and beauty are only produced in the hard scatter. This
scheme is expected to work well in regions where the transverse
momentum of the outgoing $c$ quark is of the order of the quark
mass. At higher transverse momenta, the ``massless'' scheme should be
applicable in which charm and beauty are regarded as active flavours
(massless partons) in the structure functions of the proton and photon
and are fragmented into massive quarks only after the hard process. A
calculation which matches the two schemes, called FONLL~\cite{Cacciari:1998it},
is also available for inclusive $D^*$ meson production at HERA but has
not yet been compared to published data.

\begin{figure}[htp]
\begin{center}
~\epsfig{file=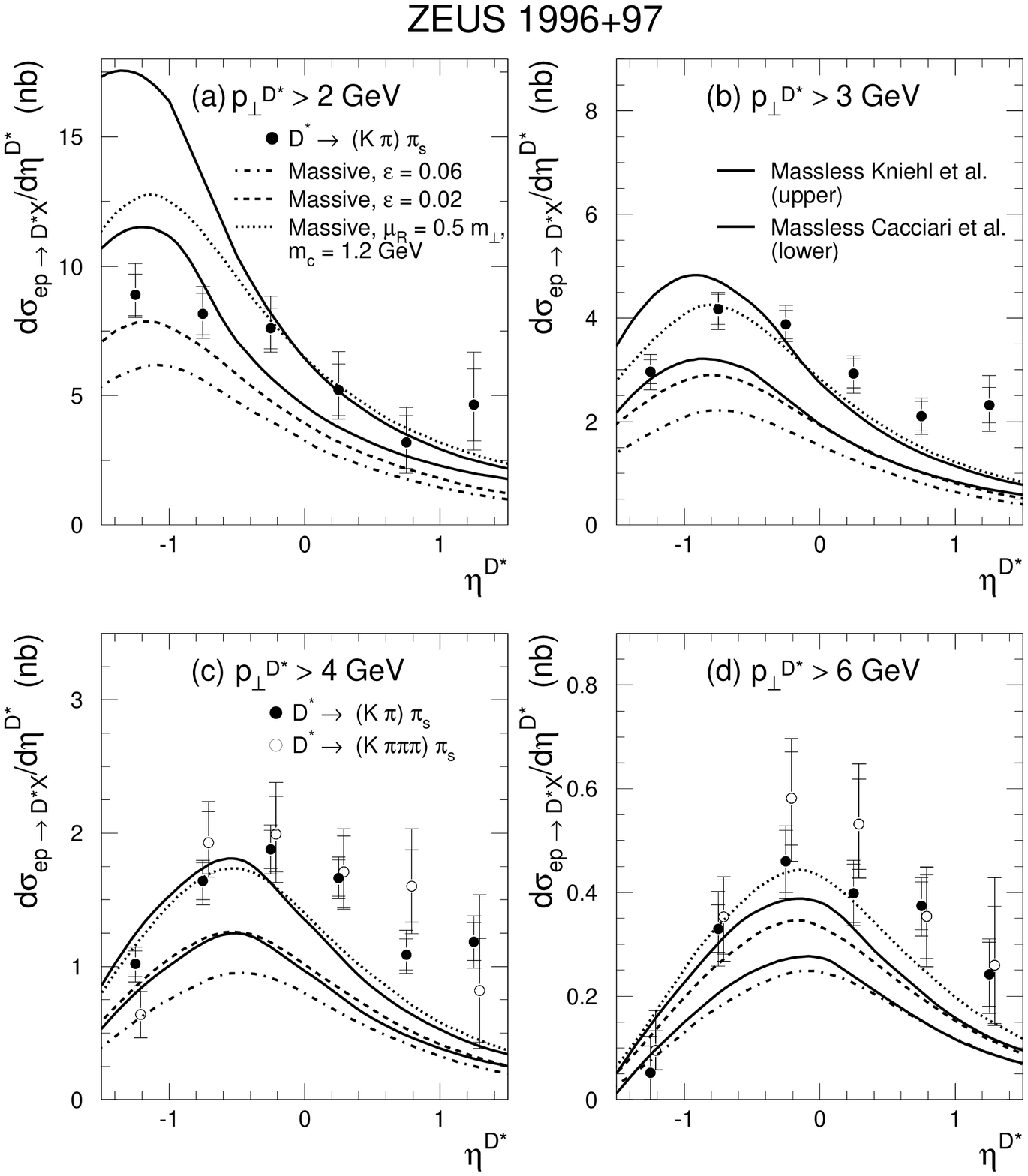,height=10.3cm}
~\epsfig{file=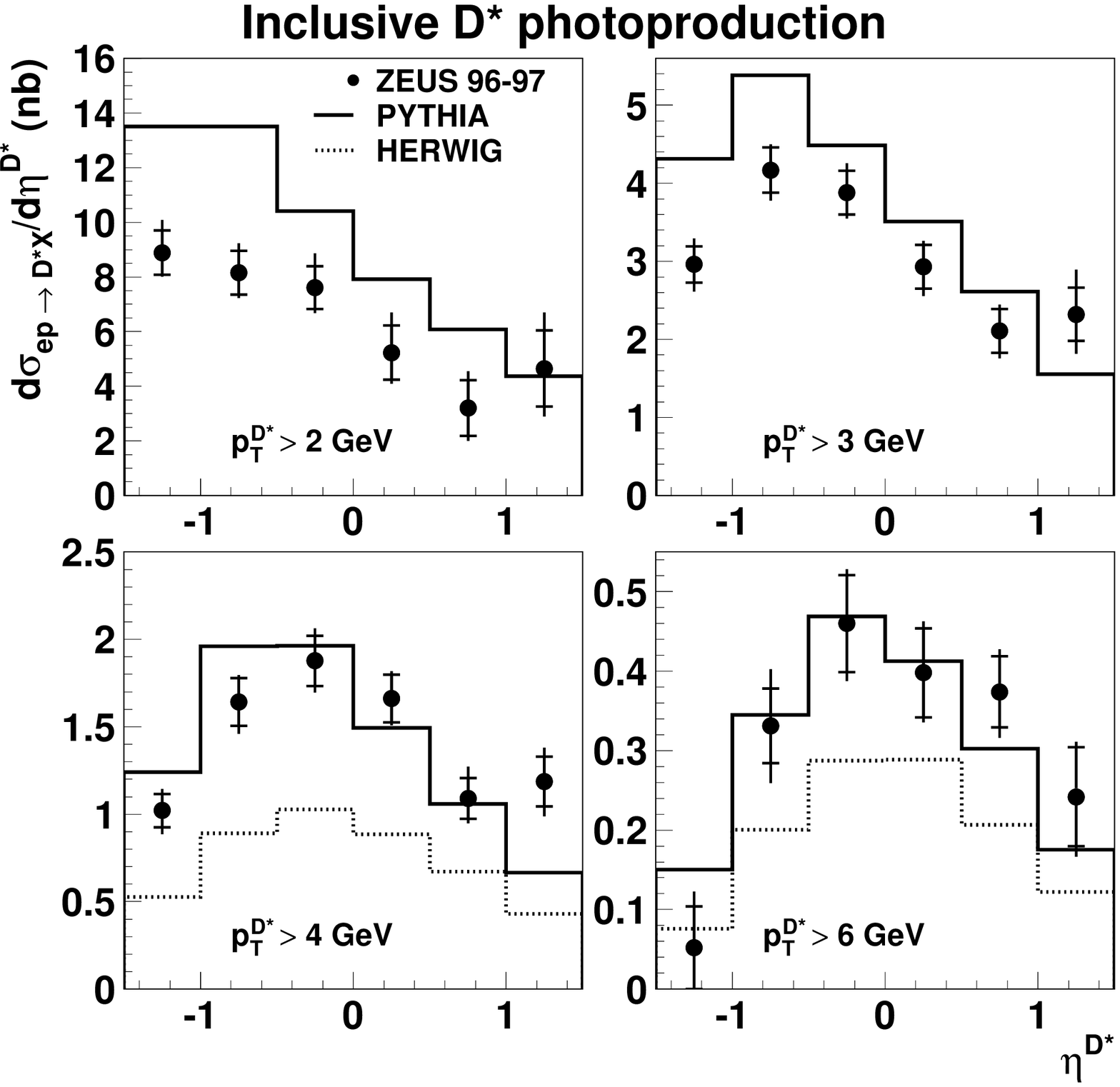,height=9.2cm}
\end{center}
\vspace{-0.5cm}
\caption{Cross-sections $d\sigma/d\eta^{D^*}$ for different regions of
$p_T^{D^*}$ in charm photoproduction at HERA. The data are compared to
calculations of NLO QCD in the massive and massless schemes. The data 
are also compared to predictions from {\sc Pythia} and {\sc Herwig} Monte 
Carlo programs. Note the {\sc Herwig} prediction is not shown for the 
two lowest $p_T^{D^*}$ regions as there is an implicit minimum parton 
$p_T$ cut in the calculation which would make the comparison invalid.}
\label{fig:c_hera} 
\end{figure}

Data as a function of the pseudorapidity of the $D^*$ meson are shown
in Fig.~\ref{fig:c_hera} compared to two different types of NLO QCD
calculations for charm production. Although roughly compatible with
the data, none of the calculations fully reproduce the trends of the
data, with discrepancies concentrated at high values of $\eta^{D^*}$
and medium $p_T^{D^*}$. This could hint at, for example, a low value
of the charm-quark mass or inadequacies in the fragmentation
model. 

In making the predictions, the fragmentation of the charm quark into a
$D^*$ meson is performed using the Peterson function~\cite{pr:d27:105}
fitted to $e^+e^-$ data. The applicability of this is unclear at HERA
particularly at high $\eta^{D^*}$, {\it i.e.} in the direction of the
proton, where the proton remnant may affect the production of the
$D^*$ meson. Clearly more phenomenological work on fragmentation is
needed before stronger conclusions can be drawn. Combining NLO
calculations with sophisticated models of hadronisation available in
Monte Carlo models, such as the MC@NLO
programme~\cite{jhep:0206:029,*jhep:0308:007} is one promising path to
understanding charm production in more detail. Another is to check the
assumptions which go into the calculations in more detail. For
example, the universality of hadronisation may be tested by measuring
the production fractions for different charmed hadrons, as has been
done very recently in DIS~\cite{Aktas:2004ka} and
photoproduction~\cite{zeusfrag}. In general the results are in
agreement and, in the case of the photoproduction data, of competitive 
precision with $e^+e^-$ data. However, in photoproduction the fraction
of charm quarks fragmenting to $\Lambda_c^+$ is around $3\sigma$
higher and the $D^{*+}$ fraction lower by a similar amount than the
corresponding fractions in $e^+e^-$. About half of the difference in
the $D^{*+}$ fraction is due to the $\Lambda_c$ difference. Since
there is no asymmetry between $\Lambda_c^+$ and $\Lambda_c^-$, this
effect, if real, is unlikely to be due simply to baryon number
flow. However, one may speculate that it might be a hint of some
non-trivial effects in fragmentation due to the incoming proton.

\begin{figure}[htp]
\begin{center}
~\epsfig{file=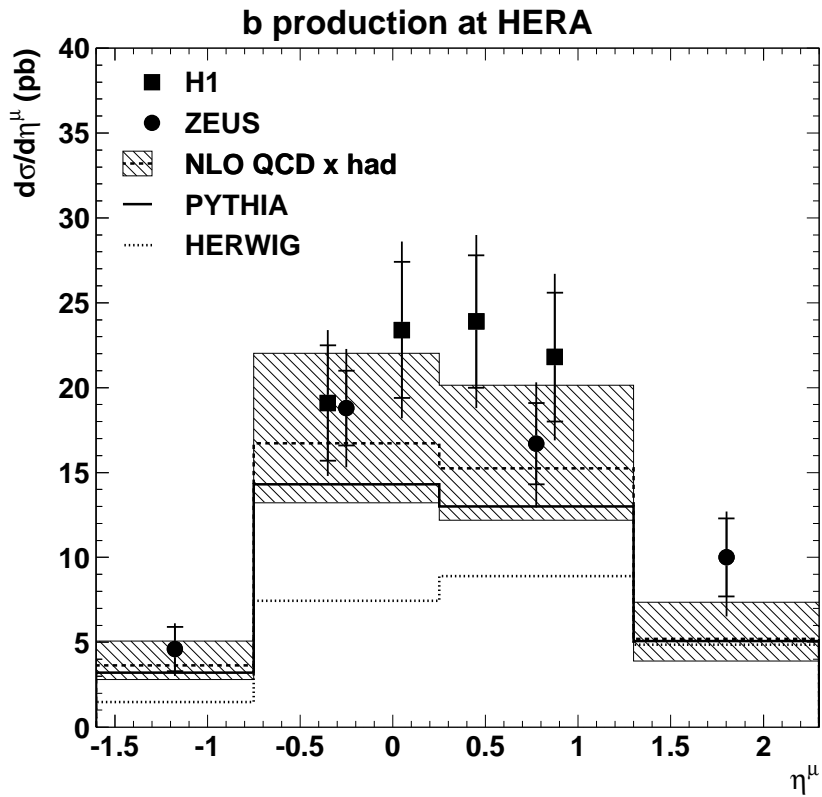,height=10.95cm}
\end{center}
\caption{Cross-section $d\sigma/d\eta^\mu$ for beauty photoproduction
at HERA. The data are compared to predictions of NLO QCD in the
massive scheme and from {\sc Pythia} and {\sc Herwig} Monte
Carlo programs.}
\label{fig:b_hera}
\end{figure}

The first measurements~\cite{pl:b467:156,*epj:c18:625} of beauty
production at HERA were higher than the NLO prediction by a similar
factor to those from the Tevatron. However, given the large
uncertainties on the ZEUS result it was consistent with the
theory. The precision of the data subsequently improved, and by
measuring cross-sections in kinematic regions where the detector
acceptance is good (rather than relying on models to extrapolate into
unmeasured regions), the estimates of the systematic errors also
became more reliable. The cross-section as a function of the
pseudorapidity of the decayed muon for beauty events is shown in
Fig.~\ref{fig:b_hera}. Data from both H1~\cite{epj:c41:453} and
ZEUS~\cite{pr:d70:012008} collaborations are shown compared with NLO
QCD predictions performed in the massive scheme. The data are
well-described by the theoretical calculation. Similar measurements
have been made in deep inelastic
scattering~\cite{pl:b599:173,epj:c41:453}; here the data are above
the NLO QCD predictions but consistent within the quoted uncertainties.

\begin{figure}[htp]
\begin{center}
~\epsfig{file=Figures/cdfpsi.eps,height=6.5cm}
\put(-27,172){\makebox(0,0)[tl]{\large (a)}}
\put(164,122){\makebox(0,0)[tl]{\large (b)}}
~\epsfig{file=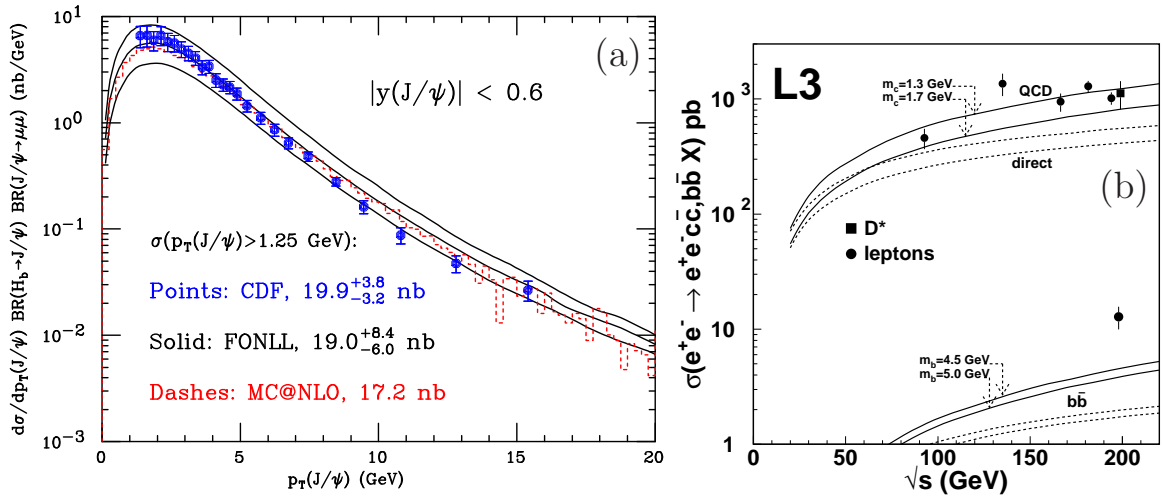,height=6cm}
\end{center}
\caption{Cross-sections (a) $d\sigma/dp_T$ for $J/\psi$ mesons from
beauty decays at the Tevatron and (b) beauty production at LEP. The
data are compared to the different NLO QCD calculations. }
\label{fig:b_tev_lep}
\end{figure}

Recent theoretical improvements and their comparison with measured,
rather than extrapolated, cross-sections have also led to an improved
description of beauty production in $p\bar{p}$ collisions. Two new
calculations have become available.  The FONLL calculation discussed
previously, matches the massive and massless schemes and should
therefore be applicable for all energies. This prediction also has an
improved fitting technique to $e^+e^-$ data for the fragmentation
function.  The second, MC@NLO, matches a NLO calculation to
parton showers and hadronisation in the {\sc Herwig} MC
programme. These calculations are shown in Fig.~\ref{fig:b_tev_lep}
compared with recent CDF data~\cite{Acosta:2004yw}; both describe the
data well. The rate of beauty production has also been measured in
$\gamma \gamma$ collisions by the L3 collaboration~\cite{Acciarri:2000kd} at
LEP. The data are measured within the acceptance of the detector and
then extrapolated by factors of $\sim 20-30$ to the full phase space
for comparison with NLO QCD predictions. This measurement is 
experimentally challenging due to the poor separation power between 
beauty decays to leptons and background. The NLO QCD prediction, shown
in Fig.~\ref{fig:b_tev_lep}, is a factor of three below the data
although the difference is below three standard deviations. Were this
result to persist with measurements from other LEP collaborations it
would certainly be very puzzling as a large part of the cross-section
comes from electromagnetic coupling of two photons to two beauty
quarks.

As well as NLO calculations, full final-state Monte Carlos can also
give a reasonable description of most heavy flavour cross-sections
which have been measured in photoproduction and other high-energy
processes. Given the importance of heavy flavour production
cross-sections, particularly beauty, in identifying interesting
processes at LHC and a future linear collider, detailed validation of
such models across all available measurements is of considerable
interest and importance, and is
ongoing~\cite{gutsche,Butterworth:2004mu,*Butterworth:2002ts,heralhc}.

\subsection{Charmonium production}
\label{sec:ineljpsi}

The $J/\psi$ meson was discovered over 30 years ago and thereby
confirmed the existence of the charm quark. However, although the
particle has been extensively measured, the nature of its production
is poorly understood. Quasi-elastic production of the $J/\psi$ and
other vector mesons was briefly discussed in
Section~\ref{sec:totalxsec}. The mechanism by which $J/\psi$ mesons are
``inelastically'' produced is expected to be predictable in QCD. The
inelastic domain arises from the presence of a large transfer of
energy between the projectile and the target particle.

In the so-called colour-singlet model~\cite{Berger:1980ni}, $c\bar{c}$
pairs produced in a colour-singlet (CS) state are identified with physical
$\psi$ states. This model has been extended in the context of
non-relativistic
QCD~\cite{Bodwin:1994jh,*Cho:1995vh,*Cacciari:1996dg,*Ko:1996xw} to
account for the production of $c\bar{c}$ pairs which are not
colourless. The charm quark pair can emit gluons (and hence change
colour) until the pair is colourless. This model is referred to as the
colour-octet (CO) model. Measurements by the CDF collaboration in
$p\bar{p}$ collisions~\cite{prl:79:572} indicate that the
colour-singlet model greatly underestimates the production rate of
$\psi$ mesons, by factors of between 10 and 80. However, it has been
claimed that calculations including colour-octet contributions can
account for this difference.  A test of this claim is the requirement
that the colour-octet matrix elements extracted from $\psi$ cross-section 
measurements in $p\bar{p}$ collisions can also account for
photoproduction data.

\begin{figure}
\begin{center}
~\epsfig{file=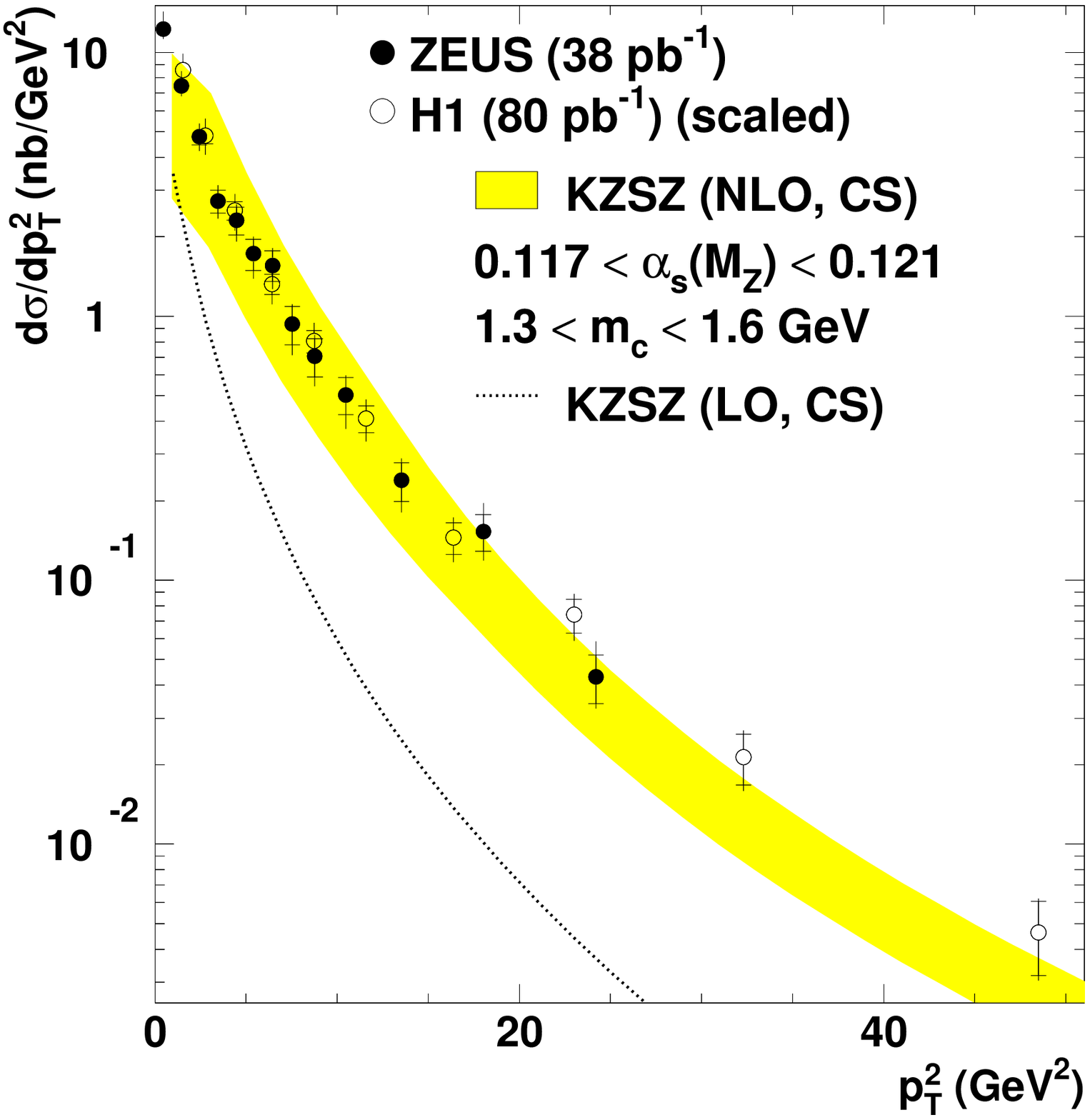,height=7.5cm}
\put(-27,199){\makebox(0,0)[tl]{\large (a)}}
\put(174,199){\makebox(0,0)[tl]{\large (b)}}
~\epsfig{file=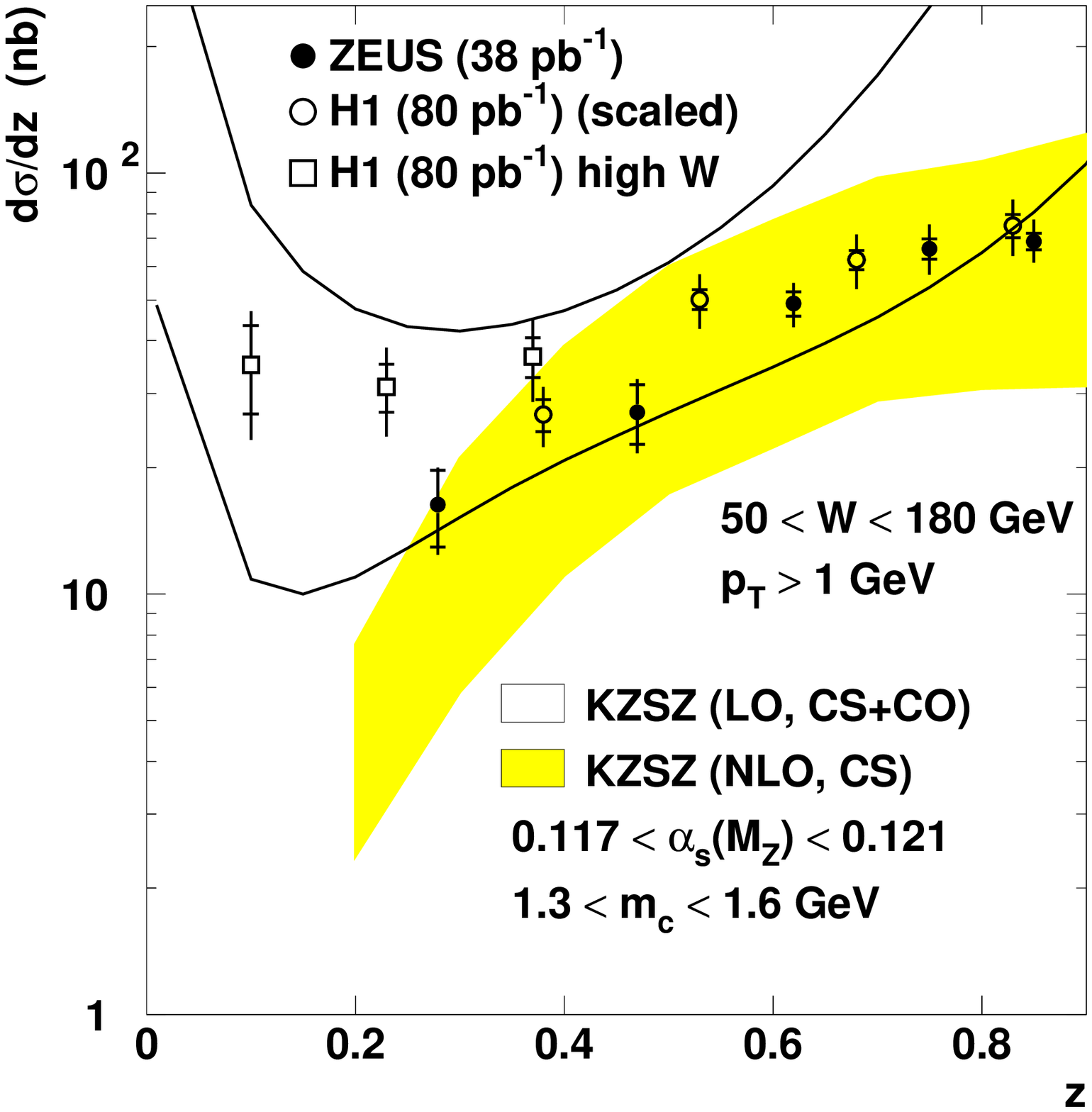,height=7.5cm}
\end{center}
\caption{Cross-sections (a) $d\sigma/dp_T^2$ and (b) $d\sigma/dz$ for
$J/\psi$ photoproduction at HERA. Data from both H1 and ZEUS
collaborations are compared with predictions at (a) LO and NLO in the
framework of the colour-singlet model and (b) NLO in the framework of
the colour-singlet model and LO with the addition of the
colour-octet model.}
\label{fig:jpsi_hera}
\end{figure}

The cross-sections $d\sigma/dp_T^2$ and $d\sigma/dz$ have been
measured in photoproduction~\cite{Chekanov:2002at,Adloff:2002ex},
where $z$, the ``inelasticity'', is the fraction of the incoming
photon's energy carried by the $J/\psi$ meson. The distributions of
these two variables are shown in \mbox{Fig.~\ref{fig:jpsi_hera}}
compared with theoretical calculations incorporating the models
mentioned previously. In Fig.~\ref{fig:jpsi_hera}a, a prediction from
the colour-singlet model alone \mbox{(LO, CS)} clearly does not
describe the data being a factor of over 10 below at high
$p_T^2$. After the inclusion of NLO
corrections~\cite{Kramer:1994zi,*Kramer:1995nb}, however, the theory
agrees well with the data. This calculation is shown again in
Fig.~\ref{fig:jpsi_hera}b for the inelasticity variable and again
describes the data reasonably well. At low $z$, the NLO prediction is
below the data although in this region more background from
e.g. resolved photons is expected. Also shown are calculations which
have both colour-singlet and colour-octet contribution, shown as
\mbox{(LO, CS+CO)}\cite{Kramer:2001hh}. They also describe the data
well. However, other models incorporating colour-octet
processes~\cite{Beneke:1999gq,*Kniehl:1998qy} give different
results~\cite{Chekanov:2002at}. The amount of colour-octet in
different calculations was extracted from the same $p\bar{p}$ data;
thus it is evident these calculations suffer from large uncertainties
from other sources, and hence a lack of predictive power.

In summary, there seems to be a real issue with the theory here.  The
very large NLO corrections to the CS calculations indicate, at least
naively, poor convergence of the perturbative expansion. These
corrections have not been calculated for the CS+CO model; the LO
predictions for this model differ substantially from each other. Due
to the large theoretical uncertainties affecting all predictions the
presence of colour-octet processes in photoproduction cannot be
excluded or definitely confirmed, and the matrix elements extracted
from hadroproduction are consistent with photoproduction data within
very large uncertainties. More precision comparisons with theory
require a better theoretical understanding of the uncertainties,
and/or the measurement of more accurately predicted variables, such as
perhaps angular distributions.

\subsection{Jet substructure and fragmentation functions} 
\label{sec:substructure}

Measurement of jet substructure allows the possibility of distinguishing 
jets originating from quarks and gluons due to their different radiation 
amplitudes. This permits a stringent test of QCD, provides information 
on the properties of quark- and gluon-initiated jets and, by tagging the 
final state, reveals details of the initial state and hence proton and 
photon structure (see Section~\ref{sec:nature_of_photon}). At 
sufficiently high jet transverse energy, where the influence of 
hadronisation becomes negligible, the internal structure of a jet is 
calculable in pQCD. Such calculations predict that gluon-initiated jets 
are broader than quark-initiated jets due to the larger colour charge of 
the gluon. The jet shape~\cite{prl:69:3615} and subjet
multiplicity~\cite{np:b383:419,*np:b421:545,*pl:b378:279,*jhep:9909:009}
have been used to study the internal structure of jets in 
photoproduction. The integrated jet shape, $\psi(r)$, using only those
particles belonging to the jet, is defined as the fraction of the jet
transverse energy that lies inside a cone in the $\eta-\phi$ plane of
radius, $r$, concentric with the jet axis:

\begin{equation}
\psi(r)  = \frac{E_T(r)}{E_T^{\rm jet}},
\label{eq:jet_shape}
\end{equation}

where $E_T(r)$ is the transverse energy within the given cone of radius 
$r$. The mean integrated jet shape, $\langle \psi(r) \rangle$, is defined 
as the averaged fraction of the jet transverse energy inside the cone
$r$:

\begin{equation}
\langle \psi(r) \rangle = \frac{1}{N_{\rm jets}} \sum_{\rm jets} \frac{E_T(r)}{E_T^{\rm jet}},
\label{eq:int_jet_shape}
\end{equation}

where $N_{\rm jets}$ is the total number of jets in the sample.

\begin{figure}[ht]
\begin{center}
~\epsfig{file=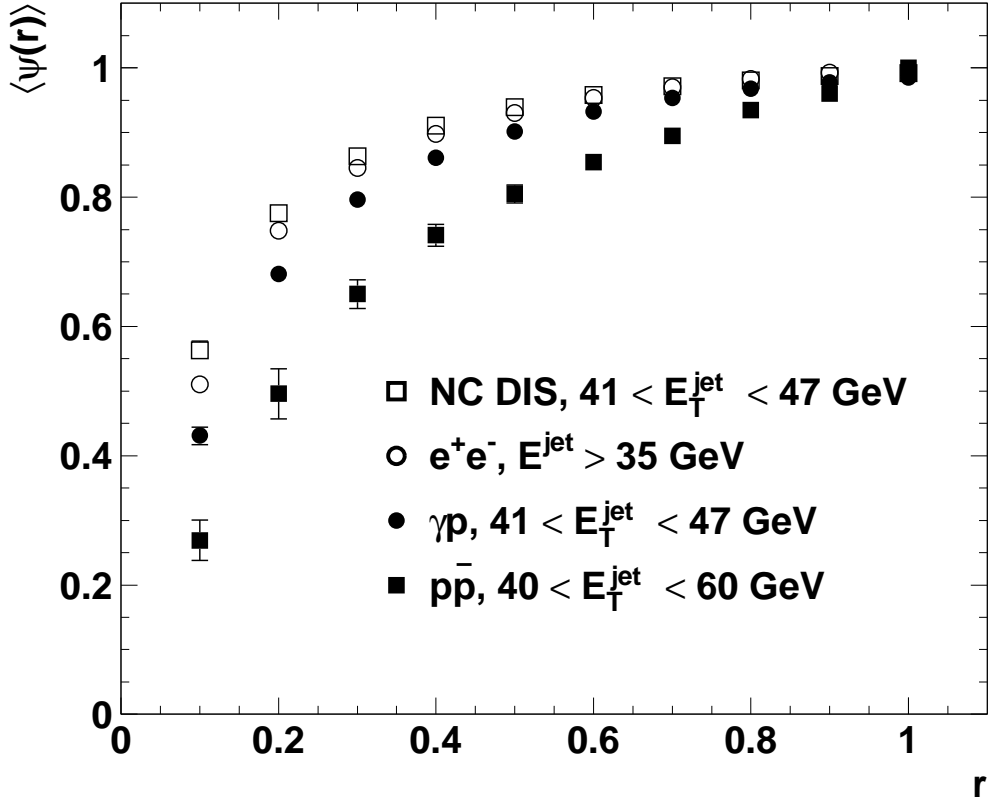,height=11cm}
\end{center}
\caption[*]{Measured jet shapes in $e^+e^-$~\cite{Akers:1994wj},
DIS~\cite{Breitweg:1998gf}, photoproduction~\cite{epj:c2:61} and
$p\bar{p}$~\cite{Abe:1992wv,*Abachi:1995zw}.}
\label{fig:jet_shapes}
\end{figure}

The internal structure of a jet
is expected to depend mainly on the type of primary parton from which
it originated and to a lesser extent on the particular hard scattering
process. This expectation is supported by comparisons such as that
shown in Fig.~\ref{fig:jet_shapes}. Here it is seen that in those
processes which at LO give rise only to quark jets ($e^+e^-$
annihilation, DIS), the jets are narrow, and very similar to each other. In
photoproduction at these transverse energies, direct processes
dominate. However, the presence of QCD Compton and resolved photon
processes means that some gluon jets are present in the final state at LO,
and the average jet shape is correspondingly broader. In $p\bar{p}$
events, where the gluon admixture is even stronger, the average jet
shape is broader still.

\begin{figure}[ht]
\begin{center}
~\epsfig{file=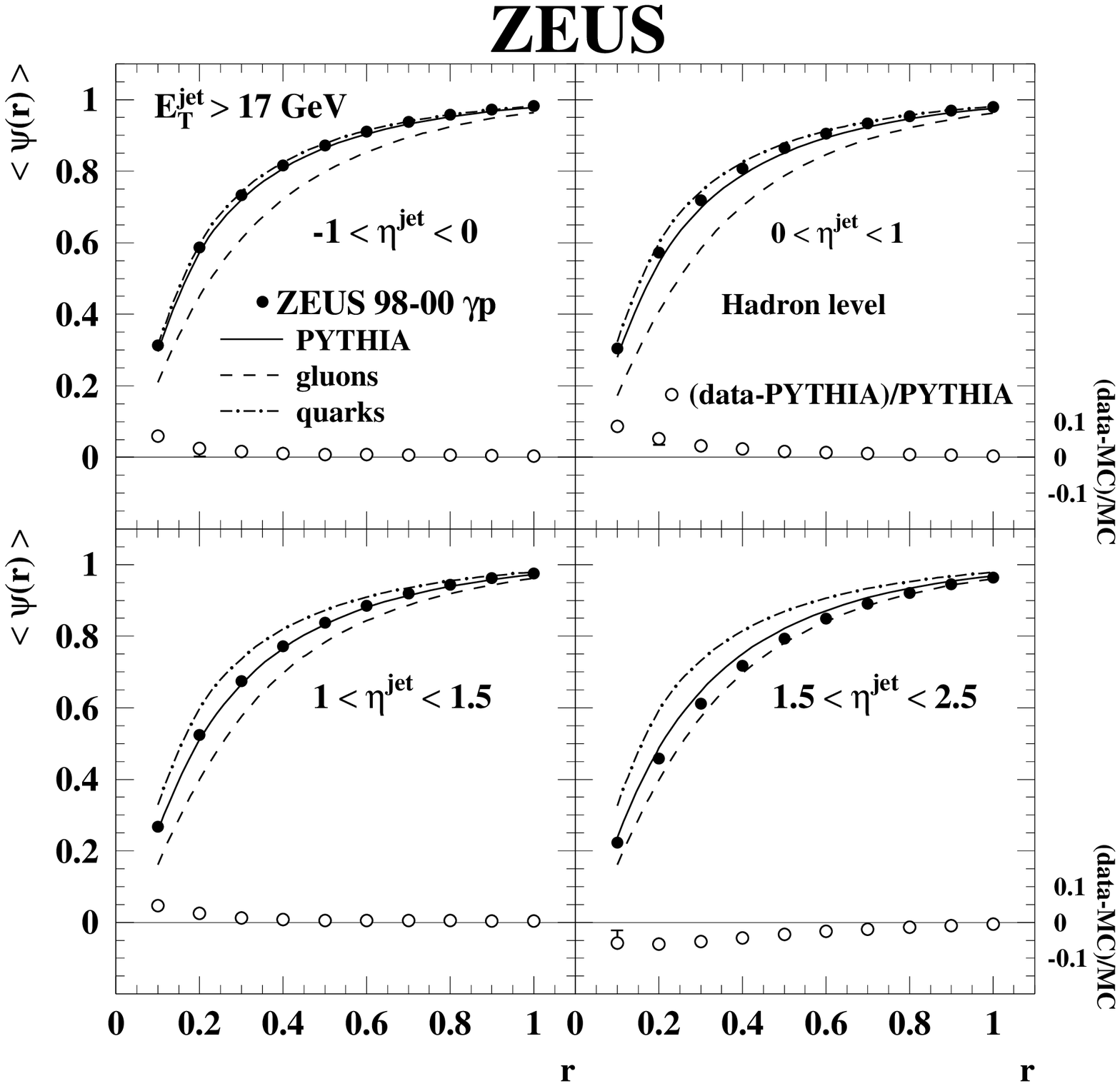,height=13cm}
\end{center}
\caption{Measured mean integrated jet shape in different regions of 
$\eta^{\rm jet}$. Predictions of the {\sc Pythia} Monte Carlo program 
for quark (dot-dashed lines), gluon (dashed lines) and all (solid lines) 
jets are shown. The fractional difference of the data to the predictions 
of {\sc Pythia} for all jets is also shown (open dots).}
\label{fig:q_g_shapes}
\end{figure}

The measured mean integrated jet shape for $E_T^{\rm jet} > 17$\,GeV
in different regions of $\eta^{\rm jet}$ is shown in
Fig.~\ref{fig:q_g_shapes}~\cite{Chekanov:2004kz}. The jets broaden as
$\eta^{\rm jet}$ increases. Leading-logarithmic parton-shower
predictions from {\sc Pythia} for resolved plus direct processes and
gluon- and quark-initiated jets are compared to the data. The
description of the data is generally good, although the data are
somewhat broader than the predictions. From the comparison with the
predictions of gluon- and quark-initiated jets, it is seen that the
measured jets are quark-like for $-1 < \eta^{\rm jet} <0$ and become
increasingly gluon-like as $\eta^{\rm jet}$ increases. Jets become
narrower as $E_T^{\rm jet}$ increases (not shown) consistent with the
dominance of direct processes, and hence quark jets, at higher
$E_T^{\rm jet}$. These effects have also been seen in measurements of
the subjet multiplicity~\cite{Chekanov:2004kz}.

The differences in gluon- and quark-initiated jets are used to select samples 
enriched in such jets to study their properties and the dynamics of the 
hard subprocesses in more detail. The two samples of jets were selected on a 
statistical basis and classified as follows:

\begin{itemize}

\item gluon-enriched sample (``broad jets''), defined as those jets with 
      $\psi(r=0.3) < 0.6$;

\item quark-enriched sample (``narrow jets''), defined as those jets with 
      $\psi(r=0.3) > 0.8$.

\end{itemize}

\begin{figure}[ht]
\begin{center}
~\epsfig{file=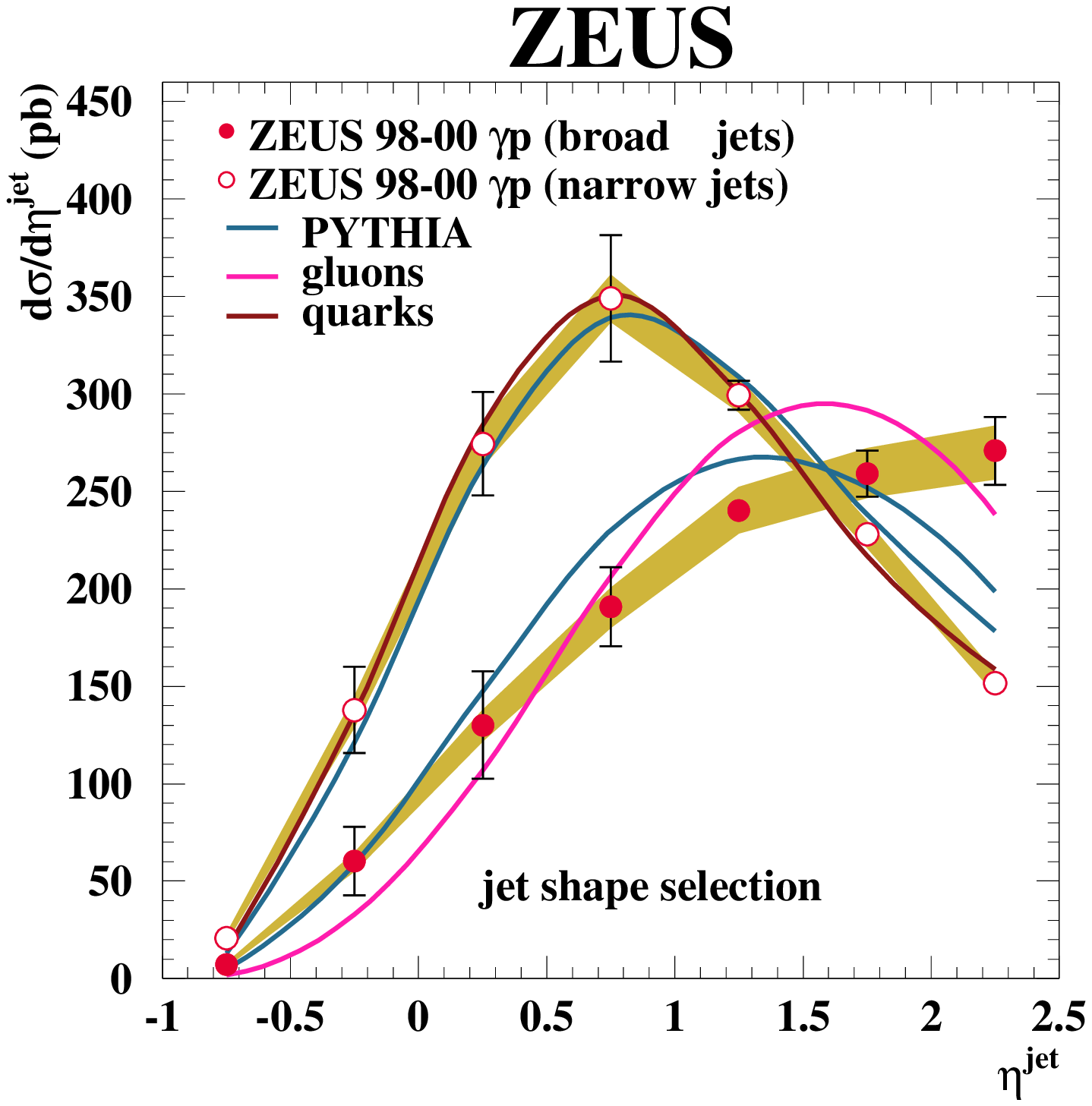,height=7cm}
\put(-27,177){\makebox(0,0)[tl]{\large (a)}}
\put(174,177){\makebox(0,0)[tl]{\large (b)}}
~\epsfig{file=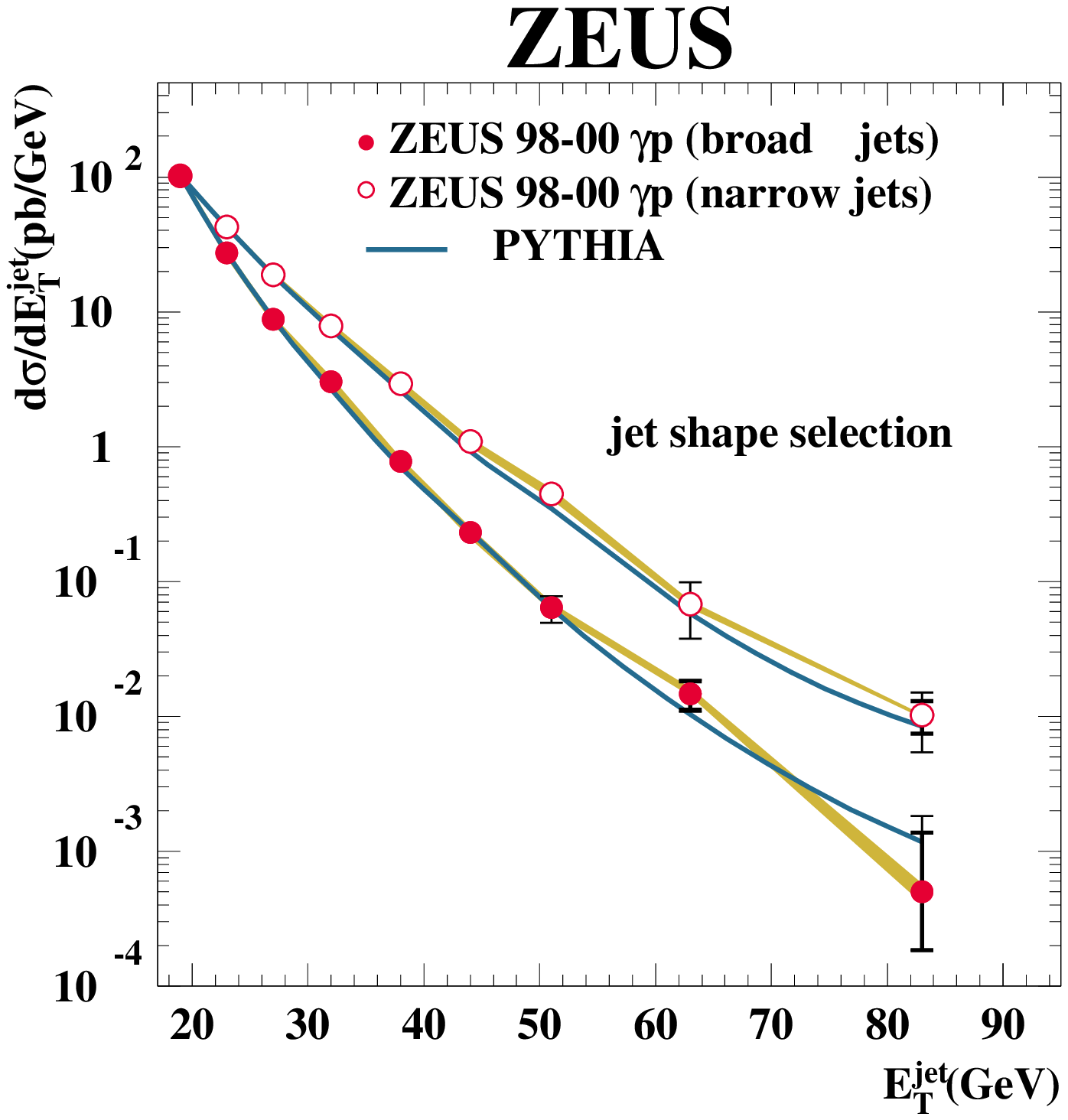,height=7cm}
~\epsfig{file=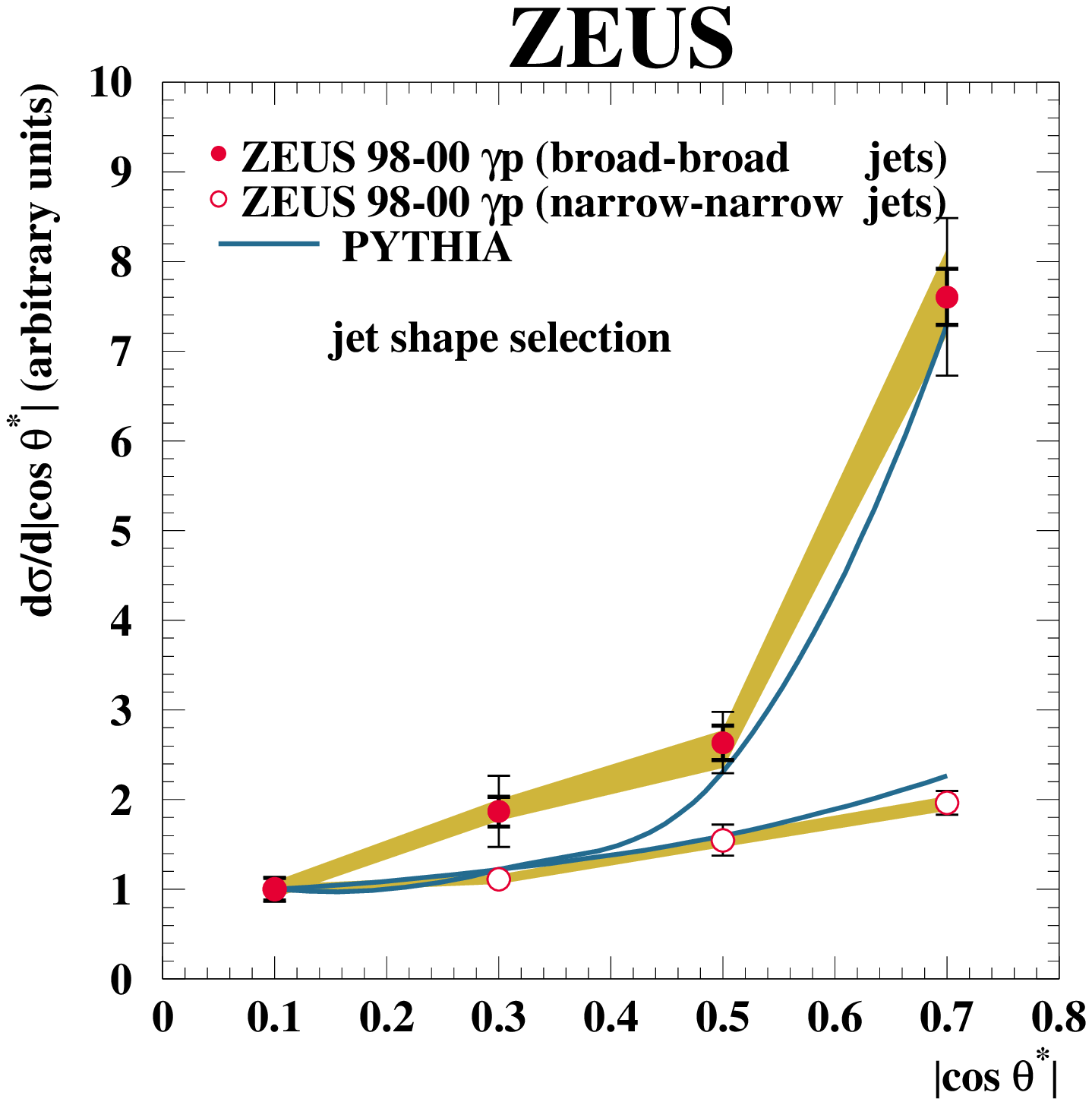,height=7cm}
\put(-27,177){\makebox(0,0)[tl]{\large (c)}}
\put(174,177){\makebox(0,0)[tl]{\large (d)}}
~\epsfig{file=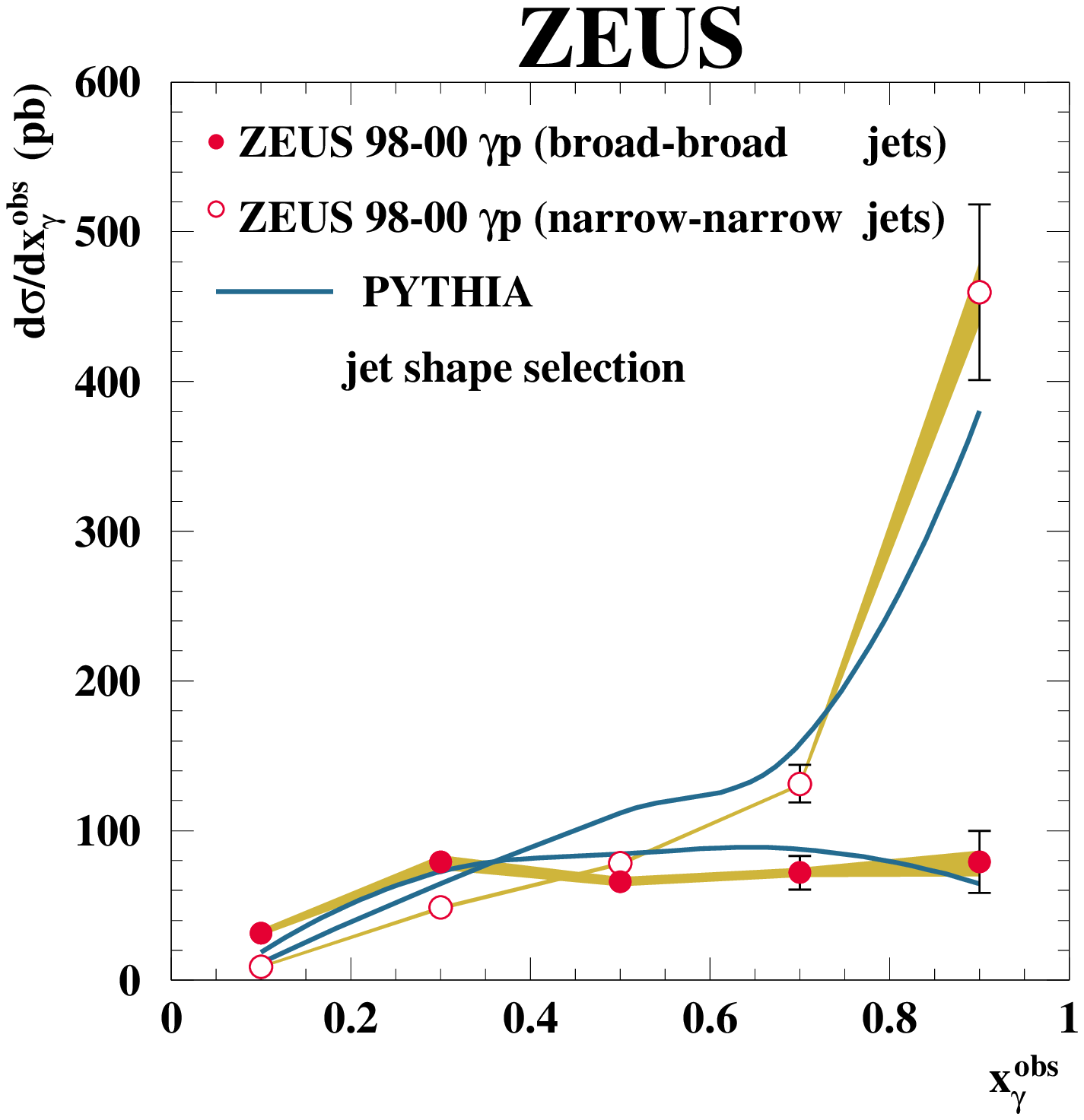,height=7cm}
\end{center}
\caption{Cross-sections (a) $d\sigma/d\eta^{\rm jet}$ and 
(b) $d\sigma/dE_T^{\rm jet}$ for broad and narrow jets and 
(c) $d\sigma/d|\cos\theta^*|$ and (d) $d\sigma/dx_\gamma^{\rm obs}$ for 
events with two broad jets or two narrow jets. The data are compared with 
{\sc Pythia MC} predictions and in (a) for {\sc Pythia} expectations for 
quark and gluon jets.}
\label{fig:q_g_xsecs}
\end{figure}

Non-overlapping regions were chosen to suppress migration effects and the 
cuts values represent a compromise between purity and statistics. In dijet 
events, samples with two broad jets, two narrow jets and one broad and one 
narrow jet can also be selected. In Fig.~\ref{fig:q_g_xsecs}, cross-sections 
for broad and narrow jets (Fig.~\ref{fig:q_g_xsecs}a and~\ref{fig:q_g_xsecs}b) 
and events with two broad or two narrow jets (Fig.~\ref{fig:q_g_xsecs}c 
and~\ref{fig:q_g_xsecs}d) are shown.

The measured cross-sections for broad and narrow jets exhibit a different 
behaviour: the $\eta^{\rm jet}$ distribution for broad jets increases up to 
the highest $\eta^{\rm jet}$ value measured whereas the distribution for 
narrow jets peaks at $\eta^{\rm jet} \sim 0.7$; and the distribution in 
$E_T^{\rm jet}$ for narrow jets has a harder spectrum than for broad jets. 
The Monte Carlo prediction, normalised to the total cross-section, gives 
a good description of the narrow-jet sample but gives a poorer description 
of the broad-jet sample. The {\sc Pythia} programme predicts that the 
broad-jet event sample consists of 15\% $gg$, 50\% $gq$ and 35\% $q\bar{q}$. 
The selection of narrow jets gives a reasonably pure sample of quark-initiated 
jets; the {\sc Pythia} programme predicts 62\% $q\bar{q}$, 34\% $qg$ and 
4\% $gg$. In Fig.~\ref{fig:q_g_xsecs}a, predictions for quark-initiated jets 
describe the narrow-jet sample well, as expected due to its high quark purity. 
The prediction for gluon-initiated jets describes the broad-jet sample less 
well due to the higher impurity from quark-initiated jets. However, the 
result supports the expectation that the broad- and narrow-jet samples are 
dominated by gluon-  and quark-initiated jets, respectively.

Dijet cross-sections in which either two broad or two narrow jets are
tagged are shown in Fig.~\ref{fig:q_g_xsecs}c
and~\ref{fig:q_g_xsecs}d. As these jets are indistinguishable, only
the absolute value of $\cos\theta^*$ can be measured (in contrast to
the case when one broad and one narrow jet are tagged, as in
Fig.~\ref{fig:costheta}). The measured and predicted cross-sections
were normalised at $|\cos\theta^*| = 0.1$. The cross-section for
events with two broad jets exhibits a faster rise to high
$\cos\theta^*$. The predictions from the {\sc Pythia} Monte Carlo
programme give an adequate description of the data.  For events with
two broad jets, {\sc Pythia} predicts the parton final state to
consist of 16\% $gg$, 52\% $qg$ and 32\% $qq$. As in the sample with only
one broad jet, the impurity is relatively large. The sample with two
narrow jets is a much purer sample of quark jets; {\sc Pythia}
predicts the parton final state to consist of 71\% $qq$, 28\% $qg$ and
1\% $gg$. The differences in the measured distributions can be
understood in terms of the dominant two-body processes: the resolved
subprocess $q_\gamma g_p \to qg$, mediated by gluon exchange for the
broad-broad dijet sample and the direct subprocess $\gamma g \to
q\bar{q}$, mediated by quark exchange for the narrow-narrow dijet
sample. The dominance of resolved and direct processes for broad-broad
and narrow-narrow dijet events, respectively is also confirmed by the
cross-section $d\sigma/dx_\gamma^{\rm obs}$. The narrow-narrow dijet
sample is strongly peaked at high $x_\gamma^{\rm obs}$ whereas the
broad-broad dijet sample is roughly flat in $x_\gamma^{\rm obs}$.
 

\subsection{Multijet production and colour coherence}

Jet photoproduction has so far mostly been discussed in terms of the
LO diagrams (Fig.~\ref{fig:feyn}) where two jets are produced. In
reality, a combination of QCD radiation and hadronisation will mean
that there is a distribution of energy around the final state outside
the two leading jets. This may be in the form of low energy particles
from soft processes, or further high $\ETJ$ jets. The distribution of
these jets, and to some extent even of softer particles, is predicted
by QCD.

\begin{figure}
\centering
\leavevmode
~\epsfig{file=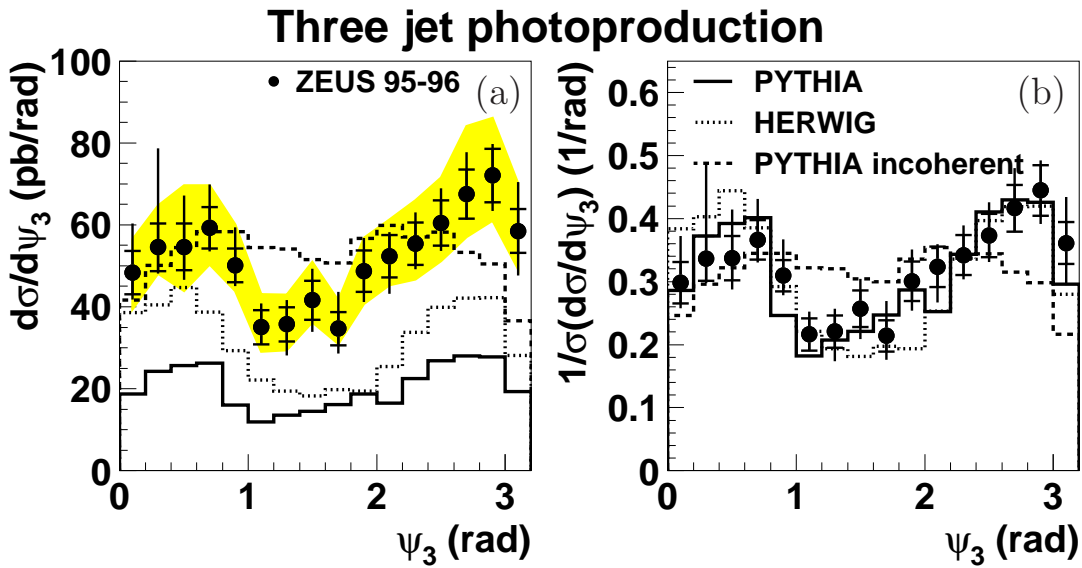,height=8cm}
\put(-234,191){\makebox(0,0)[tl]{\large (a)}}
\put(-29,191){\makebox(0,0)[tl]{\large (b)}}
\caption{(a) The cross-section $d\sigma / d\psi_3$  
and (b) the area-normalised distribution of $\psi_3$.  
The correlated systematic uncertainty due to the jet
energy-scale is shown as the shaded band in (a). The solid
histogram shows the default {\sc Pythia} prediction. In (a) and (b) 
the dotted and dashed histograms show the predictions of {\sc Herwig} 
and of {\sc Pythia} with colour coherence switched off, respectively.
\label{fig:3jet}}
\end{figure}

The three-jet photoproduction cross-section has been measured by the
ZEUS collaboration~\cite{pl:b443:394}. The measurement was made for
three-jet masses above 50~GeV and with a requirement that at least two
jets satisfy $\ETJ > 6$~GeV while the third has $\ETJ > 5$~GeV. QCD
calculations at ${\cal O}(\alpha\as^2)$ (which is LO for three jet
production) are in good agreement with the data. In addition, since
this cross-section is dominated by configurations in which the third
jet has substantially lower $\ETJ$ than the two leading jets,
leading-logarithmic parton-shower models are expected to describe the
cross-sections rather well. They do indeed, as may be seen in
Fig.~\ref{fig:3jet}. An interesting feature here is that coherent
gluon radiation must be implemented in the Monte Carlo in order to
reproduce the shape of the $\psi_3$ distribution (the angle, in the
three-jet centre-of-mass system, between the plane containing the leading jet and the
beam, and the plane containing the three jets). For coherent
radiation, these planes tend to line up such that the region $\psi_3
\approx \pi/2$ is depopulated. If a large amount of incoherent
radiation is allowed (see appendix), this effect is destroyed.

\subsection{Energy flow, rapidity gaps and forward particles} 
\label{sec:eflow_rapidity_gaps}

As discussed in Section~\ref{sec:totalxsec}, total hadronic cross-sections 
in the limit of large $s/t$ increase by some small (but
positive!) power of $s$. In Regge theory this is an exchange of vacuum
quantum numbers, known as the Pomeron. This is presumably an emergent
behaviour of the QCD Lagrangian, but making a connection with the
partonic language of perturbative QCD is a challenge which continues
to throw up surprises in both data and theory\footnote{See
\cite{forshaw:1997:pomeron} for a discussion and
bibliography.}.

The high energy regime probed at HERA and the Tevatron means that the
large $s/t$ limit may be approached even when $t$ is sufficiently
large for perturbative techniques to be applied. It is also possible,
even when the momentum exchange, $t$, with the proton is small, for
relatively high-mass final states to be produced, which opens up the
phase space for short distance, high energy scale processes in the
diffractively produced system. These two distinct cases both offer the
possibility of describing some or all of ``Pomeron'' exchange in terms
of perturbative quarks and gluons. Both have been measured in high
energy photoproduction.

\subsubsection{Hard colour singlet exchange}

In configurations where the two leading jets are separated by a large
rapidity interval, it is possible in high energy photoproduction to
approach the regime where $t \approx \pt^2$ is of the order a few tens
of GeV, but $\hat{s}$, the partonic centre-of-mass energy squared, is
a few hundred GeV. Thus the scattering will be
dominated by the exchange of vacuum quantum numbers, but yet the high
$t$ means that pQCD calculations should be applicable. The lowest
order colour-singlet exchange in QCD is a pair of gluons. However,
such a diagram is subject to large logarithmic corrections, which are
resummed in the BFKL~\cite{jetp:45:199,sovjnp:28:822} approach. This is
a topic in QCD with a very extensive literature (see for example,
references within~\cite{forshaw:1997:pomeron}), but for our purposes
here it is sufficient to say that there is a possible enhancement in
the cross-section for colour-singlet exchange between jets separated
by a large rapidity interval in hadronic collisions.

If no colour is exchanged between the hadrons (in the case of
photoproduction, the proton and a hadronic photon), colour coherence
leads to a suppression of QCD radiation, and thus energy flow, between
the jets. Hence the large rapidity interval is very likely to be a
rapidity gap. This is the experimental signature searched for by the
HERA collaborations. A key observable is the ``gap fraction''
measured as a function of the rapidity interval, $f(\Delta\eta)$. This
is the ratio of the cross-section for dijet events with a rapidity
gap between the jets (however the gap is defined) to the total number
of dijet events.

\begin{figure}
\centering
\leavevmode
~\epsfig{file=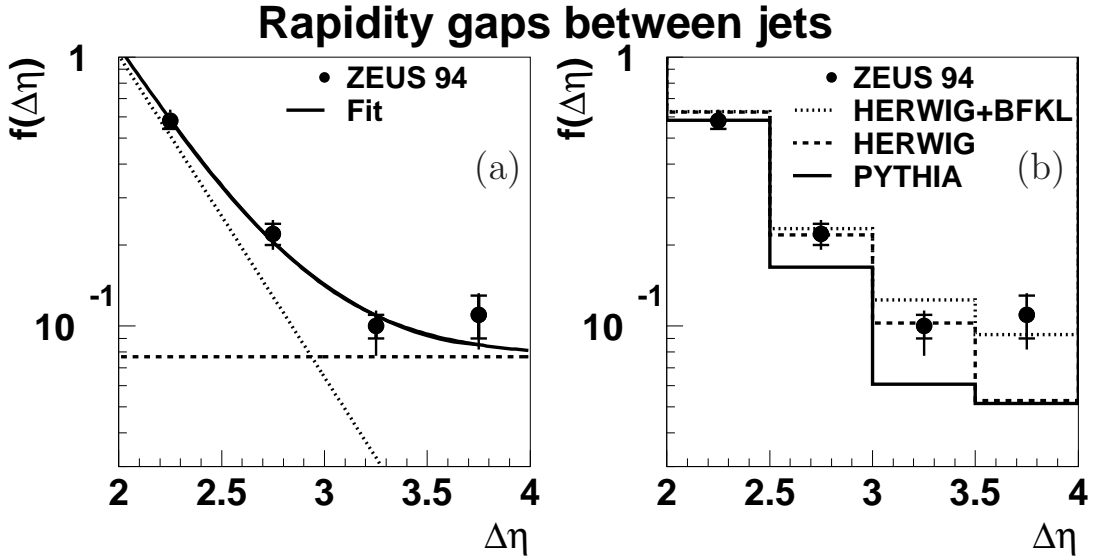,height=8cm}
\put(-234,161){\makebox(0,0)[tl]{\large (a)}}
\put(-29,161){\makebox(0,0)[tl]{\large (b)}}
\caption[*]{ZEUS gap fraction measurement (solid points). 
In (a) the gap fraction is compared
to a fit to the sum of an exponential (non-colour-singlet) and a flat
(assumed colour singlet) contribution. In (b) the gap fraction is
redisplayed and compared to {\sc Pythia} and {\sc Herwig}
non-diffraction MC, and to a {\sc Herwig} sample which includes a
colour singlet exchange component from a BFKL-based
calculation~\cite{pl:b284:123}.
\label{fig:gbj1}
}
\end{figure}

In the first measurement~\cite{pl:b369:55}, a gap was defined by the
requirement that no particle with transverse energy $>300$~MeV lay in
the interval between the jets. In ``normal'' QCD events, the
probability of zero radiation into the gap falls exponentially as the
interval increases, and thus so does $f(\Delta\eta)$. This is seen in
the first bins of Fig.~\ref{fig:gbj1}a and \ref{fig:gbj1}b. However,
for colour singlet exchange no such suppression is expected, and
indeed the final bin is higher than one would expect given the fall in
the first three. A simple fit to a flat fraction for colour singlet
exchange combined with the exponential fall gives good description of
the data. This was the first evidence for strongly interacting
colour-singlet exchange in jet photoproduction. Such measurements have
also been made at the
Tevatron~\cite{Abachi:1995gz,*Abbott:1998jb,*Abe:1994de,*Abe:1997ie}
and show similar evidence for such processes in $p\bar{p}$ events,
albeit at a much lower rate.

\begin{figure}[ht]
\begin{center}
~\epsfig{file=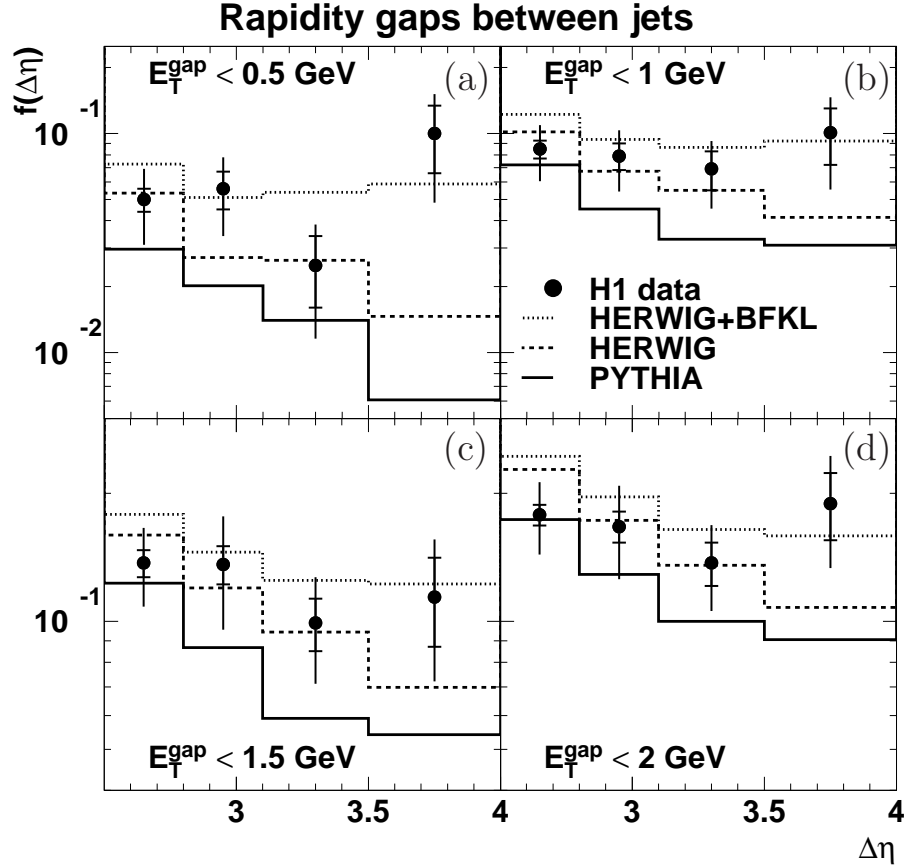,height=12cm}
\put(-176,310){\makebox(0,0)[tl]{\large (a)}}
\put(-25,310){\makebox(0,0)[tl]{\large (b)}}
\put(-176,170){\makebox(0,0)[tl]{\large (c)}}
\put(-25,170){\makebox(0,0)[tl]{\large (d)}}
\end{center}
\caption{H1 gap fraction measurement (solid points) differential in
$\Delta \eta$. Gap events are defined for four values of
$\etgap$, shown in (a-d). The gap fractions are compared to the
prediction of \herwig~(dashed line) and \pythia~(solid line).  The
dotted line shows \herwig~+ BFKL colour singlet exchange (see
text).\label{fig:h1gbj}}
\end{figure}

The main problem with this measurement, apart from the limited
statistics and rapidity reach available at the time, is that the
method used to define a rapidity gap is not infrared safe. That is, a
definition of activity in terms of particles is not amenable to
calculation in perturbative QCD. A more robust definition can be made
in terms of energy flow within the interval, and by using minijets
rather than particles to calculate this energy
flow~\cite{Cox:1999dw,*wyatt,*mhs,Oderda:1999kr,Oderda:1999ta}.  Such
a definition was used by the H1 collaboration in a subsequent
measurement~\cite{Adloff:2002em}. In this analysis, the $\kt$ jet
algorithm~\cite{np:b406:187} was run over the final state in the
inclusive mode. This produces a list of ``jets'', some of which have
very low transverse energy and consist of very few particles.
However, since the algorithm is infrared safe, quantities based on
these objects are, in principle at least, calculable. The highest
$\ETJ$ pair define the rapidity interval, as usual. The cross-sections
and gap fraction in the paper are defined in terms of $\etgap$, the
transverse energy of all the jets within the rapidity interval between
the jets. The requirement $\etgap < E_T^{\rm cut}$ was applied, for
values of $E_T^{\rm cut}$ ranging between 0.5~GeV and 2~GeV.  For all
values of $E_T^{\rm cut}$ studied, (Fig.~\ref{fig:h1gbj}) there is an
excess of gap events at large rapidity intervals compared to the
expected fall-off from normal QCD events.

The exponential fall of the gap fraction expected for
non-colour-singlet processes is also seen in the non-colour singlet
{\sc Herwig} and {\sc Pythia} curves shown in Fig.~\ref{fig:gbj1}b and
in Fig.~\ref{fig:h1gbj}. The data are also compared to {\sc
Herwig} predictions~\cite{Butterworth:1995jf} implementing a
leading-logarithmic BFKL-based calculation of hard colour-singlet
exchange between two partons as computed by Mueller and
Tang~\cite{pl:b284:123}. In the limit $\Delta\eta \gg 1$ the 
cross-section for \mbox{quark--quark} scattering may be approximated as
\begin{equation}
\frac{d \sigma(q q \to q q)}{d \hat t} \approx (C_F \as)^4
\frac{2\pi^3}{\hat t^2} \frac{\exp(2 \omega_0 y)}{(7 \as C_A
\zeta(3) y)^3}
\label{mullertang}
\end{equation} 
where 
\begin{equation}
\omega_0 = C_A (4 \ln 2/\pi) \as.
\label{intercept}
\end{equation}
and
\begin{equation}
y = \Delta\eta = \ln \left( \frac{\hat{s}}{-\hat t} \right)
\end{equation}
Here, $1+\omega_0$ is the perturbative Pomeron intercept, $C_F =
\frac{4}{3}$ is the usual colour factor for \mbox{quark--quark} scattering,
$C_A$ is the number of colours, $-t \approx (\ETJ)^2$, and $\zeta$ is
the Riemann $\zeta$-function. In the leading-logarithm approximation, the 
values of $\as$ in Eqs.\ref{mullertang} and \ref{intercept} are free parameters. 
In the H1 analysis~\cite{Adloff:2002em}, where the exact
rather than asymptotic form of Eq.\ref{mullertang} was
used~\cite{Cox:1999dw}, these were both set to 0.18. This corresponds
to a choice of Pomeron intercept of $1 + \omega_0 = 1.48$, which was
used for the asymptotic {\sc Herwig} curves shown here. For both 
ZEUS~(Fig.~\ref{fig:gbj1}) and H1 (Fig.~\ref{fig:h1gbj}), good
agreement with the data is obtained for this model.

For this kind of measurement in a real experiment with finite rapidity
and non-zero $\etgap$, the large logarithms in rapidity, or $x$, which
are summed in the BFKL approach are not necessarily the only, or even
the most important, terms which should be included. There are also
terms with logarithms in $\etgap/\ETJ$ to be considered. Such
logarithms have been resummed for various jet
definitions~\cite{Oderda:1999kr,Oderda:1998en,Berger:2001ns,Appleby:2002ke,*Appleby:2003sj},
and there is progress towards an approach which sums all important
logarithms of both types~\cite{Forshaw:2005sx}.

Two other photoproduction processes which give access to hard colour
singlet exchange are the high-$|t|$ diffractive photoproduction of vector
mesons, and of photons.

The main advantage that diffractive photoproduction of vector mesons
at high $|t|$ has over jet photoproduction is the experimental
cleanliness of the measurement. At HERA, for example, the final state
consists of; the decay products of the vector meson, an electron
scattered through a very small angle (which may escape down the beam
pipe), and a low-mass hadronic system in the forward (proton)
direction. The $\q2$ is constrained to be low by the low
scattering-angle of the electron, and thus $t$ may be determined from
the vector meson decay products.  Theoretically, the unknown
fragmentation and hadronisation effects involved in the formation of
jets are absent, but the transition probability of a photon to a
vector meson is introduced.

Calculations of high-$|t|$ vector meson exchange based upon BFKL
resummation have been
performed~\cite{zfp:c68:137,pl:b375:301,pl:b478:101,Enberg:2003jw}.
In this approach, as with the calculations of rapidity gaps between
jets, a ladder of gluons (``hard Pomeron'') couples to a single parton
in the proton at one end. At the other end things are very different;
the photon splits into a $q\bar{q}$ pair. The probability to form a
vector meson is calculated from the overlap between the photon and
vector meson wavefunctions caused by the coupling to the two gluons at
the top of the ladder to the dipole formed by the $q\bar{q}$ pair.
The cross-section is predicted to fall as $1/|t|^N$, and to rise
strongly with the photon-proton centre of mass energy, $W_{\gamma p}$.
The calculations are able to describe the measured magnitude of the
cross-sections for $\rho$, $\phi$ and $J/\psi$
production~\cite{epj:c14:213,Chekanov:2002rm,Aktas:2003zi}. They also
describe the dependence on $W_{\gamma p}$ for $J/\psi$ production for
$|t| > 5$~GeV~\cite{Aktas:2003zi}.

The transition of a quasi-real photon (helicity -1 or +1) to a vector
meson (helicity -1, 0 or +1) is characterised by three independent
helicity flip amplitudes. These may be extracted from measurements of
the angular distribution of the vector meson decay
products~\cite{Chekanov:2002rm}. In lowest order QCD, $N = 4$ for transversely
polarized mesons, and $N = 3$ for longitudinally polarized mesons. For
massless quarks this helicity amplitude is suppressed, since the
photon only couples to $q\bar{q}$ pairs of even chirality. 

However, there are approximations involved at this stage since
for light quarks the formation of the $q\bar{q}$ dipole cannot be
perturbatively calculated, and nor can the vector meson wave
function. With some flexibility introduced to allow for such effects,
the calculations can also describe the $|t|$ dependence of the
data~\cite{Poludniowski:2003yk}, as well as the helicity structure of $J/\psi$
production (where the charm mass adds reliability to the perturbative
approximations). They still fail to describe the helicity structure of
$\rho$ and $\phi$ production. Nevertheless, the agreement with the
cross-section measurements is further evidence that some understanding
of diffraction has been obtained within QCD.

High-$t$ photon production shares the advantages of vector meson
production, with the added advantage that the hard subprocess is
completely calculable in pQCD - there is no need for a Vector Meson
form factor. The disadvantage is that the cross-section is
smaller. However, given the enhancements expected from the large $s/t$
logarithms, the process was predicted to be observable at
HERA~\cite{Ginzburg:1996vq,*Evanson:1999zb,*Ivanov:1998jw,*Cox:1999kv,*Hoyer:2000mb},
and has indeed been observed~\cite{beckingham,*Favart:2003kw},
providing further compelling evidence, from photoproduction, for the
understanding of diffraction within QCD.

Understanding hard colour singlet exchange is not only of great
theoretical importance for our understanding of high energy QCD, but
is of importance for critical measurements at hadron colliders. In
particular, both forward ({\it i.e.} high rapidity) jet rates and a
suppression of QCD radiation between jets will play a key part in
measurements of $WW$ scattering (see for
example~\cite{atlastdr,*atlasptdr,*cmstdr} and references therein). This
is an important search channel for the Standard Model Higgs, and is
also the process where perturbative unitarity is violated within the
LHC energy range in the absence of a light Higgs.

Photoproduction has not has its final say on this topic. Preliminary
jet and $J/\psi$ data with higher statistics have already been
presented~\cite{ZEUS-prel-04-001,*ZEUS-prel-03-018}. In addition,
there is a possibility of extending the rapidity reach of the
jet measurement by using the ``forward plug
calorimeter''\cite{Bamberger:1999xz,Butterworth:1995jf} which was
installed during part of HERA I data taking at ZEUS.

\subsubsection{Forward rapidity gaps and forward particles}

The other class of rapidity gap events observed in photoproduction is
that in which the gap is not within the jet system, but is between the
jets and the forward (proton) direction. In fact, in these cases the
proton remains intact, or dissociates into a low-mass baryonic
state. Thus, an alternative signature for these processes is a tag of
a fast forward going proton or neutron. In these events, the moment
transferred across the gap is small, and so the coupling between the
exchanged colour singlet and the proton is not perturbatively
calculable. However, in a subset of events there can be a large
momentum scale involved in the scattering between the photon and the
exchanged colour singlet, and thus partonic language may be useful for
discussing some aspects of the event. In particular, one can assume
factorisation, and discuss conditional parton distributions in the
proton, {\it e.g.} diffractive parton distributions.  Often a further
assumption is made in which the exchange is treated as a particle with
a momentum distribution and its own parton densities (Regge
factorisation for diffraction, or the one pion exchange model for
forward neutron production). While these models are
known~\cite{Chekanov:2005vv}, or expected, to fail in detail, they
remain a useful approximation over a wide range of phase space.

\begin{figure}[htp]
\begin{center}
~\epsfig{file=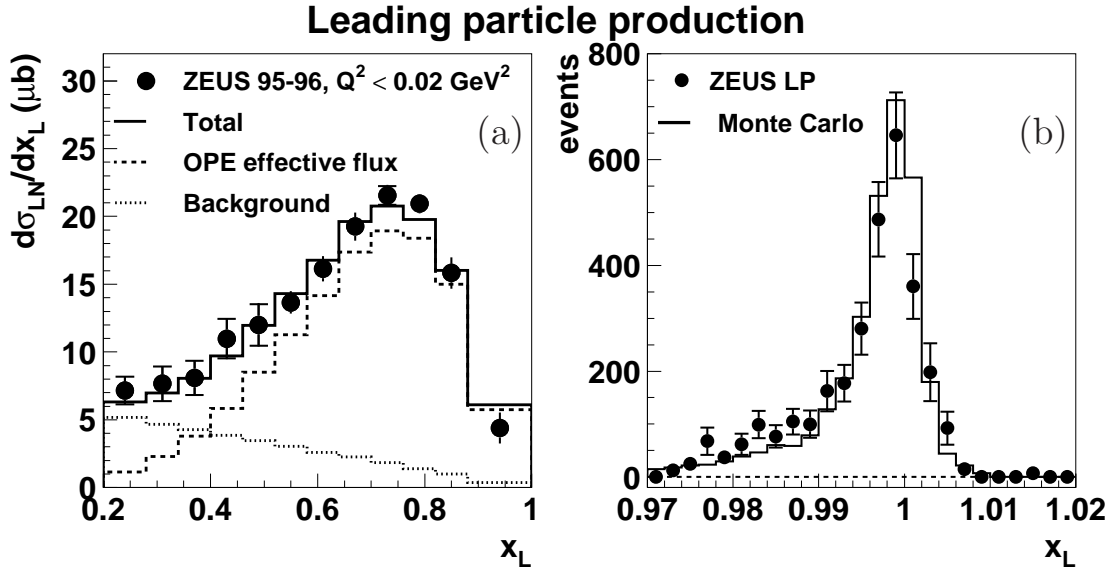,height=8cm}
\put(-240,176){\makebox(0,0)[tl]{\large (a)}}
\put(-35,176){\makebox(0,0)[tl]{\large (b)}}
\caption[*]{\label{fig:xl} The distribution $x_L$, of fraction of the
incoming proton momentum carried by a tagged forward 
(a) neutron~\cite{Chekanov:2002pf} 
and (b) proton~\cite{epj:c2:237}. Note the different horizontal scales!}
\end{center}
\end{figure}

The distribution of the momentum fraction $x_L$ for forward protons
and neutrons is shown in Fig.~\ref{fig:xl}. In the case of protons, a
clear peak due to diffractive processes is seen at $x_L \approx 1$. In
general, those processes where the proton or low-mass dissociated
system carries more than around 99\% of the initial proton energy are
dominated by diffraction (this may even be seen as a definition of
diffraction). In the case of neutrons, no such diffractive peak is
seen, though the distribution does peak at lower values, around $x_L =
0.7$, as expected from single pion exchange models.

Of particular interest is the use of photoproduction to study
factorisation and the breaking of factorisation in diffractive
processes. There is a factorisation theorem for diffraction in
DIS~\cite{Collins:1997sr,*Collins:2001ga}; that is, one can define and
use a set of diffractive parton densities for the proton and use them
in different diffractive DIS processes as one would with the normal
inclusive PDFs. However, this explicitly does not apply in
hadron-hadron interactions because of the possibility of soft
rescattering between the hadrons. At least naively then, one might
expect direct photoproduction to exhibit factorisation, since there is
only one incoming hadron, and resolved photoproduction to exhibit
factorisation breaking, since it looks like a hadron-hadron collision.
It is important to remember, however, that the terms resolved and
direct photoproduction are short hand for kinematic extremes, the no
unique separation between the two processes is possible (see
Section~\ref{sec:dual}). Specifically then, a lower fraction of
diffractive processes might be expected at low $\xgo$ compared to high
$\xgo$.

Dijet photoproduction in association with a forward rapidity gap has
been measured~\cite{cpaper:ichep2004:177,*cpaper:ichep2004:249} and
compared to LO Monte Carlo models as well as NLO QCD
calculations~\cite{Klasen:2001sg,np:b507:315,*np:b467:399,*np:b507:295}, 
using diffractive PDFs extracted from DIS data. The LO Monte Carlos, which 
in this case do not include any remnant-remnant interactions, describe the 
shape of the data distributions well without any need for a rescattering 
correction. However, in the NLO calculations, agreement is only seen if a 
rescattering correction of around 0.5 is applied. Additionally, it seems 
that this must be applied over all $\xgo$ values. The predicted 
suppression~\cite{Kaidalov:2003xf} is around 0.34 for hadronic events. The 
validity or not of QCD factorisation here is, however, complicated by the 
uncertainty in the diffractive PDFs~\cite{dpdfs:levy,*dpdfs:newman} and 
their application in dijet production in deep inelastic 
scattering~\cite{cpaper:eps2005:342}. Fits to different data sets lead 
to significantly different diffractive parton densities which when used 
to calculate the diffractive dijet cross section in deep inelastic 
scattering differ by a factor of two. Only by further understanding of 
the inclusive diffractive data and the techniques used can this be resolved 
and the question of factorisation breaking in jet production in hadron-hadron 
collisions, photoproduction and deep inelastic scattering be addressed.

The leading neutron energy spectrum in photoproduction (as well as
DIS) at HERA has been shown to be well described by single pion
exchange~\cite{epj:c6:587,Chekanov:2002pf}. Some of the most
interesting aspects of these interactions concern what is variously
described as rescattering, absorption, or the multiple interaction
probability.

\subsubsection{Survival and rescattering}

In Section~\ref{sec:ue}, underlying event models were discussed, with
a particular emphasis on multiple hard scatters. However, soft
scattering may also be eikonalised ({\it i.e.} repeated and
exponentiated). For instance, multiple soft Pomeron exchange preserves
unitarity in Regge based models which make no reference to partonic
scattering. The natural ``soft'' counterpart of multiple parton
scattering is soft rescattering, or absorption, which will affect the
forward proton or neutron rate even if not enough momentum transfer is
involved to throw particles into a detector and destroy a
(pseudo)rapidity gap.

In hard multiple scattering models in hadron-hadron collisions, the
presence of a hard scale is correlated to an increased probability of
multiple interactions, since both are more likely in central
collisions.

In inclusive photoproduction, where there is no hard scale, the photon
is dominantly hadronic, and the forward neutron rate is expected to be
reduced by soft rescattering. In DIS, two things are different. First
there is a hard scale, and second, the photon is small, (that is,
pointlike at the scale of the interaction). Obviously in DIS these are
both due to the large photon virtuality. However, the statements are
not equivalent.  Because it is pointlike, the photon has no remnant to
undergo rescattering. The forward neutron rate is correspondingly
higher, even though if the photon were considered as a hadron, one
might expect a hard scale to increase rescattering.

In charm photoproduction~\cite{Chekanov:2004dk}, there is again a hard scale,
provided by the charm mass. Some contribution from the hadronic photon
is expected to be present, but this is suppressed with respect to the
inclusive case, at least for inclusive dijet charm
events~\cite{epj:c6:67}. There is no evidence for rescattering in these
events, with the measured neutron fraction of $9 \pm 1\%$ being in
good agreement with the DIS rate, and inconsistent with the rate for
inclusive photoproduction. 

\begin{figure}[ht]
\begin{center}
~\epsfig{file=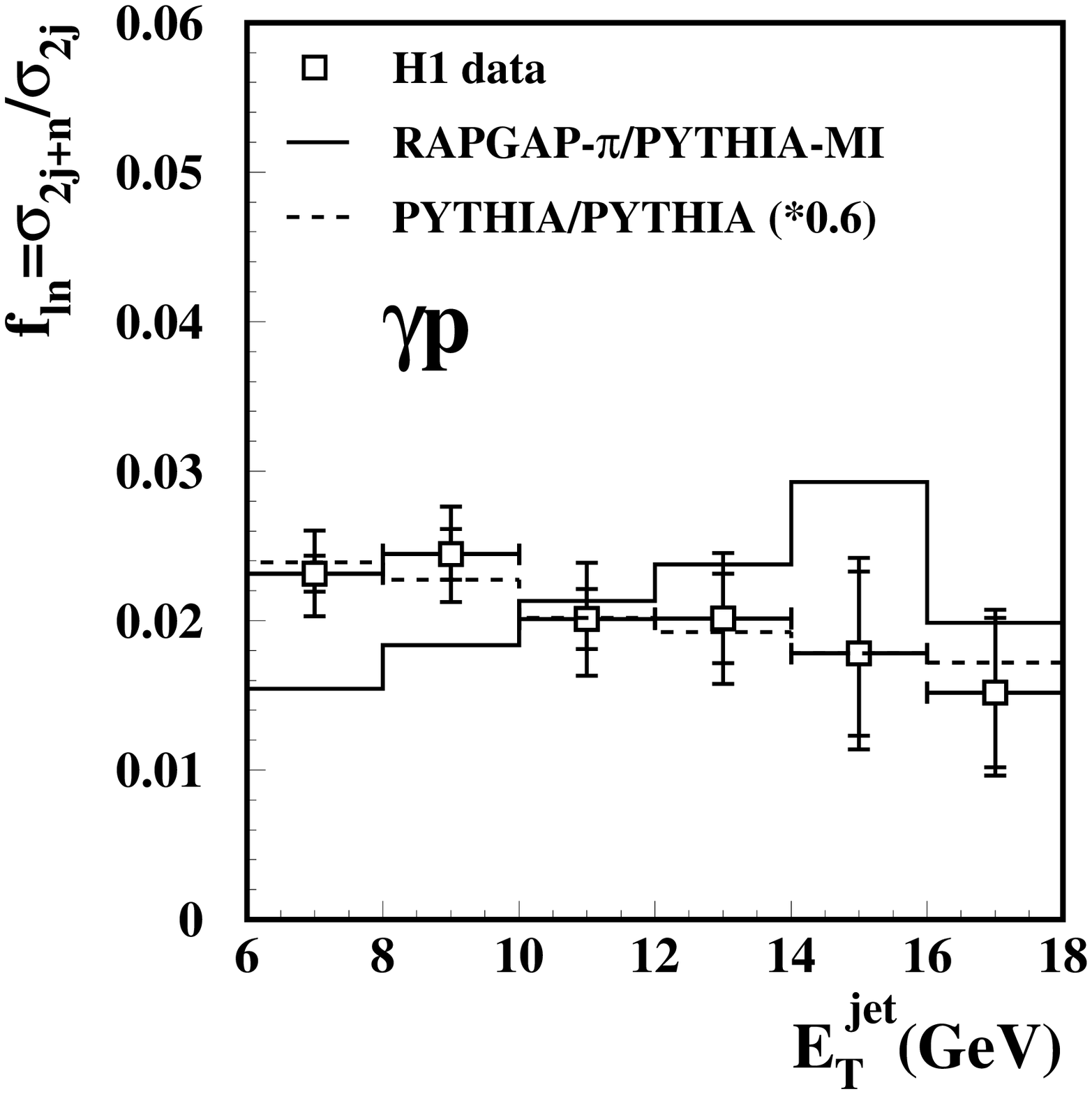,height=6.8cm}
\put(-27,182){\makebox(0,0)[tl]{\large (a)}}
\put(168,182){\makebox(0,0)[tl]{\large (b)}}
~\epsfig{file=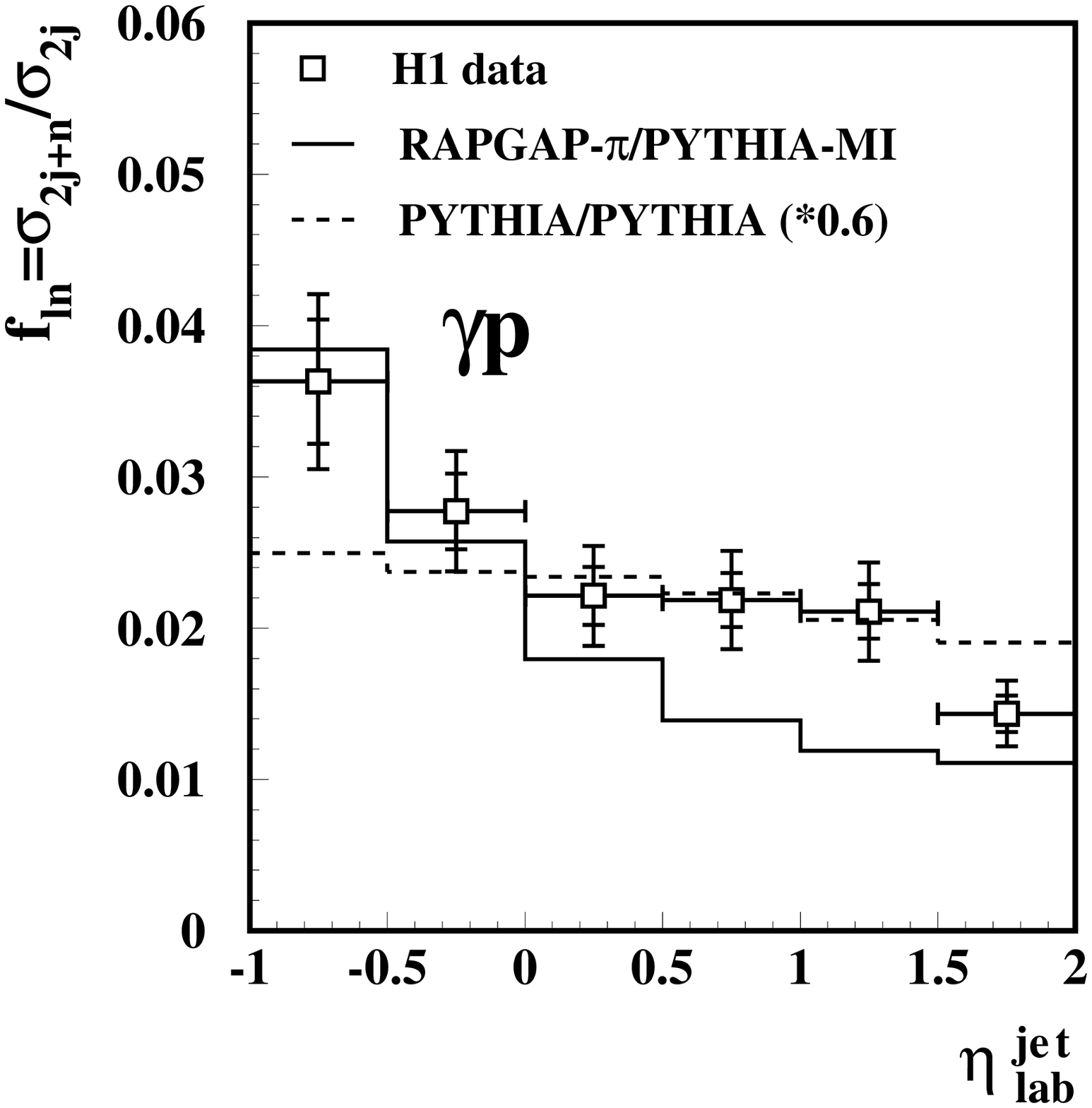,height=6.8cm}
~\epsfig{file=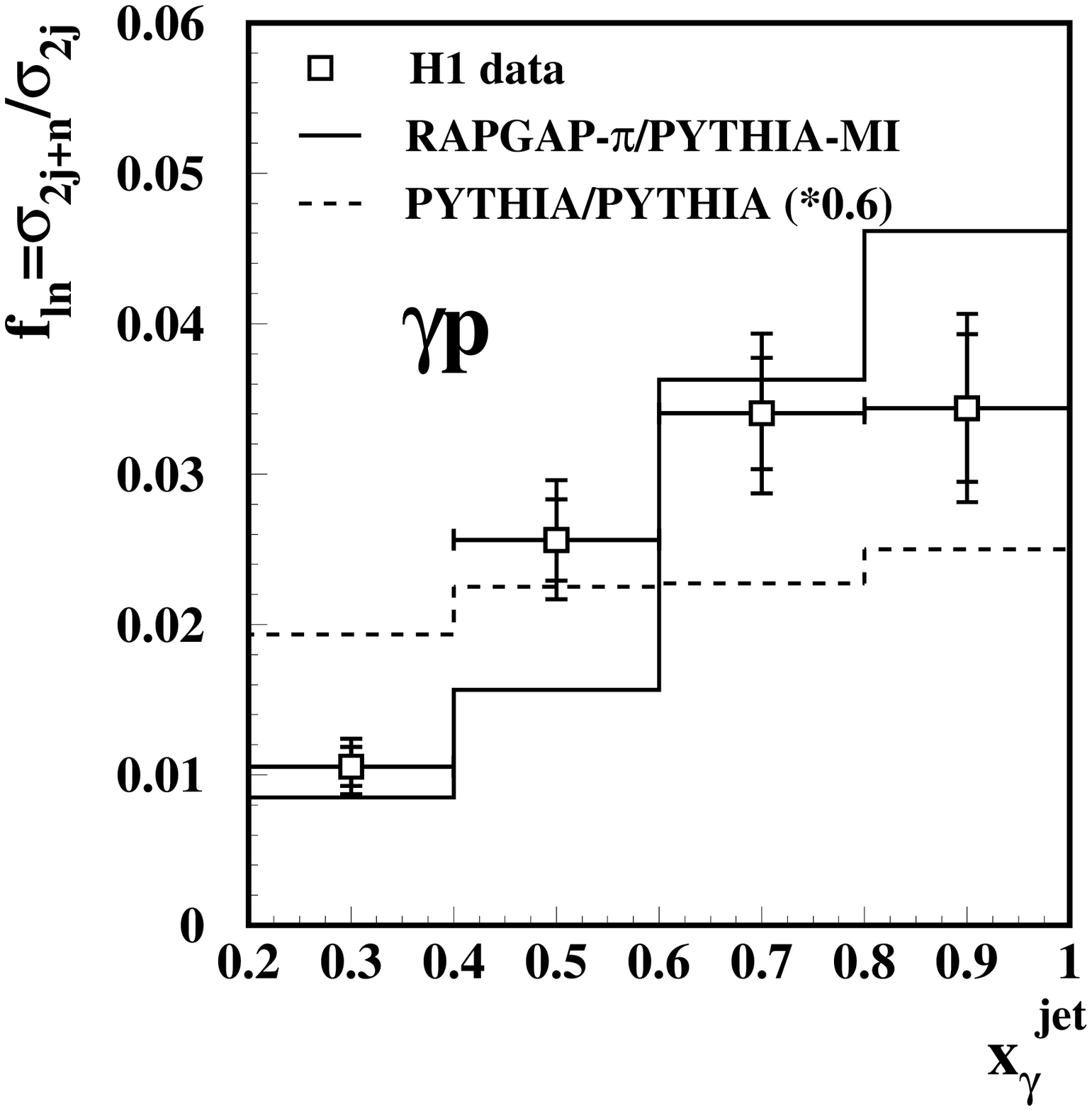,height=6.8cm}
\put(-27,182){\makebox(0,0)[tl]{\large (c)}}
\put(168,182){\makebox(0,0)[tl]{\large (d)}}
~\epsfig{file=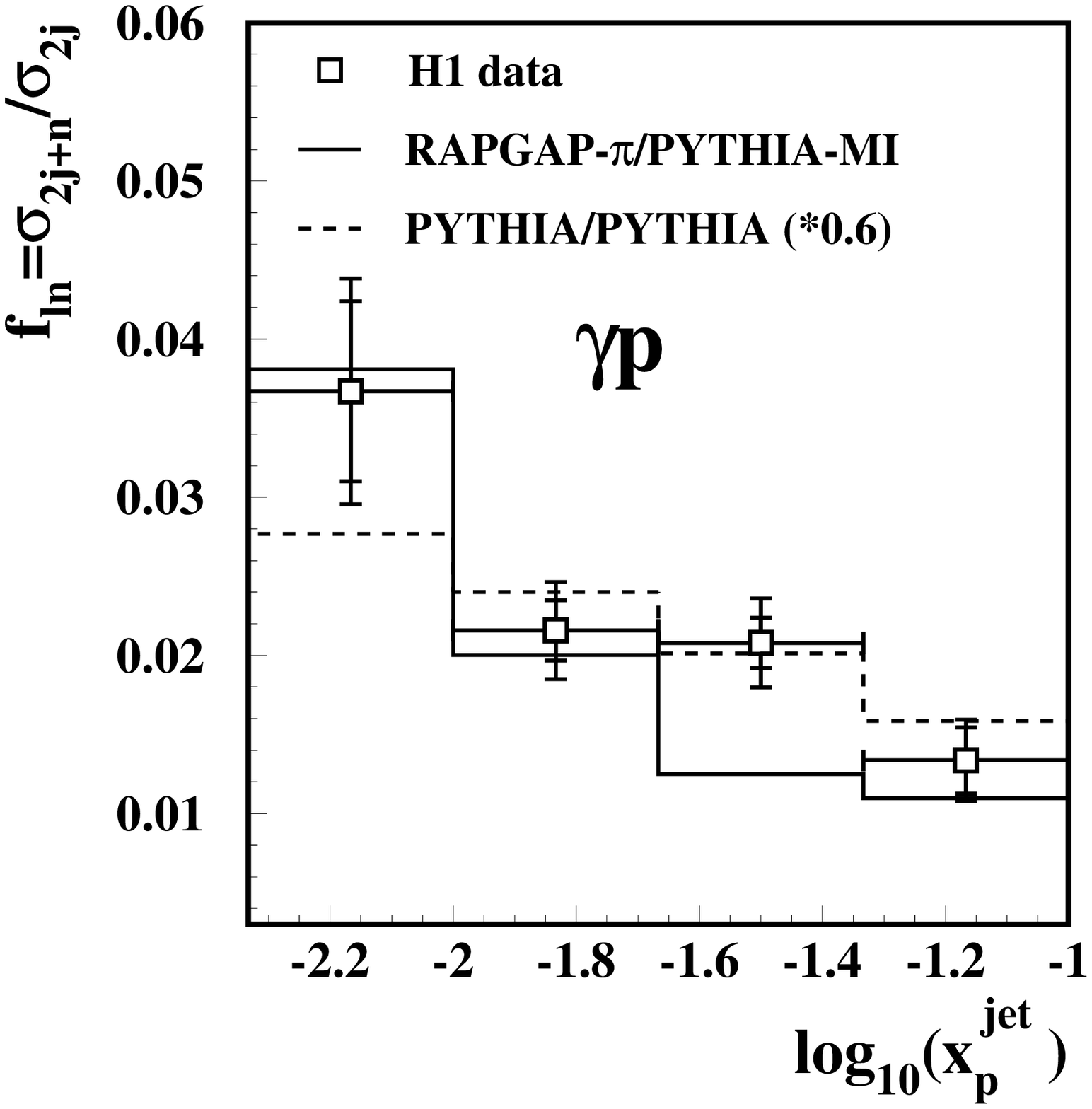,height=6.8cm}
\end{center}
\caption{The ratio of the cross section for dijet photoproduction with a
         leading neutron to that for inclusive dijet photoproduction as 
         a function of (a) $E_T^{\rm jet}$, (b) $\eta^{\rm jet}$, 
         (c) $x_\gamma^{\rm jet}$ and (d) $x_p^{\rm jet}$. The Monte Carlo 
         predictions are shown for {\sc Pythia} for inclusive production 
         and either {\sc Pythia} or {\sc Rapgap} with one-pion exchange 
         for neutron production.}
\label{fig:ln_jets}
\end{figure}

Finally one can consider dijet photoproduction. Here a hard scale is
present, but one can select between hadronic and pointlike photons
using the $\xgo$ variable.  Measurements of dijet photoproduction as a
function of $\xgo$ help clarify the situation. All jet cross-sections
self-evidently involve a hard scale, but the $\xgo$ variable allows
the ``size'' of the photon to be deconvoluted from this, in that
larger, hadronic photons are more likely to give rise to events with
low $\xgo$. The H1 measurement~\cite{epj:c41:273} of the ratio of
forward-neutron-tagged to inclusive dijet photoproduction in 
Fig.~\ref{fig:ln_jets} confirms
that, independent of the scale, there is a lower neutron rate in
resolved photoproduction events, as expected from rescattering models
in hadronic collisions.

One major source of interest in the measurements of forward proton and
neutron production, and the understanding of rescattering effects, is
that exclusive diffractive production may be an important search
channel for new physics at
LHC~\cite{Schafer:1990fz,*Bialas:1991wj,*Lu:1994ys,*Cudell:1995ki,*Khoze:1997dr,*Khoze:2000cy,*Khoze:2001xm,*DeRoeck:2002hk,*Cox:2003xp,*Boonekamp:2004nu,*Cox:2005if,*Boonekamp:2005yi}. The
phenomenological predictions require a good understanding of
diffractive processes, particularly diffractive PDFs and factorisation
breaking.  Comparison to high energy photoproduction data, as well as
to $p\bar{p}$ data, is a critical factor in developing and
demonstrating this understanding.

\section{Photoproduction and the search for new physics}
\label{sec:searches}

As discussed in Section~\ref{sec:intro}, photon beams generally derive
from electron or positron beams. Thus they carry some fraction the
lepton energy, and so in general do not compete in terms of sheer
centre-of-mass energy when it comes to searching for new physics.
However, in some specialised cases photon-proton interactions do have
an enhanced sensitivity to new physics which compensates for this
energy disadvantage. 

Of course, it is always possible that some new physics, or a model for
new physics, will arise which is expected to show up uniquely in
photoproduction due for instance to something unique in the
photon-quark coupling. Therefore it is worthwhile, at the highest
photoproduction energies, doing a general search for new signatures
regardless of the existing theoretical prejudices. An example of such
an analysis is the search for new heavy resonances (${\cal P}$) in the
channel $e^+p \ra e^+ {\cal P} X$ where the heavy resonance decays to
two jets~\cite{Chekanov:2001if}. Since the search is made in the
kinematic region $\q2 < 1~\gev2$, this corresponds to the
photoproduction of ${\cal P}$ from almost-real photons. The upper limit 
for the production of such a resonance was measured to be about 1~pb for 
masses up to 155~GeV.

\begin{figure}[ht]
\begin{center}
~\epsfig{file=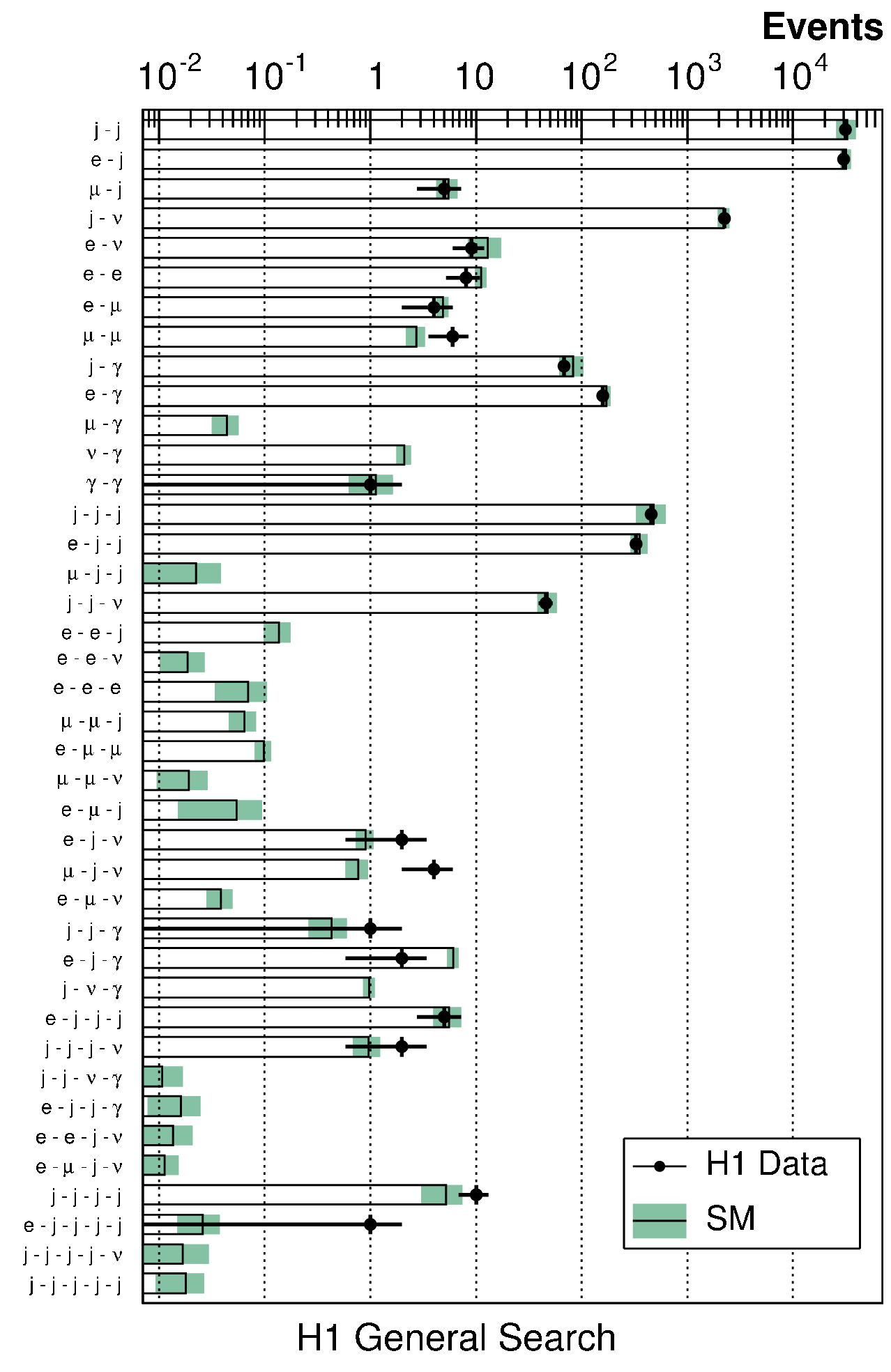,height=15cm}
\end{center}
\caption{Data (integrated luminosity of 117 pb$^{-1}$) compared with the 
         Standard Model expectation for all event classes with a Standard 
         Model expectation of more than 0.01 events.}
\label{fig:generic}
\end{figure}

Taking an even more general approach, the H1 collaboration have made a
general search for new phenomena in $ep$ scattering at
HERA~\cite{Aktas:2004pz} shown in Fig.~\ref{fig:generic}. This involves a 
systematic comparison of a
wide variety of signatures (leptons, jets, missing transverse energy)
with SM expectations, and thus includes high energy photoproduction as
well as DIS. Most event classes show good agreement with the SM, but
there is an excess of events containing an isolated muon, missing
transverse momentum and a jet. This corresponds to the observation,
for transverse momentum of the hadronic system, $\pt > 40$~GeV, of 3 events 
where the expected background is $0.54 \pm 0.11$, which was already reported in
a more specific analysis~\cite{pl:b561:241}. In this analysis there is
also an excess in the channel with an isolated electron instead of
muon (3 events for an expected background of $0.55\pm 0.12$). A similar
search has been carried out by ZEUS~\cite{}, where, disappointingly, no
such excess is seen. The $\tau$ channel has also been studied by
ZEUS~\cite{Chekanov:2003bf}, where one event is seen with an expected
background of $0.07\pm 0.02$.  The dominant contribution to SM
backgrounds quoted above is the photoproduction of $W$
bosons~\cite{np:b325:253}. This cross-section peaks at lower missing
$\pt$, however, where it is consistent with the observed number of
events.

\begin{figure}[htp]
\begin{center}
\vspace{0.8cm}
~\epsfig{figure=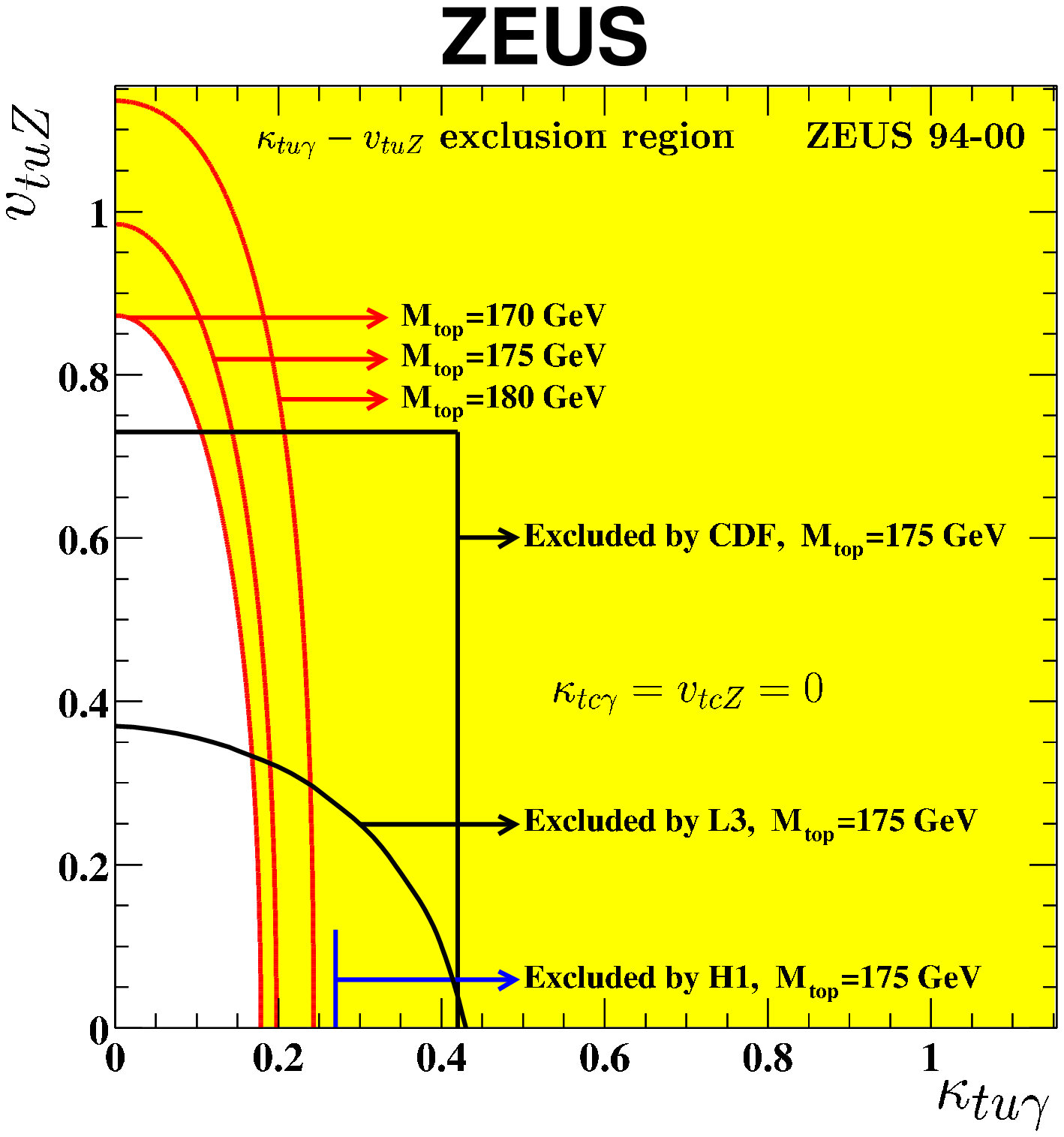,width=15cm,clip=}
\end{center}
\caption{Exclusion regions at $95\%$ CL in the $\kappa_{tu\gamma}-v_{tuZ}$ 
         plane for three values of $M_{\rm top}$ (170, 175 and 180~GeV) 
         assuming $\kappa_{tc\gamma}=v_{tcZ}=0$. The CDF, L3 and H1 exclusion 
         limits are also shown.}
\label{fig:stop}
\end{figure}

If the events are due to new physics, the initially favoured explanation 
would be single top quark 
production~\cite{np:b454:527,*pl:b426:393,*pr:d58:073008,*pr:d60:074015,*pl:b457:186},
arising from an anomalous $tu\gamma$ (flavour-changing
neutral-current) coupling, $\kappa_{tu\gamma}$. However, the lack of
an excess in the electron and muon channels from ZEUS sets a stringent
limit on such a coupling~\cite{Chekanov:2003yt}. Even H1 data, if the
observed excess is taken to be a statistical
fluctuation~\cite{Aktas:2003yd}, sets a limit on $\kappa_{tu\gamma}$
which is more stringent than Tevatron or LEP limits in the case that
the anomalous $Z$ coupling is small (Fig.~\ref{fig:stop}).
Supersymmetric-top production in R-parity-violating SUSY
models~\cite{Kon:1991ad,*np:b397:3,*Porod:1998yp} followed by either
R-parity violating two-body decay or R-parity conserving cascade
decays are another source of such signatures. Neither the SUSY or FCNC
case involves photoproduction as the signal process. Nevertheless, if
the excesses are confirmed in HERA-II data, and the observations of
the two experiments are consistent, there is no doubt other new
physics models will be proposed, and this represents perhaps the most
exciting frontier in high energy photoproduction. At the time of
writing, H1 have reported the observation of two further events, in the
electron channel, with hadronic $\pt > 40$~GeV from HERA II data, on an
expected background of around 1~\cite{cpaper:eps2005:637}.

One sometimes exotic area we have not considered in this review is
that of hadron spectroscopy.  Recently, several low-energy
photoproduction experiments reported evidence for 5-quark (pentaquark)
states~\cite{Nakano:2003qx,*Barth:2003es,*Stepanyan:2003qr,*Kubarovsky:2003fi,*Airapetian:2003ri}. Possible
candidates were also seen in DIS~\cite{Chekanov:2004kn},
proton-proton~\cite{Abdel-Bary:2004ts,*Alt:2003vb},
proton-nucleus~\cite{Aleev:2004sa}, 
neutrino-nucleus~\cite{Asratyan:2003cb} and
Kaon-nucleus~\cite{Barmin:2003vv} interactions, but several negative
results have also been published~\cite{Bai:2004gk,*Longo:2004gd,*Chekanov:2005at,*Abt:2004tz,*Airapetian:2004mi,*Antipov:2004jz} or presented. The only
{\it high-energy} photoproduction observation of such a
state~\cite{Aktas:2004qf} is less significant than the observation in
the same publication in DIS. The observation is contradicted in both
photoproduction and DIS by other measurements~\cite{epj:c38:29}
and to date remains unconfirmed.

Deuteron photoproduction has been measured, and limits set on the
photoproduction cross-section for other heavy stable charged
particles~\cite{Aktas:2004pq}. This, and other studies of identified
hadrons and resonances in DIS at HERA may indicate a potential for a
major contribution in this area from high-energy photoproduction as
well.

\section{Summary: past, present and future} 

In the past few years, high energy photoproduction has lead to major
advances in our understanding of the photon, the proton and QCD in
general. Substantial progress has been made, with a great synergy
between the HERA experiments and an active theory and phenomenology
community. 

In a review of resolved photoproduction in 1995~\cite{Drees:1995wh},
Drees and Godbole concluded with a list of open issues in hard
photon-proton scattering. It is interesting and fun to return to this
and see how much progress has really been made since that year, as
well to suggest what progress might be made in the next ten years. We
quote from their paper:

\begin{itemize}

\item
{\bf 1995:} {\it ``No complete NLO treatment of di--jet production
exists. Recall that one needs to measure the rapidities of both
high$-p_T$ partons/jets in a hard event in order to reconstruct the
Bjorken$-x$ variables.''}

{\bf 2005:} Several complete leading order programs are now
available(Section~\ref{sec:jetpheno}). They have been used extensively
in many of the results discussed
herein~(e.g. Sections~\ref{sec:pdfjets}, \ref{sec:nature_of_photon},
\ref{sec:proton_structure}, \ref{sec:as}). Of course, physicists are
never satsified. To make best use of the data, it would be great if
jet photoproduction processes could be made available within the Monte
Carlo programs which include NLO terms, which are now becoming
available for other
processes~\cite{Dobbs:2004qw}.

\item
{\bf 1995:} {\it ``The measured jet cross--sections should be extended
both in rapidity and in $p_T$. The former increases the sensitivity to
the interesting region of small $\xg$, while the latter should allow to
test theory cleanly, since a detailed understanding of the underlying
event (see below) is less crucial at high $p_T$, and differences
between parametrizations of photonic parton densities are small at
large $\xg$ and large momentum scale.''}

{\bf 2005:} Both of these have been achieved experimentally. Going to
high $\ETJ$ has indeed meant that model uncertainties can be reduced
to the point that jet cross-section may be used with confidence to
extract NLO parton densities for the proton
(Section~\ref{sec:proton_structure}). This kind of approach is the best
way to extract information on the photon structure as
well. Unfortunately, the need to stay a reasonably high $\ETJ$ means
it is unlikely that high energy photoproduction at HERA will ever
produce strong constrains at $\xg$ below 0.05 or so.

\item
{\bf 1995:} {\it ``It might be interesting to try to correlate
properties of the photonic remnant jet with those of the high$-p_T$
jets. In the usual ``nonperturbative + anomalous" description of the
hadronic photon, the nonperturbative component should always have a
remnant jet with very small $k_T$; this component is also
characterized by soft parton densities. In this picture one therefore
expects nontrivial correlations between $\xg$ and $k_T$.''}

{\bf 2005:} This is one where no data have been published since
1995 (Section~\ref{sec:res_dir}). However, the
anomalous and non-perturbative aspects of the photon structure are
elucidated somewhat by studies of dijet and charm production from
virtual photons~(Section~\ref{sec:nature_of_photon}).

\item
{\bf 1995:} {\it ``Studies of heavy flavour production hold great
potential. We do not think it very interesting to try to derive total
cross--sections from measurements covering only a limited region of
phase space, which contains only a small fraction of all produced
heavy quarks. It might be more fruitful to attempt to extract the
resolved photon contribution, which is sensitive to the as yet poorly
constrained gluon density in the photon. At high $p_T$, ``excitation"
contributions from the charm in the photon have to be taken into
account. An important open problem is the fragmentation of rather soft
(low$-p_T$) charm quarks, which contribute most to the total charm
cross--section. Theoretical predictions are more reliable for $b$
production, but it might be difficult to find a clean signal.''}

{\bf 2005:} On this issue, Drees and Godbole were particularly
prescient. The field of heavy flavour photoproduction has exploded
since 1995 (see Section~\ref{sec:hf}), and one of the main reasons has
been the fact that sophisticated programs were developed which could
make QCD calculations of measured cross-sections in realistic
kinematics regions, rather than being limited to total cross-sections. 
In addition, despite the lower cross-section, good
measurements of beauty production have been made, and these have also
benefitted from similar calculational advances.  The result is a
vastly improved confidence in our understanding of charm and beauty
production in QCD.

\item
{\bf 1995:} {\it ``The production of direct photons is by now quite
well understood, although an NLO calculation of photon + jet
production would certainly be welcome.  Realistic background studies
are also needed, but can presumably only be performed by members of
HERA experiments.''}

{\bf 2005:} NLO calculations of $\gamma$+jet have indeed been made
(Section~\ref{sec:jetpheno}) as have further measurements at HERA
(Section~\ref{sec:pdfothers}). These have provided a convincing test
of factorisation, as well as information on the effective transverse
momentum of quarks in the proton~\cite{pl:b511:19}.

\item
{\bf 1995:} {\it ``In spite of the recent NLO calculation \cite{Kramer:1994zi}
of direct $J/\psi$ production in the colour singlet model, much needs
to be done here: The resolved photon contribution is only known to
leading order in the colour singlet model.  Nothing is known about the
contribution from the colour octet component of the wave function of
the $J/\psi$~\cite{Bodwin:1994jh}, which is accessible to resolved photons
already in LO. In addition, there are contributions at high $p_T$
coming from charm and gluon fragmentation.''}

{\bf 2005:} Inelastic $J/\psi$ photoproduction has been measured, and
calculated, with increasing precision
(Section~\ref{sec:ineljpsi}). Certainly something is now known about
colour-octet contributions, but more precision in the calculations,
including NLO terms, is highly desirable.  On the issue of high $p_T$
production, fragmentation production may still have an effect in high
$\pt$ photoproduction data from HERA II, but no clear sign of it is
seen in current data.

\item
{\bf 1995:} {\it ``It is important to test our understanding of the
hadronic photon in as many different channels as possible. The
production of Drell--Yan lepton pairs, two photon final states, and
associate $J/\psi + \gamma$ final states are all plagued by rather
small cross--sections, but this should at least partly be compensated
by the cleanliness of the final states.''}

{\bf 2005:} Thus far in most of these particular cases, we are still
waiting for data (although inelastic photoproduction of $\psi^\prime
\rightarrow J/\psi + \gamma$ has been
measured~\cite{Chekanov:2002at}). Towards the end of HERA II there
should be enough luminosity to measure Drell-Yan lepton pairs
(Section~\ref{sec:pdfothers}). Several other interesting final states,
not obvious in 1995, have emerged. These include the whole class of
rapidity gap events, both forward gaps and gaps between jets, which
were measured in photoproduction very shortly after the 1995 review
(Section~\ref{sec:eflow_rapidity_gaps}). Along with high-$|t|~J/\psi$
photoproduction and forward neutron production, these processes have
led to great advances in the understanding of high energy, low-$x$ QCD
and related areas.

\end{itemize}

In addition to the above areas specific to photoproduction, several of
the more generic areas highlighted by Drees and Godbole (minimum bias
events, multiple parton interactions...) have seen major advances
during the past decade. 

As to the where the next photoproduction measurements will come from;
HERA II still has much to say. Data taking is currently planned to
stop in 2007, and data analysis is likely to be complete within a few
years after that date. Following that, the future is less clear. At
the LHC, the photoproduction of heavy quarks and vector
mesons~\cite{Baron:1993nk,*Greiner:1994db,*Klein:2002wm,*Klein:2003vd},
including even $t\bar{t}$ pairs~\cite{Klein:2000dk}, in peripheral
proton-proton or heavy ion collisions will have a significant 
cross-section and may provide valuable information on high-density, low-x
QCD. Beyond that, the only high-energy photoproduction for the
forseeable future seems likely to be that from ultra-high-energy
cosmic rays.

There is a necessary arbitrariness in choosing a particular process to
review, and photoproduction in particular touches upon a wide variety
of topics in high-energy physics. The inclusion of jet photoproduction
in QCD fits for the proton structure represents something of a
``coming of age'' for the area, showing as it does that our
understanding of these processes is comparable now to our knowledge of
deep inelastic scattering. This clearly fits as well in a discussion
of proton structure as in a review of photoproduction. Hopefully the
reader is convinced by now that diffraction and low-$x$ physics, heavy
flavour physics and searches for new physics are all areas on which
high-energy photoproduction has an impact. Because we learn about
these areas of physics, photoproduction data has value for physicists
preparing for data analysis at the LHC. The nature and high energy
reactions of the photon itself are of huge fundamental interest; they
also have an impact on astrophysics and physics at the ILC. The study
of the interactions of the photon remains one of the more fruitful
areas within high-energy physics.

\section*{Appendix A: Monte Carlo models used in this review}

Most of the Monte Carlo curves shown in the body of the text have been
remade by the authors using the {\sc HZTool}~\cite{Bromley:1995np}
package, released as part of CEDAR~\cite{Butterworth:2004mu}. Further models
and comparisons are available from the CEDAR/JetWeb pages.

For these MC distributions, unless otherwise stated explicitly, the
proton PDF is CTEQ5 (LO)~\cite{epj:c12:375} and the photon PDF is GRV
(LO)~\cite{pr:d45:3986}.

The standard {\sc Herwig}~\cite{jhep:01:010,*Corcella:2002jc} curves use:\\ {\sc Herwig
6.507}, IPROC=15000 for direct photoproduction and 11500 for resolved,
12400 (hard colour singlet exchange).  The underlying event is
simulated using {\sc Jimmy}~\cite{jimmy}, with PTJIM=3.0 GeV, and the
JMRAD(73)=JMRAD(75)=2.13 GeV. For the colour singlet exchange, the
changes OMEGA0=0.48 and ASFIXD=0.18 were made. For the photon remnant
curve with intrinsic $\kt$ (Fig.~\ref{fig:photon_remnant}), PTRMS was set to 1 GeV.

The standard {\sc Pythia}~\cite{cpc:135:238} curves use:\\ {\sc Pythia
6.206}, MSTP(14) = 30 (mixture of resolved and direct QCD processes).
To obtain the ``incoherent'' model of Fig.~\ref{fig:3jet}, the
following parameters were changed from their defaults: 
PARP(67)=4.0, MSTP(62)=2, MSTP(63)=2, MSTP(67)=0, 
MSTJ(42)=1,\\ MSTJ(46)=1, MSTJ(50)=0.




{
\def\bibname{\Large\bf References}
\def\refname{\Large\bf References}
\pagestyle{plain}

\bibliographystyle{./BiBTeX/bst/l4z_default}

{\raggedright
\bibliography{./BiBTeX/user/syn.bib,%
              ./BiBTeX/bib/l4z_articles.bib,%
              ./BiBTeX/bib/l4z_books.bib,%
              ./BiBTeX/bib/l4z_conferences.bib,%
              ./BiBTeX/bib/l4z_h1.bib,%
              ./BiBTeX/bib/l4z_misc.bib,%
              ./BiBTeX/bib/l4z_old.bib,%
              ./BiBTeX/bib/l4z_preprints.bib,%
              ./BiBTeX/bib/l4z_replaced.bib,%
              ./BiBTeX/bib/l4z_temporary.bib,%
              ./BiBTeX/bib/l4z_zeus.bib}}
}
\vfill\eject





\end{document}